\documentclass[twocolumn,prb,aps,showpacs,superscriptaddress,floatfix]{revtex4}
\usepackage{graphicx}
\usepackage{dcolumn}
\usepackage{bm}
\usepackage{amsmath}
\usepackage{amsfonts}

\begin{document}

\title{Control of electron spin decoherence caused by electron-nuclear spin dynamics in a quantum dot}
\author{Ren-Bao Liu}
\affiliation{Department of Physics, University of California San Diego, La Jolla, California 92093-0319}
\affiliation{Department of Physics, The Chinese University of Hong Kong, Shatin, N.T., Hong Kong, China}
\author{Wang Yao}
\affiliation{Department of Physics, University of California San Diego, La Jolla, California 92093-0319}
\author{L. J. Sham}
\affiliation{Department of Physics, University of California San Diego, La Jolla, California 92093-0319}
\date{\today}

\begin{abstract}
Control of electron spin decoherence in contact with a mesoscopic
bath of many interacting nuclear spins in an InAs quantum dot is
studied by solving the coupled quantum dynamics. The nuclear spin
bath, because of its bifurcated evolution predicated on the
electron spin up or down state, measures the which-state
information of the electron spin and hence diminishes its coherence. The
many-body dynamics of nuclear spin bath is solved with a
pair-correlation approximation. In the relevant timescale,
nuclear pair-wise flip-flops, as elementary excitations in the
mesoscopic bath, can be mapped into the precession of
non-interacting pseudo-spins. Such mapping provides a geometrical
picture for understanding the decoherence and for devising
control schemes. A close examination of nuclear bath dynamics
reveals a wealth of phenomena and new possibilities of
controlling the electron spin decoherence. For example, when the
electron spin is flipped by a $\pi$-pulse at $\tau$, its
coherence will partially recover at $\sqrt{2}\tau$ as a
consequence of quantum disentanglement from the mesoscopic bath.
In contrast to the re-focusing of inhomogeneously broadened
phases by conventional spin-echoes, the disentanglement is
realized through shepherding quantum evolution of the bath state
via control of the quantum object. A concatenated construction of
pulse sequences can eliminate the decoherence with arbitrary accuracy,
with the nuclear-nuclear spin interaction
strength acting as the controlling small parameter.
\end{abstract}

\maketitle

\section{Introduction}
\label{S_Introduction}

A quantum system, unlike a classical one, can be in a coherent
superposition of constituent states. This coherence is the
wellspring of quantum properties and key to quantum technology.
The contact of a quantum object with a macroscopic system causes
loss of state
coherence.\cite{Zurek_decoherence_RMP,Decoherence_classicalWorld,Schlosshauer_decoherence}
Advances towards quantum technology have effectively substituted the
macroscopic environments by mesoscopic ones. Usually, the contact
interaction between the quantum object and a ``particle'' of the
bath weakens with increasing the bath size (defined by
the number of particles, $N$), while the interaction between
particles inside the bath is independent of the bath size. When
the bath size is in the mesoscopic regime such that the
object-bath interaction dominates their interaction with the rest
of universe, the quantum object and the mesoscopic bath evolves
as a closed system in the relevant timescales. Thus, a proper
description of the quantum object in such a mesoscopic bath is given
by the full quantum mechanical solution of the coupled object-bath
evolution,\cite{Yao_Decoherence,Yao_DecoherenceControl} as
opposed to the semiclassical treatment in conventional decoherence
studies.\cite{Anderson_Spectral,Schulten:1978}

Spins of single electrons confined in semiconductor quantum dots
are paradigmatic systems in mesoscopic physics\cite{ImryMesoscopic,Leggett_Two_state_Dynamics} and in spin-based
quantum technology\cite{Loss_QDspinQC,Imamoglu_CQED_Spin,SpintronicsQC}. Spin decoherence
is a main limiting factor to quantum properties and has been extensively
studied both in theories\cite{Merkulov_decoherence_nuclei,
Loss_decoherence_nuclei,Khaetskii_nuclear,Espin_HF_1_Loss,Semenov_Nuclear,
Nazarov_Spinflip1,Reinecke_T1_phonon,Loss_SpinT1T2_phonon,Semenov_SpinT2_phonon,
deSousa_Spectral1,deSousa_Spectral2,Shenvi_scaling,Shenvi_bounds,Witzel_Quantum1,
Yao_Decoherence,Espin_HF_Hu,Yao_DecoherenceControl,Witzel_Quantum2}
and in experiments.\cite{Fujisawa_T1,Readout_spin_Kouwenhoven,Finley_spin_memory,Marcus_T1,
Marie_NucleiQD,Imamoglu_cooling,Gurudev,Gammon_opticalPump,
Kouwenhoven_singlet_triplet,Marcus_T2,Greilich_lock}
It has been well established by theories\cite{Nazarov_Spinflip1,Reinecke_T1_phonon,Loss_SpinT1T2_phonon,Semenov_SpinT2_phonon}
and experiments\cite{Fujisawa_T1,Readout_spin_Kouwenhoven,Finley_spin_memory,Marcus_T1}
that the electron spin decoherence caused by phonon scattering is negligible at a temperature lower than a few Kelvins.
The relevant bath at low temperature then is a spin bath\cite{Stamp_spinbath} composed of lattice nuclear spins in a
quantum dot (QD). The electron spin is coupled to a nuclear spin through the contact hyperfine interaction
with magnitude inversely proportional to the total number $N$ of nuclei in the QD.
For the QD size of interest, the hyperfine coupling is much stronger than the mutual interaction
between nuclear spins. Therefore, a mesoscopic bath consisting of all nuclear spins within the QD
(i.e., in direct contact with the electron) is identified. Electron spin decoherence
at low temperatures is determined by the quantum dynamics of the mesoscopic spin system.

Under a moderate to strong magnetic field ($B\gtrsim 0.1$~T), the
Zeeman energy of the electron is by orders of magnitude larger
than the hyperfine coupling, so that the longitudinal electron
spin relaxation by off-diagonal hyperfine interaction is
virtually suppressed.\cite{Yao_Decoherence} The Hamiltonian of the
electron-nuclear spin system can then be reduced to the form
$\hat{H}=\sum_{\pm}|\pm\rangle \hat{H}_{\pm} \langle\pm|$ where
$|\pm\rangle$ denote the electron spin eigenstates in the external
field and $\hat{H}_{\pm}$ are the nuclear bath Hamiltonians
depending on the electron spin states. The pure dephasing
(decoherence) of the electron spin is caused by the
electron-nuclear spin entanglement\cite{Schliemann:2002} as
follows. Let the electron-nuclear spin system start from a
product (i.e. unentangled) state, $\left(C_+ |+\rangle+
C_-|-\rangle\right)\otimes |{\mathcal J}\rangle$. The nuclear
spin state $|{\mathcal J}\rangle$ would be driven by the
Hamiltonians $\hat{H}_{\pm}$ to the states $|{\mathcal
J}^{\pm}(t)\rangle$ corresponding to the electron states
$|\pm\rangle$, respectively. Thus the electron-nuclear spins
would evolve into an entangled state, $C_+|+\rangle\otimes
|{\mathcal J}^+(t)\rangle+ C_-|-\rangle\otimes |{\mathcal
J}^-(t)\rangle$. The off-diagonal element of the reduced density
matrix of the electron spin $\rho^e_{+,-}(t)= C^*_{-} C_{+}
\left\langle\mathcal{J}^{-}(t)|\mathcal{J}^{+}(t)\right\rangle$
measures the electron spin coherence. Therefore, the bifurcated
evolution of the nuclear bath leads to the loss of electron spin
coherence. From the viewpoint of quantum measurement, the
electron spin states are registered by different nuclear bath
states and a measured quantum object sustains no coherence in the
measurement basis.\cite{SchrodingerCat,Griffiths_consistent}

\begin{figure}[t]
\begin{center}
\includegraphics[width=7cm]{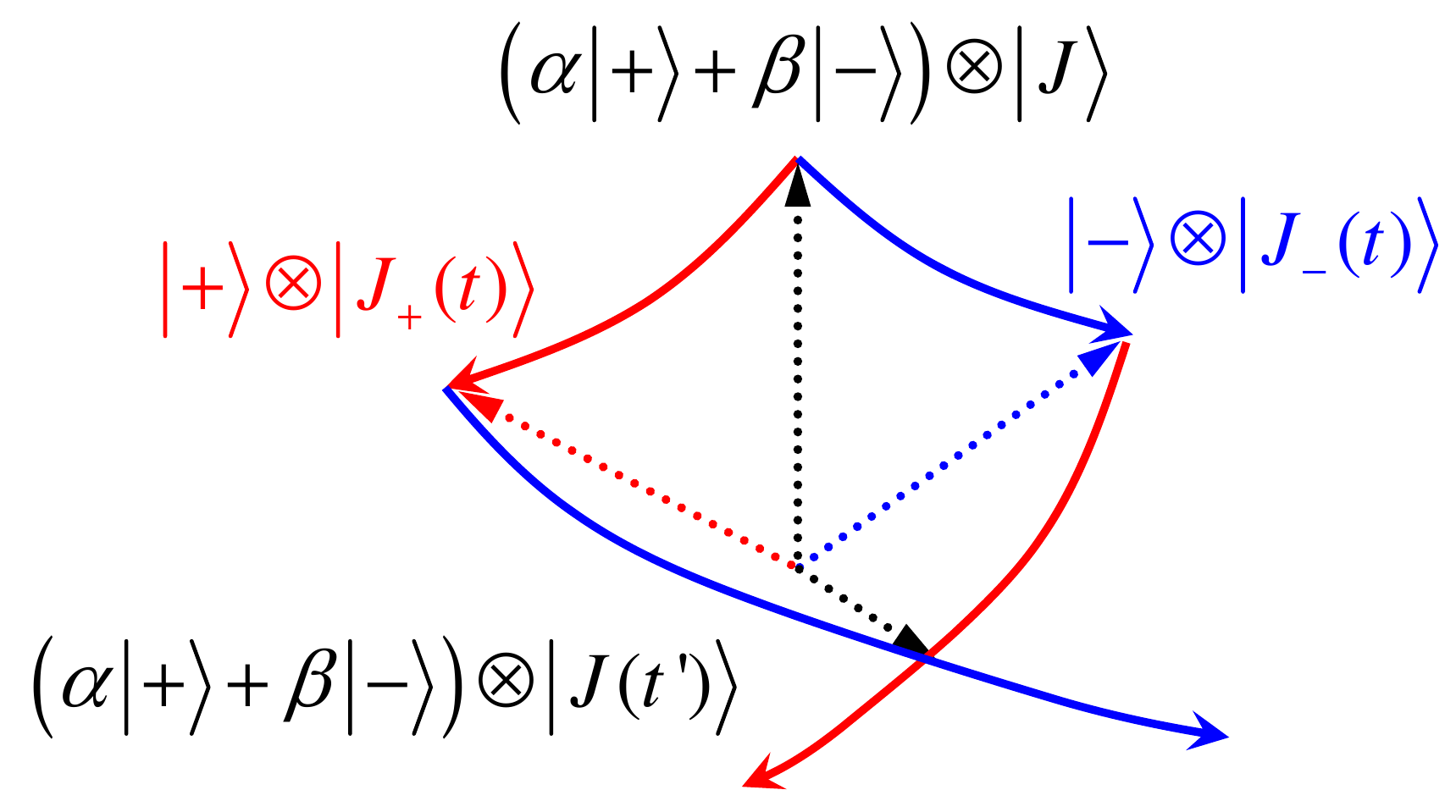}
\end{center}
\caption{Schematic illustration of the bifurcation of
the bath state evolution conditioned on the electron spin up or down state and
the exchange of the evolution direction of the bath pathways when the
electron spin is flipped.} \label{Fig_pathways}
\end{figure}

Within the timescale when the irreversible leakage of the quantum coherence
of a mesoscopic bath into the macroscopic environment is negligible,
it is still possible to control the mesoscopic bath dynamics\cite{Lukin_NucleiControl}
and hence the decoherence of the quantum object embedded in the bath.\cite{Yao_DecoherenceControl}
As illustrated in Fig.~\ref{Fig_pathways}, the bath evolution pathways in the
Hilbert space corresponding to opposite electron spin states will exchange their evolution directions
when the electron spin is flipped:
\begin{eqnarray}
 && \left(C_+ |+\rangle+ C_- |-\rangle\right)\otimes |{\mathcal J}\rangle \nonumber \\
 &\longrightarrow&
C_+ |+\rangle\otimes|{\mathcal J}^+(t)\rangle+ C_- |-\rangle\otimes |{\mathcal J}^-(t)\rangle \nonumber \\
& \longrightarrow & C_+ |-\rangle\otimes|{\mathcal
J}^+(t')\rangle+ C_- |+\rangle\otimes |{\mathcal J}^-(t')\rangle.
\nonumber
\end{eqnarray}
Thereafter, the two bath pathways
$|{\mathcal J}^+(t')\rangle$ and $|{\mathcal J}^-(t')\rangle$ intersect at some later time, i.e.,
$|{\mathcal J}^+(t')\rangle\cong |{\mathcal J}^-(t')\rangle=|{\mathcal J}(t')\rangle$.
At the intersection, the electron spin is disentangled from the bath and
its lost coherence is recovered (by the controlled erasure of the quantum
information registered in the bath). Such a discovery of the reversal
of coherent dynamics can be traced back to the early theoretical and experimental
findings on the Loschmidt echo in NMR.\cite{Pines_DD,Ernst_polarizationecho}

The recovery of the lost coherence (recoherence) by quantum disentanglement is
fundamentally different from the conventional spin echo realized by the
refocusing of the random phase in an inhomogeneously broadened
ensemble. In general, the disentanglement can occur at a time different
from the spin echo time. For example, by a single flip of the electron
spin at $t=\tau$, there will be a prominent recoherence
at a magic time $\sqrt{2}\tau$ as a consequence of disentanglement\cite{Yao_DecoherenceControl}
while the conventional spin echo would occur at $2\tau$. When the electron spin is observed in ensemble
measurement (realized either by using many similar QDs or by cycling
measurements on a single dot), the disentanglement-induced recoherence will be
concealed by the inhomogeneous broadening unless it is forced to take place at a
spin echo time by proper design of pulse sequences.

The recoherence by disentanglement can be exploited as a
coherence protection scheme in lieu of the dynamical decoupling
schemes developed for NMR spectroscopies\cite{Haeberle,Pines_DD,Slichter}
and recently for quantum computation.\cite{DD_VL,DD_VKL1,Viola_Random,Lidar_CDD,Lidar_DD_longpaper,Kern_DD}
While the dynamical decoupling schemes seek to eliminate the
object-bath interaction through rapid rotation of a quantum object,
the disentanglement focuses on controlling the wavefunction evolution
of the bath and in general does not lead to a vanishing object-bath
coupling. It will be shown that in the disentanglement scheme, the controlling
small parameter for coherence protection is determined by the interactions
within the bath instead of the object-bath coupling, in contrast with the
dynamical decoupling schemes.

This paper is organized as follow:
Sec.~\ref{S_Theory} describes a specific model system (a self-assembled InAs QD)
and summarizes the theoretical ingredients for solving the electron-nuclear spin dynamics.
The free-induction decay (FID) in single-system dynamics (without ensemble
average) and the Hahn echo in ensemble dynamics are studied in Sec.~\ref{S_FID} and
Sec.~\ref{S_echo}, respectively. Sec.~\ref{S_disentanglement} presents
a close examination of the single-system dynamics in the single-pulse Hahn-echo
configuration, revealing a coherence recovery due to disentanglement at
a magic time which would otherwise be invisible in ensemble-averaged signals.
Sec.~\ref{S_concatenation} studies the disentanglement under control of
pulse sequences, in particular, the concatenated sequences.
Further discussions including the comparison between the quantum
disentanglement and the dynamical decoupling are presented in the
summary Section. Some technical details are given in the Appendices.

\section{Theory}
\label{S_Theory}

\subsection{The Model}
\label{SS_Model}

\begin{figure}[t]
\begin{center}
\includegraphics[width=7cm]{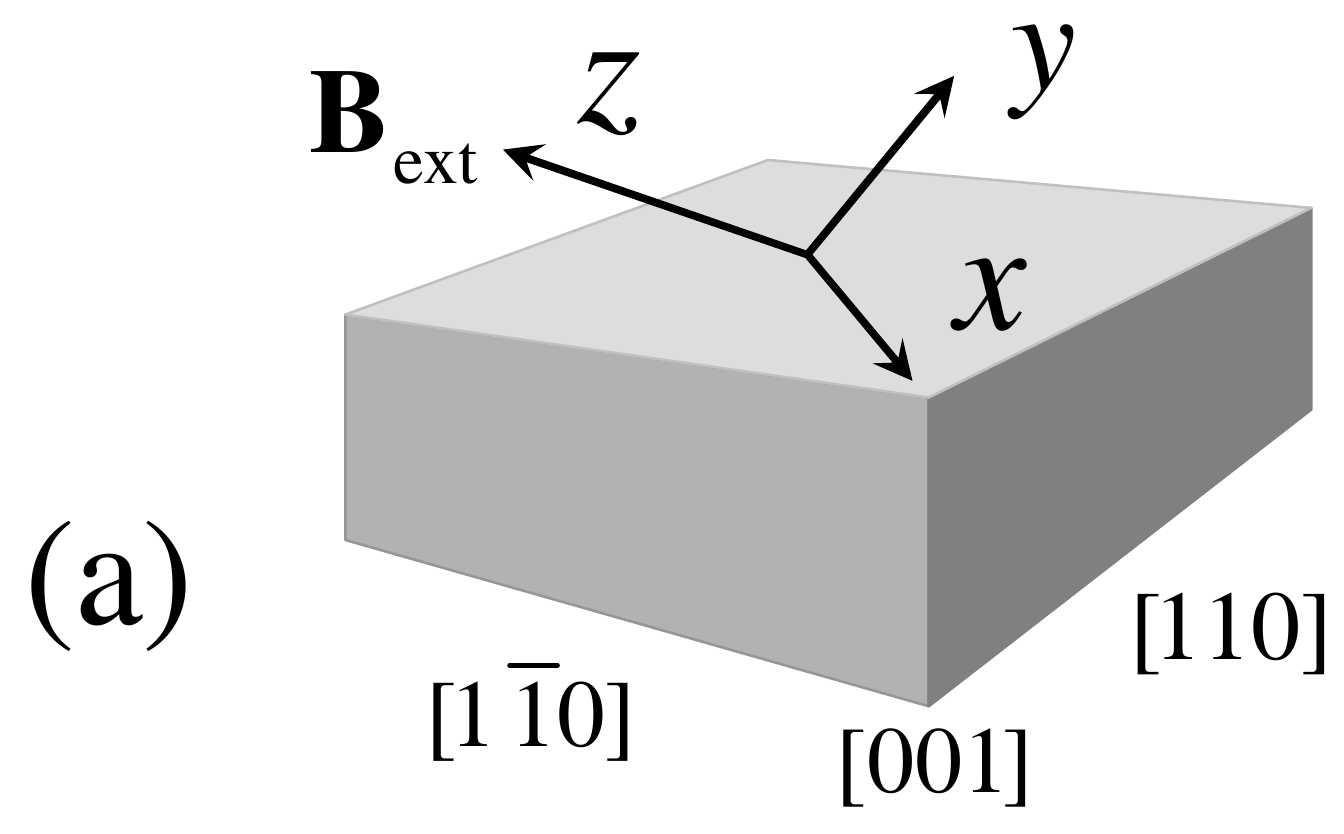}
\\
\vskip 0.5cm
\includegraphics[width=7cm]{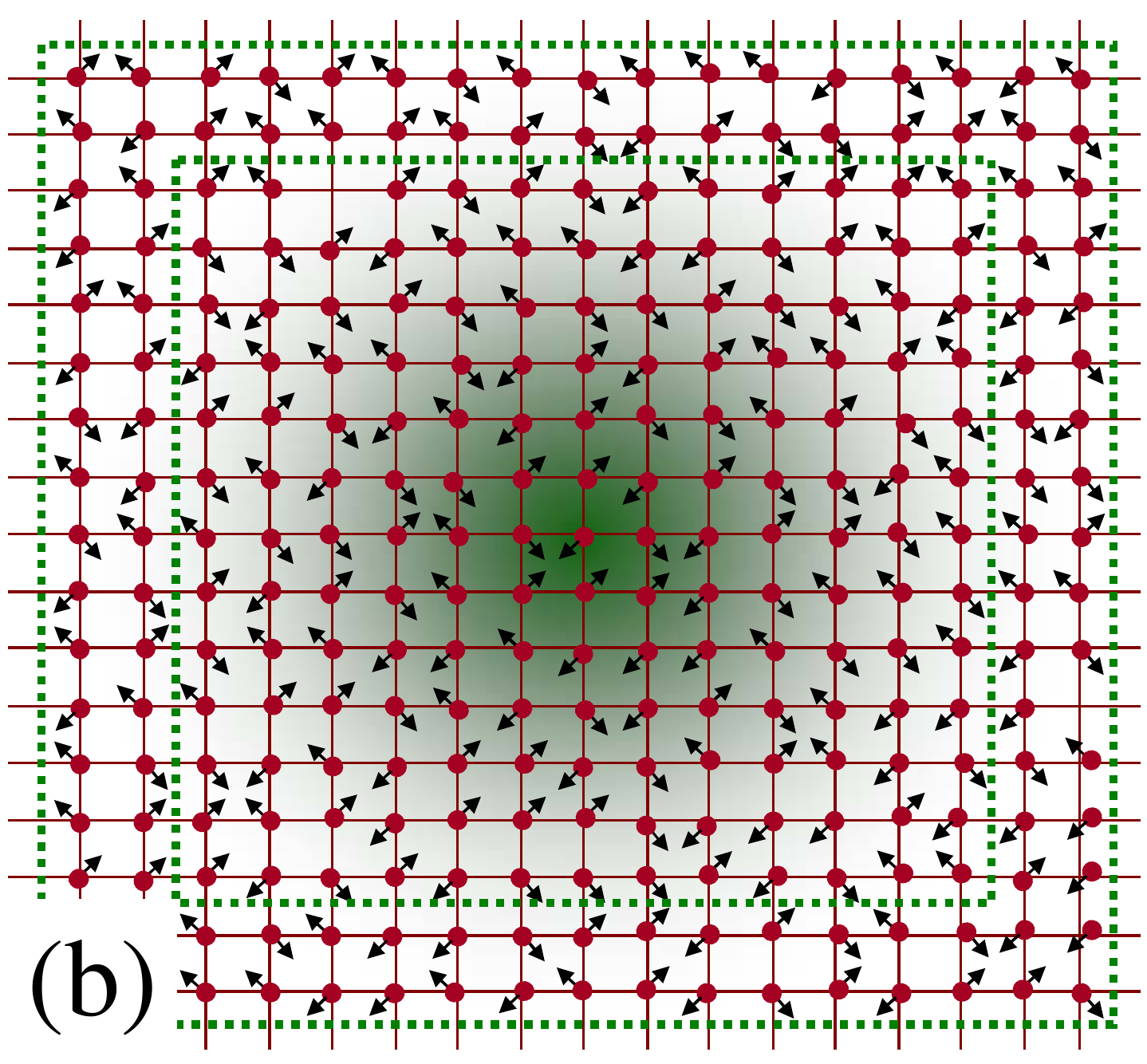}
\end{center}
\caption{(a) Schematics of a quantum box model for an InAs QD.
(b) Schematics of an electron (the shadow) and one layer of nuclear
spins in the QD. The two boxes in dotted lines indicate two
possible choices of boundary of the nuclear spin bath, which are
relatively arbitrary due to the interaction between nuclei within
and without the boundary. When the hyperfine interaction dominates
over the nuclear spin interaction, such arbitrariness has negligible
effects on calculation of the electron spin decoherence as long as
all the nuclei in direct contact with the electron spin have been
enclosed.} \label{Fig_structure}
\end{figure}

The system consists of an electron with spin-$1/2$
$\hat{{\mathbf S}}_{\rm e}$ and $N$ nuclear spins, $\hat{{\mathbf J}}_{n,\alpha}$,
with Zeeman energies $\Omega_{\rm e}$ and
$\omega_{\alpha}$ under a magnetic field $B_{\rm ext}$,
respectively, where $n$ denotes the position and $\alpha$ denotes
the isotope type. Hereafter the subscript $\alpha$ will be absorbed
into $n$ unless it is needed for clarity.
InAs has the Zincblende structure, with In and As ions
located in two interpenetrating face-centered cubic lattices. The natural
isotope abundance in InAs materials is $100\%$, $4.28\%$ and $95.72\%$ for $^{75}$As,
$^{113}$In, and $^{115}$In, respectively. All the isotopes have
non-zero nuclear spin moments, namely, $J_{^{75}{\rm As}}=3/2$ and
$J_{^{113}{\rm In}}=J_{^{115}{\rm In}}=9/2$. The self-assembled InAs QD
under typical growth condition is modeled as a rectangular quantum box with
the growth direction along $[001]$ and the in-plane extension
directions $[110]$ and $[1\bar{1}0]$.\cite{InAsQDReview} The
electron is assumed to be confined by hardwall potential and the
envelope wavefunction of the ground state is
\begin{eqnarray}
f \left( {\mathbf r}\right) =\prod_{\mathbf a}
\sqrt{\frac{2}{L_{\mathbf a}}}\cos \frac{\pi r_{\mathbf
a}}{L_{\mathbf a}}
 \theta \left( L_{\mathbf a}-2\left| r_{\mathbf a} \right| \right),
 \label{Eq_envelope_f}
\end{eqnarray}
where $r_{\mathbf a}$ is the coordinate in the ${\mathbf a}$
direction($[110]$, $[1\bar{1}0]$, or $[001]$), and $L_{\mathbf a}$
is the dimension of the dot along the indicated direction. The
model system is illustrated in Fig.~\ref{Fig_structure}.

\begin{figure}[b]
\begin{center}
\includegraphics[width=7cm]{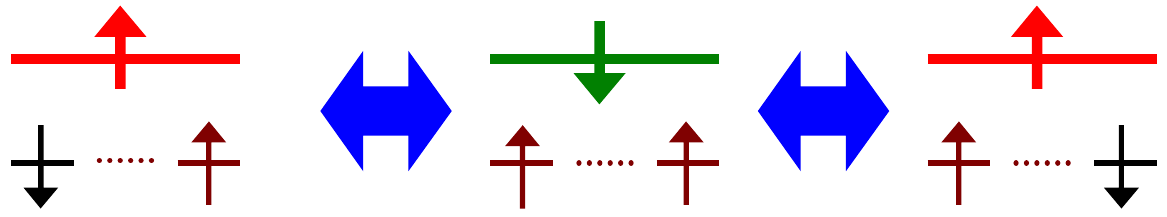}
\end{center}
\caption{Hyperfine-mediated interaction between two distant nuclear spins
via the electron (top line).}
\label{Fig_mediated}
\end{figure}

\begin{table}[t]
\caption{InAs material parameters, where $\gamma_{\alpha}$ is the
nuclear gyromagnetic ratios, $\gamma^{\rm ex}_{\alpha}$ the effective nuclear gyromagnetic ratios
for the pseudo-exchange interaction, $\gamma_e$ the electron gyromagnetic ratio, $\gamma^*_e$
the effective electron gyromagnetic ratio in the QD,
and ${\mathcal A}_{\alpha}$ the hyperfine constants.
}
\begin{tabular}{|c||c|c|c|} \hline
                    & $^{75}$As & $^{113}$In & $^{115}$In \\ \hline
  Abundance         & 100\%     &  4.28\%    &    95.72\% \\ \hline
  Spin moment ${J}_{\alpha}$ & 3/2      & 9/2        & 9/2       \\ \hline
  $\gamma_{\alpha}$ ($10^6$~s$^{-1}$T$^{-1}$)
                    & $+45.8$   & $+58.5$    & $+58.6$    \\ \hline
 $\gamma^{\rm ex}_{\alpha}$
 ($10^6$~s$^{-1}$T$^{-1}$)\footnote{See details in Ref.~\onlinecite{Yao_Decoherence},
 \onlinecite{indirect_exchange_Bloembergen}, and \onlinecite{indirect_exchange_Sundfors}.}
                    & $+34.0$   & $+70.1$    & $+70.2$    \\ \hline
${\mathcal A}_{\alpha}$ ($10^9$~s$^{-1}$)\footnote{Estimated with the method in Ref.~\onlinecite{Paget}.}
                    &   +69.8   &  +85.1     &  +85.3     \\ \hline
 \multicolumn{4}{|c|}{$\gamma_e=+0.176$,
                      $\gamma_e^*=-0.132$ ($10^{12}$~s$^{-1}$T$^{-1}$)\footnote{$g$-factor taken to be $-1.5$.}} \\ \hline
  \multicolumn{4}{|c|}{lattice constant $c_0=6.06$~\AA} \\ \hline
\end{tabular}
\label{Table_Parameters}
\end{table}

\begin{table}[t]
\caption{The characteristic energy scales in an InAs QD with dimensions
$35\times 35\times 6$ nm$^{3}$ ($N\sim 0.3\times 10^6$), under a field of $B_{\rm ext}\sim 10$~Tesla.
}
\label{Table_scales}
\begin{tabular}{|l||l|l|l|l|l|}
\hline
 Units\footnote{Throughout this paper, the units are chosen such that the Plank constant $\hbar$ and the Boltzmann
constant $k_B$ are unity.}
                 & $\Omega_e$ & $\omega_{\alpha}$   & $E_n$      & $B_{n,m}$, $D_{n,m}$     & $A_{n,m}$      \\ \hline
 $10^6$~s$^{-1}$ & $10^{6}$   & $500 $              & $1$      & $10^{-4}$        & $10^{-6}$  \\ \hline
 $\mu$eV         & $10^3$     & $0.5$               & $10^{-3}$  & $10^{-7}$        & $10^{-9}$ \\ \hline
 mK              & $10^4$     & $5$                 & $10^{-2}$  & $10^{-6}$        & $10^{-8}$  \\ \hline
\end{tabular}
\end{table}

The first-principles Hamiltonian for the electron-nuclear spin
system includes the electron-nuclear hyperfine interaction and
various intrinsic nuclear-nuclear interactions. A detailed
discussion was given in Ref.~\onlinecite{Yao_Decoherence}.
Under a moderate magnetic field ($B_{\rm ext}> 0.1$~T),
the electron spin flip is virtually suppressed due to the large Zeeman
energy mismatch between the electron and nuclei. Virtual flip-flops of
the electron spin, however, could mediate an extrinsic interaction between
nuclear spins even when they are well separated in space (as illustrated in
Fig.~\ref{Fig_mediated}).\cite{Yao_Decoherence}
Furthermore, the non-secular part of the nuclear spin interaction
which does not conserve the total Zeeman energy, including
flip-flops of hetero-nuclear pairs and single-spin flips,
can be neglected. Thus the total effective Hamiltonian is reduced to the form
\begin{equation}
\hat{H}=\hat{H}_{\rm e}+\hat{H}_{\rm N}+\sum_{\pm}|\pm\rangle
\hat{H}_{\pm} \langle\pm|, \label{Hamil_reduced}
\end{equation}
in the limit of long longitudinal relaxation time ($T_1 \rightarrow \infty$),
where $|\pm\rangle$ are the eigenstates of $\hat{S}^z_{\rm e}$,
$\hat{H}_{\rm e}\equiv\Omega_{\rm e} \hat{S}^z_{\rm e}$,
$\hat{H}_{\rm N}\equiv \omega_{n}\hat{J}^z_{n}$, and the nuclear spin interaction
\begin{equation}
\hat{H}_{\pm}=\pm \hat{H}_A+\hat{H}_B+\hat{H}_D\pm \hat{H}_E,
\end{equation}
with
\begin{subequations} \begin{eqnarray}
\hat{H}_A &\equiv&
  {\sum_{n\ne m}}'\frac{a_n a_m}{4  \Omega _{\rm e}}
  \hat{J}_n^{+}\hat{J}_m^{-}\equiv {\sum_{n \ne m}}' A_{n,m} \hat{J}_n^{+}\hat{J}_m^{-}, \ \ \ \   \label{HA} \\
\hat{H}_B &\equiv&{\sum_{n \ne  m}}' B_{n,m} \hat{J}_n^{+}\hat{J}_m^{-}
\label{HB}  \\
\hat{H}_D &\equiv&{\sum_{n < m}} D_{n,m} \hat{J}_n^{z}\hat{J}_m^{z}
\label{HD}  \\
\hat{H}_E &\equiv&
 \sum_{n}\left(a_n/2\right)\hat{J}_n^{z}\equiv \sum_{n}E_n \hat{J}_n^{z},   \label{HE}
\end{eqnarray} \label{Hamiltonian}
\end{subequations}
Here the summation with a prime runs over only the homo-nuclear pairs,
$\hat{H}_A$ denotes the extrinsic interaction mediated by the hyperfine interaction,
$\hat{H}_B$ denotes the off-diagonal part of the intrinsic nuclear interaction
and $\hat{H}_D$ the diagonal part, and $\hat{H}_E$ is the diagonal part of the contact
hyperfine interaction with amplitude
$a_n={\mathcal A}_{\alpha}c^3_0\left|f({\mathbf R}_n)\right|^2$.
The material parameters are given in Table~\ref{Table_Parameters}
and the typical energy scales are estimated in Table~\ref{Table_scales}.
The hyperfine interaction has a typical energy scale $E_n\sim a_n \sim 10^6$~s$^{-1}$
for a dot with about $10^6$ nuclei.\cite{Paget} The intrinsic nuclear spin-spin interaction
is effectively finite-ranged with the near-neighbor coupling
$B_{n,m}\sim {D}_{n,m} \sim 10^2$~s$^{-1}$. The extrinsic interaction
is ``infinite-ranged'' (coupling any two nuclear spins within the QD) and
has opposite signs for opposite electron spin states.
The extrinsic interaction, in contrast with the intrinsic one, depends on the
external field, and has an energy scale $A_{n,m}\sim 1$ to $10$~s$^{-1}$ for field
strength $B_{\rm ext}$ varying from $10$ to $1$~Teslas.

\subsection{General formalism}
\label{SS_Thery_general}

The electron-nuclear spin system is assumed initially prepared in a product state,\cite{Note_on_product}
\begin{eqnarray}
\hat{\rho}(0)=\hat{\rho}^{\rm e}(0)\otimes \hat{\rho}^{\rm N}.
\end{eqnarray}
where the nuclear spins are in a thermal state with temperature $T$.
The electron spin can start from a pure state or a mixed state.\cite{Note_on_InitialState}
The state evolution is determined by the Louville equation
\begin{eqnarray}
\partial_t\hat{\rho}(t)= -i[\hat{H},\hat{\rho}(t)].
\end{eqnarray}
The reduced density matrix of the electron spin is obtained
by partial trace over the nuclear spins as
\begin{eqnarray}
\hat{\rho}^{\rm e}(t)={\rm Tr}_{\rm N} \hat{\rho}(t),
\end{eqnarray}
which is related to the initial state through the correlation superoperator $\hat{\mathcal L}$, namely
\begin{eqnarray}
\rho^{\rm e}_{\mu,\nu}(t)=\sum_{\mu',\nu'}{\mathcal L}_{\mu,\nu; \mu',\nu'}(t) \rho^{\rm e}_{\mu',\nu'}(0),
\end{eqnarray}
where $\rho^{\rm e}_{\mu,\nu}\equiv \langle \mu|\hat{\rho}^{\rm e}|\nu\rangle$, and
$|\mu\rangle$, $|\nu\rangle\in\{|+\rangle,\ |-\rangle\}$.
The electron spin relaxation is quantified by the correlation function
${\mathcal L}_{\mu,\nu; \mu',\nu'}$.

Due to the block diagonal form of the reduced Hamiltonian
Eq.~(\ref{Hamil_reduced}), the correlation function has following
properties
\begin{subequations}
\begin{eqnarray}
{\mathcal L}_{\mu,\nu; \mu',\nu'}(t)&=& {\mathcal L}_{\mu,\nu}(t)\delta_{\mu,\mu'}\delta_{\nu,\nu'}, \\
{\mathcal L}_{\mu,\mu}(t)&=&1, \\
{\mathcal L}_{+,-}(t)&=& {\mathcal L}^*_{-,+}(t) \nonumber \\
&=&  e^{-i\Omega_e t}{\rm Tr}_{\rm N} \left[\hat{\rho}^N e^{+i\hat{H}_- t}e^{-i\hat{H}_+ t}\right], \ \ \
\end{eqnarray}
\label{electroncoherence}
\end{subequations}
where the last equation has been expressed for the free-induction decay and can be straightforwardly
extended to the dynamics under pulse control. No longitudinal relaxation remains ($T_1=\infty$)
after the elimination of the electron spin flip processes by the magnetic field. Since
the Zeeman energy $\hat{H}_0$ results in only a global phase-factor $e^{-i\Omega_e t}$, from now on
we will consider only the interaction Hamiltonian $\hat{H}_1\equiv \hat{H}-\hat{H}_0$ by dropping the trivial phase-factor.

Since the nuclear Zeeman energy under the external field dominates over the
hyperfine interaction and the nuclear spin interaction and the temperature
to be considered is much higher than the interaction energy, the thermal nuclear
spin state can be taken as
\begin{eqnarray}
\hat{\rho}^{\rm N} \cong e^{-\hat{H}_{\rm N}/T}
= \sum_{\mathcal J}P_{\mathcal J}|{\mathcal J}\rangle\langle {\mathcal J}|,
\label{thermal}
\end{eqnarray}
where $|{\mathcal J}\rangle\equiv \bigotimes_{n}|j_{n}\rangle$ is
an eigenstate of $\hat{H}_{N}$ and $P_{\mathcal
J}=\prod_{n}p_{j_{n}}$ is the probability distribution with
\begin{eqnarray}
p_{j_{n,\alpha}}\equiv \frac{e^{-j_{n,\alpha}\omega_{n} / T}}
{\sum_{j=-{J_{\alpha}}}^{+{J_{\alpha}}}e^{-j\omega_{n} /T}},
\label{distribution}
\end{eqnarray}
giving the population of the single-spin state $|j_{n}\rangle$.
We note that such a nuclear spin state has no off-diagonal coherence
(between the Zeeman energy eigenstates).
The correlation function of the electron spin coherence,
expressed in terms of the wavefunction overlap of the nuclear
spins, is
\begin{eqnarray}
{\mathcal L}_{+,-}(t)=\sum_{\mathcal J}P_{\mathcal J}
\left\langle {\mathcal J}^-(t)\right|\left.{\mathcal J}^+(t)\right\rangle,
\label{ensembleaverage}
\end{eqnarray}
with $\left|{\mathcal J}^{\pm}(t)\right\rangle\equiv
\exp\left(-i\hat{H}_{\pm} t\right)\left|{\mathcal
J}\right\rangle$. The correlation function ${\mathcal L}_{+,-}(t)$
is closely related to the Loschmidt echo in
literature.\cite{LoschmidtDecoherence,Zurek_Decoherence_Spin,SunCP_Loschmidt}
Being independent of the electron spin initial state, ${\mathcal L}_{+,-}(t)$
will be equated with the electron spin coherence
without causing confusion. The decoherence by entanglement is
transparent from Eq.~(\ref{ensembleaverage}): With the nuclear spin states $|{\mathcal
J}^{\pm}(t)\rangle$ driven by different Hamiltonians
$\hat{H}_{\pm}$, the electron-nuclear spin state
$C_+|+\rangle\otimes |{\mathcal J}^+(t)\rangle
+C_-|-\rangle\otimes |{\mathcal J}^-(t)\rangle$ becomes an entangled
one when $\left\langle {\mathcal J}^-(t)\right|\left. {\mathcal
J}^+(t)\right\rangle<1$, so the electron spin coherence is lost.

\subsection{Essential assumptions}
\label{SS_theory_assumption}

The method to be used in this paper consists of two steps. First,
the electron spin decoherence in {\em ensemble dynamics} is factorized
into two parts: the dephasing due to static inhomogeneous broadening,
and the decoherence due to the dynamical entanglement between the electron
and the nuclei in {\em single-system dynamics}. Second, the nuclear spin
dynamics is solved with the pair-correlation approximation. The following
assumptions are essential for the theory:
\begin{enumerate}

\item {\em Mesoscopia}: First, the QD should be small enough to
guarantee the dominance of the hyperfine interaction over the
nuclear spin interaction so that the isolation of a mesoscopic
bath is possible. Second, the bath size (i.e., the number of
nuclear spins in the QD, $N$) should be large enough to have genuine
decoherence (i.e., the Poincar\'{e} time is much greater than
the spin relaxation time $T_1$ and $T_2$). Third,  $N\gg \sqrt{N}$ is
required so that the central limit theorem in statistics can be utilized
for the factorization of the single-system dynamics from the inhomogeneous
broadening effect. And finally, to justify the pair-correlation approximation,
the bath size $N$ is bounded so that the number of nuclear pair-flips in
the relevant timescale is small compared with $N$ (see Appendix \ref{Append_boudary}).

\item {\em Moderate to strong field}: The electron Zeeman energy $\Omega_e$
  be much greater than the hyperfine energy $a_{n}$ so that the electron spin flip is
  suppressed.\cite{Yao_Decoherence,Note_on_StrongFieldAssumption}

\item {\em Finite temperature}: The temperature be higher than or comparable to the
  nuclear Zeeman energy ($T\gtrsim \omega_{\alpha}\sim 1$~mK). Otherwise,
  in a near fully polarized nuclear spin bath under an extremely low-temperature,
  the pair flip-flops occur only among a few available nuclear spins and hence are
  highly correlated, in which the pair-correlation approximation is invalid.
  Also, the temperature should be low enough to prevent the phonon scattering effect
  ($T\lesssim 1$~K)

\item {\em Short-time dynamics}: The timescales under consideration should be short
  compared with the inverse nuclear spin interaction strength ($\sim 1$~ms)
  so that the number of pair-flips
  over time is small compared with $N$. The timescales of interest in the decoherence
  problem include the coherence memory time, pulse delay time in control sequences, and the
  total duration of a control pulse sequence.
\end{enumerate}
The other approximations in the model, including the assumptions on the QD shape,
the electron wavefunction, and the specific form of the spin interactions,
are inconsequential to the theory.

\subsection{Ensemble and single-system dynamics}
\label{SS_theory_ensemble}

The electron spin decoherence given in
Eq.~(\ref{electroncoherence}) includes two contributions, namely,
the thermal fluctuations due to the ensemble average of different
nuclear spin configurations $\{|{\mathcal J}\rangle\langle
{\mathcal J}|\}$, and the quantum entanglement due to the dynamical
evolution starting from a pure state $|{\mathcal J}\rangle$, which
is conditioned on the electron spin state. We use the term
{\em single-system dynamics} to denote the quantum evolution
governed by the Schr\"{o}dinger equation
\begin{eqnarray}
i\partial_t |{\mathcal J}^{\pm}(t)\rangle=H_{\pm}|{\mathcal J}^{\pm}(t)\rangle,
\end{eqnarray}
and {\em ensemble dynamics} to denote the ensemble average of the
single-system dynamics.

The ensemble dynamics was studied in the formalism of
density matrix with the pair-correlation approximation.\cite{Witzel_Quantum1}
Many important features of the quantum entanglement process, however,
could have been overlooked if without resolving the single-system
dynamics since the static inhomogeneous broadening usually has
much stronger effects on the decoherence than the dynamical
entanglement does.
We choose a different procedure to study the problem. Namely, we
will first solve the single-system dynamics starting from a
certain nuclear spin configuration, and then construct the
ensemble dynamics via
\begin{subequations}
\begin{eqnarray}
{\mathcal L}^{\mathcal J}_{+,-}(t)\equiv \langle {\mathcal J}^-(t)|{\mathcal J}^+(t)\rangle, \label{singlesyscoherence} \\
{\mathcal L}_{+,-}(t)=\sum_{\mathcal J}P_{\mathcal J}{\mathcal
L}^{\mathcal J}_{+,-}(t). \label{ensemblecoherence}
\end{eqnarray}
\end{subequations}
Such a construction of the ensemble dynamics, in general, would require sampling a large number of initial
states $|{\mathcal J}\rangle$ in the thermal ensemble.

When the system is sufficiently large and the temperature is
appreciable for the nuclear spins, the decoherence due to
quantum fluctuations in single-system dynamics (Eq.~\ref{singlesyscoherence})
is almost the same for all possible initial nuclear configurations
in the thermal ensemble, except for a global phase factor related to the static
Overhauser field resulting from the diagonal hyperfine interaction.
This is confirmed by numerical verification (see Appendix~\ref{Append_factorization}).
Here we give a physical argument:
For a sufficiently large number of nuclear spins far from being
fully polarized, the number of spins available for pair-wise flip-flops is large, so the electron
spin decoherence is virtually determined by the nuclear spin excitation spectra (or density of states).
By the statistical central limit theorem, the excitation spectrum of a large system is the same
(up to a relative variance $\sim 1/\sqrt{N}$) for different initial states $|{\mathcal J}\rangle$.
Then the electron spin decoherence can be factorized into a contribution
from the static inhomogeneous broadening ${\mathcal L}^{\rm inh}$ and one from
the quantum fluctuation in single-system dynamics ${\mathcal L}^{\rm s}$, defined as
\begin{subequations}
\begin{eqnarray}
{\mathcal L}_{+,-}(t)&\cong& {\mathcal L}^{\rm inh}_{+,-}(t) {\mathcal L}^{\rm s}_{+,-}(t), \\
{\mathcal L}^{\rm s}_{+,-}(t)  & \equiv & \left|\langle {\mathcal J}^-(t)|{\mathcal J}^+(t)\rangle\right|, \\
{\mathcal L}^{\rm inh}_{+,-}(t)& = & \sum_{\mathcal J} P_{\mathcal J}e^{-iE_{\mathcal J}t}=\int P\left(E\right)e^{-iEt} d E, \ \ \
\end{eqnarray}
\label{Eq_factorize}
\end{subequations}
where the hyperfine energy due to the local Overhauser field in a certain nuclear spin configuration is
\begin{eqnarray}
E_{\mathcal J}\equiv \sum_{n} j_{n}a_{n},
\end{eqnarray}
and the inhomogeneous broadening distribution is defined as
\begin{eqnarray}
P\left(E \right)\equiv \sum_{\mathcal J} P_{\mathcal J}\delta\left(E-E_{\mathcal J}\right).
\end{eqnarray}
The quantum fluctuation effect in the single-system dynamics can be evaluated for an initial
configuration $|{\mathcal J}\rangle$ randomly selected from the ensemble, with the global phase
factor absorbed into the inhomogeneous broadening.

\begin{figure}[b]
\includegraphics[width=7cm]{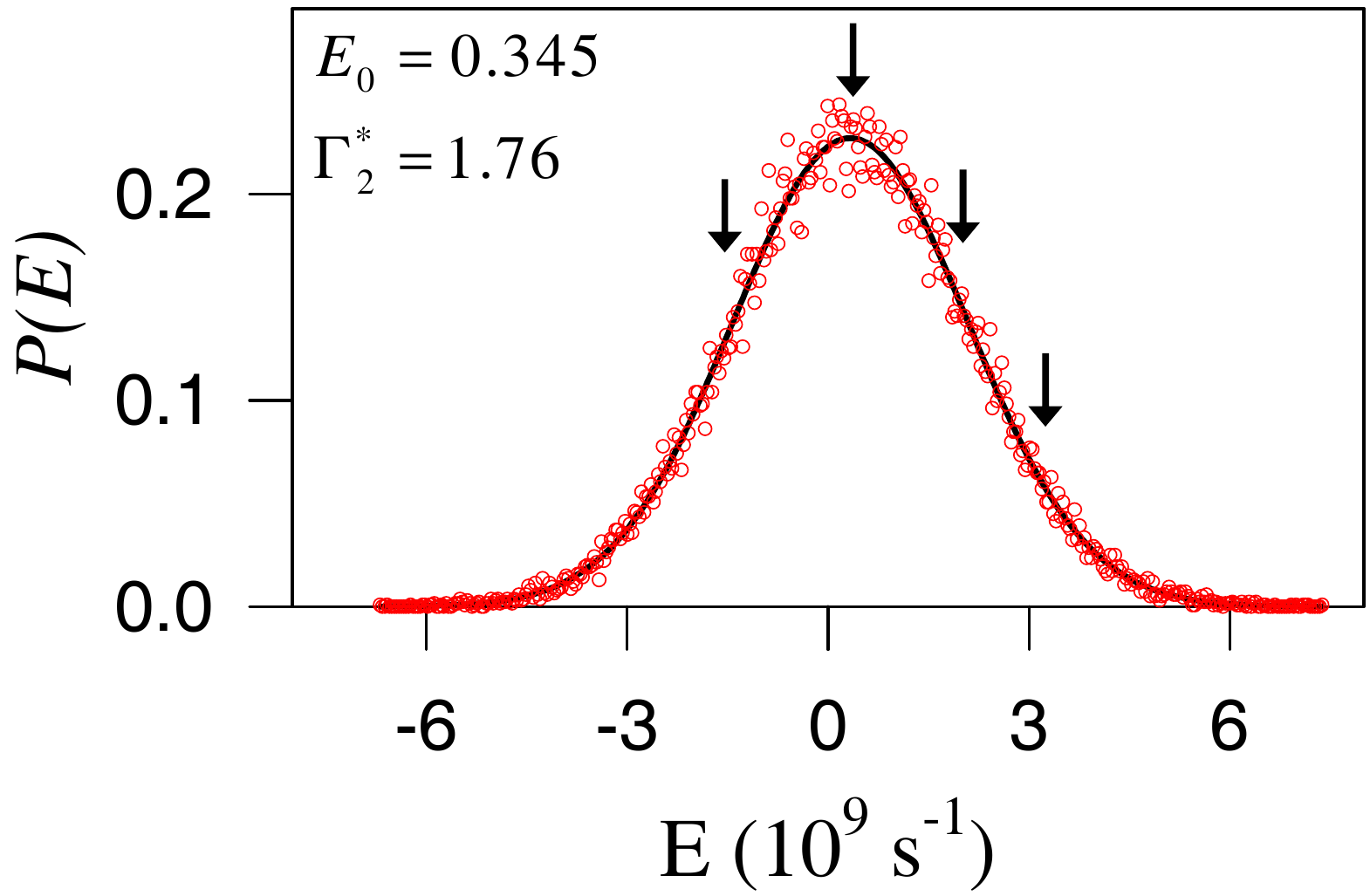}
\caption{Distribution of the Overhauser field for a thermal ensemble of nuclear spins
in an InAs QD of size $34\times 34\times 3$~nm$^{3}$ at a temperature of 1~K and under an external field
of 1~Tesla. The dots in the figure are obtained by numerical simulation of 40000 random samples,
and the curve is calculated with Eqs.~(\ref{Eq_inhomo}-\ref{Eq_inhpara}).
The arrows indicate four randomly selected nuclear spin configurations to be
used for Fig.~\ref{Fig_factorization} in Appendix~\ref{Append_factorization}.} \label{Fig_inhom}
\end{figure}

The effect of the static inhomogeneous broadening can be calculated by neglecting the
nuclear spin interaction (setting $\hat{H}_A=\hat{H}_B=0$). Without the flip-flop dynamics,
the nuclear spin ensemble would be a random distribution of frozen configurations,
which, as a standard multinomial distribution for a large system, leads to a Gaussian
distribution of the Overhauser field
\begin{eqnarray}
P\left(E_{\mathcal J}\right)=\frac{1}{\sqrt{2\pi}\Gamma^*_2}e^{-\left(E_{\mathcal J}-E_0\right)^2/\left(2\Gamma^*_2\right)^2},
\label{Eq_inhomo}
\end{eqnarray}
where the averaged local field $E_0$ and the inhomogeneous broadening $\Gamma^*_2$ are
\begin{subequations}
\begin{eqnarray}
E_0 & \cong &\sum_{\alpha}  \left({2N_{\alpha}{\mathcal A}_{\alpha}}{N^{-1}}\overline{j_{\alpha}}\right)\equiv \sum_{\alpha} E_{0,\alpha}, \\
\Gamma^*_2 & \cong & \sqrt{ \sum_{\alpha}\left({4N_{\alpha} A^2_{\alpha}}{N^{-2}}\overline{j^2_{\alpha}} -{E^2_{0,\alpha}}{N_{\alpha}^{-1}}\right)}, \ \ \ \
\end{eqnarray}
\label{Eq_inhpara}
\end{subequations}
respectively, with $\overline{j_{\alpha}}\equiv\sum_{j} j p_{\alpha,j}$, $\overline{j^2_{\alpha}}\equiv\sum_{j} j^2 p_{\alpha,j}$,
and $N_{\alpha}$ denoting the number of $\alpha$-type nuclei ($N\equiv \sum_{\alpha}N_{\alpha}$).
A typical example of the Overhauser field distribution is shown in Fig.~\ref{Fig_inhom}.
The electron spin dephasing by inhomogeneous broadening is a Gaussian decay
\begin{eqnarray}
{\mathcal L}^{\rm inh}_{+,-}(t)=e^{-iE_0 t -\left(t/T^*_2\right)^2},
\end{eqnarray}
with the effective dephasing time $T^*_2\equiv \sqrt{2}/\Gamma^*_2$.
The inhomogeneous broadening effect can be removed by spin echo.
In the rotating frame rested on the electron spin, the Overhauser field changes its sign
each time the electron spin is flipped by a short pulse. Thus under the flipping operation
of a sequence of pulses applied at $t=t_1$, $t_2$, $\ldots$, $t_n$, the dephasing by the
inhomogeneous broadening becomes
\begin{eqnarray}
{\mathcal L}^{\rm inh}_{+,-}(t)=\sum_{\mathcal J}P_{\mathcal J} e^{-iE_{\mathcal J}{\overline t}}=
e^{-iE_0 {\overline t} -{\overline t}^2/\left(T^*_2\right)^2},
\end{eqnarray}
with ${\overline t}\equiv t_1-(t_2-t_1)+(t_3-t_2)\cdots
-(-1)^n(t-t_n)$. Obviously, the phase accumulation from the random
Overhauser field is cancelled at ${\overline t}=0$, leading to the
spin echo. At the spin-echo time, the
electron spin decoherence results solely from the dynamical quantum
entanglement.

Under typical conditions considered in this paper, $T^*_2$ is found to be of
the order of nanoseconds, consistent with various experimental
measurements.\cite{Gupta_spin_Nano,Gurudev,Kouwenhoven_singlet_triplet,Marcus_T2}
As will be shown below, in agreement with the available experimental
data,\cite{Koppens_T2,Marcus_T2,Greilich_lock} the electron spin
decoherence by quantum fluctuations in single-system dynamics has
a timescale in the order of microseconds. Thus, from Eq.~(\ref{Eq_factorize}),
the ensemble dynamics will be dominated
by inhomogeneous broadening and the single-system dynamics is
virtually invisible in ensemble experiments, unless the
inhomogeneous broadening effect is removed by spin echo. The decay
of spin-echo signals is usually attributed to the pure decoherence
due to quantum fluctuations. We have shown, however, by examining
directly the single-system dynamics, the flipping pulses
have non-trivial effects on the nuclear spin dynamics and, therefore,
on the dynamical entanglement, making the spin-echo decay time significantly
different from the FID time in single-system dynamics.\cite{Yao_Decoherence}

From now on, we will concentrate on the single-system dynamics,
and show a few surprising effects which would otherwise be
concealed by inhomogeneous broadening in ensemble dynamics.
It should be emphasized that the single-system dynamics in a QD at a
finite temperature is not a mathematical idealization but has measurable
effects. For example, a projective measurement of the local Overhauser
field could be used to limit the nuclear spin configurations by post-selection
and thus to observe single-electron dynamics without spin
echo.\cite{Espin_HF_1_Loss,Giedke_spinmeasure,Klauser_spinmeasure,Imamoglu_WeakMeasurement}
The single-system dynamics is the basic to quantum technologies such as
quantum computation which can not be performed in ensembles for scalability
to large systems without exponential explosion in resource.\cite{Liu_readwrite}

\subsection{Pair-correlation approximation and pseudo-spin model for nuclear spin dynamics}
\label{SS_PCA}

With the Hamiltonian given in Eq.~(\ref{Hamiltonian}), the quantum
evolution of the nuclear spins is driven by the pair-wise homo-nuclear
spin flip-flops as the elementary excitations. The transition for
the pair-flip driven by the operator $J^+_{n}J^-_{m}$ is
\begin{eqnarray}
|j_{n}\rangle_{n}|j_{m}\rangle_{m}
\longrightarrow
|j_{n}+1\rangle_{n}|j_{m}-1\rangle_{m}.
\label{Eq_pair_flip}
\end{eqnarray}
The transition between the electron-nuclear spin many-body states through the
pair-flip is denoted by
\begin{eqnarray}
|\pm\rangle|{\mathcal J}\rangle\longrightarrow
|\pm\rangle|{\mathcal J},k\rangle, \label{Eq_J_flip}
\end{eqnarray}
where $k$ is shorthand for the pair-flip
in Eq.~(\ref{Eq_pair_flip}). The transition described in Eq.~(\ref{Eq_J_flip})
is characterized by the matrix elements
$\pm A_k+B_k$ and the energy costs $ D_k\pm E_k$, which are derived from the
microscopic models as follows
\begin{widetext}
\begin{subequations}
\begin{eqnarray}
A_k &\equiv &  \left\langle {\mathcal J},k\right| \hat{H}_A
\left|{\mathcal J}\right\rangle
 =  \frac{a_{n}a_{m}}{4\Omega_e}
\sqrt{{J}_{n}\left({J}_{n}+1\right)-j_{n}\left(j_{n}+1\right)}
\sqrt{{J}_{m}\left({J}_{m}+1\right)-j_{m}\left(j_{m}-1\right)}, \\
B_k &\equiv &  \left\langle {\mathcal J},k\right| \hat{H}_B
\left|{\mathcal J}\right\rangle =B_{n,m}
\sqrt{{J}_{n}\left({J}_{n}+1\right)-j_{n}\left(j_{n}+1\right)}
\sqrt{{J}_{m}\left({J}_{m}+1\right)-j_{m}\left(j_{m}-1\right)}, \\
D_k &\equiv &  \left\langle {\mathcal J},k\right| \hat{H}_D
\left|{\mathcal J},k\right\rangle
 -  \left\langle {\mathcal J}\right| \hat{H}_D \left|{\mathcal J}\right\rangle
 \nonumber \\ &=&
\sum_{n'}D_{n,n'}j_{n'}-\sum_{m'}D_{m,m'}j_{m'}- D_{n,m} +
D_{n,n}\left(j_{n}+1\right)-D_{m,m}\left(j_{m}-1\right), \ \ \
\label{Eq_Dk}\\
E_k & \equiv &  \left\langle {\mathcal J},k\right| \hat{H}_E
\left|{\mathcal J},k\right\rangle
 -  \left\langle {\mathcal J}\right| \hat{H}_E \left|{\mathcal J}\right\rangle
=\left(a_{n}-a_{m}\right)/2.
\end{eqnarray}
\label{elementaryexcitation}
\end{subequations}
\end{widetext}
The basic processes for the electron spin decoherence may be
described as follows: The off-diagonal part of the nuclear spin
interaction (including the intrinsic one $\hat{H}_B$ and the
extrinsic hyperfine-mediated one $\hat{H}_A$) causes the pair-wise
nuclear spin flip-flops, leading to a fluctuating local Overhauser
field and in turn a random dynamical phase for the electron spin.
In the quantum picture, the entanglement is developed because the
quantum evolution of the nuclear spins driven by the Hamiltonians
$\pm\hat{H}_A+\hat{H}_B+\hat{H}_D\pm \hat{H}_E$ is conditioned
on the electron spin state through the $\pm$ signs originating from the
hyperfine interaction.

As established in Ref.~\onlinecite{Yao_Decoherence} and discussed in
further details in Appendix \ref{Append_pairflip} for pulse sequence controls,
within a time $t$ much smaller than the inverse nuclear interaction strength,
the total number of pair-flip excitations $N_{\rm flip}$ is much smaller
than the number of nuclei $N$. The probability of having
pair-flips correlated is estimated to be $P_{\rm corr}\sim 1-e^{-q
N_{\rm flip}^2/N}$ ($q$ being the number of nearest
neighbors), which, as also shown by \textit{a posteriori}
numerical check (see Appendix~\ref{Append_pairflip}), is
bounded by $\sim 10 \%$ in the worst scenario studied in this paper.
Thus, the pair-flips as elementary excitations from the initial state can be
treated as independent of each other, with a relative error
$\epsilon \lesssim P_{\rm corr}$ (see Appendix~\ref{Append_error}).
Then the single-system dynamics $|\mathcal{J}^{\pm}
(t)\rangle$ can be described by the excitation of
pair-correlations as non-interacting quasi-particles from the
``vacuum'' state $|\mathcal{J}\rangle$, driven by the
``low-energy'' effective Hamiltonian,
\begin{equation}
\hat{H}^\pm_{\mathcal J} = \sum_k \hat{\chi}^{\pm}_k\equiv
\sum_k {\bm \chi}_k^{\pm} \cdot \hat{\boldsymbol \sigma}_k/2,
\label{pseudospin_Hamil}
\end{equation}
which has been written in such a way that each pair-flip
is treated as a pseudo-spin~$1/2$, represented by the Pauli
matrix $\hat{\boldsymbol \sigma}_k$, with $k$ labelling all
possible pair-flips. The time evolution from the initial state $|
\mathcal{J} \rangle $ can be viewed as the rotation of the
pseudo-spins, initially all polarized along the $-z$ pseudo-axis:
$\bigotimes_k |\downarrow\rangle_k $, under the effective
pseudo-magnetic field,
\begin{equation}
 {\bm \chi}^{\pm}_k \equiv (\pm
2A_k+2B_k,0, D_k \pm E_k),
\end{equation}
for the electron spin state $|\pm\rangle$, respectively.
Such a treatment of the nuclear spin correlations amounts to taking into
account all the pairwise correlations and neglecting the higher-orders,
as justified by a systematic study based on linked cluster expansion.\cite{Saiken_Cluster}

\begin{figure}[t]
\begin{center}
\includegraphics[width=5cm]{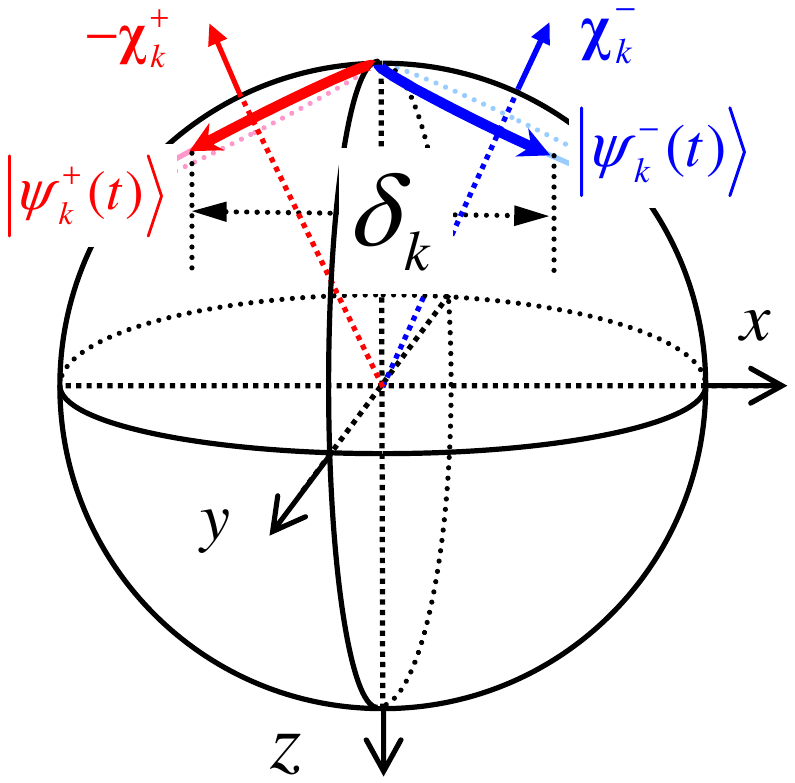}
\end{center}
\caption{The Bloch vectors for the pseudo-spin states $|{\psi}_{k}^{\pm}(t)\rangle$
precess about the pseudo-fields ${\boldsymbol \chi}_{k}^{\pm}$ for the electron spin states $|\pm\rangle$,
respectively.} \label{Fig_geometry}
\end{figure}

The entanglement between the electron spin and the pseudo-spins,
and hence the electron spin decoherence, are developed as the
pseudo-spins precess about different pseudo-fields ${\boldsymbol \chi}_{k}^{\pm}$
corresponding to different electron spin states $|\pm\rangle$.
Namely, the product state
\begin{eqnarray}
\left(C_+|+\rangle+C_-|-\rangle\right)\bigotimes_k|\downarrow\rangle_k,
\end{eqnarray}
will evolve into an entangled one
\begin{eqnarray}
 C_+|+\rangle\bigotimes_k \left|\psi_{k}^{+}(t)\right\rangle
+C_-|-\rangle\bigotimes_k \left|\psi_{k}^{-}(t)\right\rangle,
\end{eqnarray}
with
\begin{eqnarray}
\left|\psi_{k}^{\pm}(t)\right\rangle=\exp\left(-i\hat{\chi}_{k}^{\pm}t\right)\left|\downarrow \right\rangle=
e^{-i{\boldsymbol \chi}_{k}^{\pm}\cdot\hat{\boldsymbol \sigma}_k t/2}|\downarrow\rangle.
\label{kthState}
\end{eqnarray}
The electron spin coherence is determined by the overlap between the ``conjugate'' states
$|{\psi}_{k}^{+}(t)\rangle$ and $|\psi_{k}^{-}(t)\rangle$ of the pseudo-spins,
\begin{eqnarray}
{\mathcal L}_{+,-}^s(t)= \prod_k \left|\left\langle \psi_{k}^{-}(t)\right|\left. \psi_{k}^{+}(t)\right\rangle\right|.
\label{Eq_exact}
\end{eqnarray}

The pseudo-spin states have a geometrical representation as Bloch vectors,
\begin{eqnarray}
{\boldsymbol \sigma}_{k}^{\pm} \equiv \langle \psi_{k}^{\pm}|\hat{\boldsymbol\sigma}_k|\psi_{k}^{\pm}\rangle/2.
\end{eqnarray}
Fig.~\ref{Fig_geometry} shows the evolution of two conjugate pseudo-spin states
$|{\psi}_{k}^{\pm}(t)\rangle$ in two paths of the conjugate Bloch
vectors ${\boldsymbol \sigma}_{k}^{\pm}$ on a sphere.
The overlap between the conjugate pseudo-spin states is related to the distance between the
two vectors $\delta_k=\left|{\boldsymbol \sigma}_{k}^{-}-{\boldsymbol \sigma}_{k}^{+}\right|$ by
\begin{eqnarray}
\left|\left\langle \psi_{k}^{-}(t)\right|\left. \psi_{k}^{+}(t)\right\rangle\right|^2 =1- \delta_k^2.
\end{eqnarray}
Thus, the square of the distance $\delta_k^2$ is the distinguishability between the two conjugate states.
This geometrical picture gives an interpretation of the decoherence process as a measurement.
The ``measuring device'' composed of nuclear spins evolves in two pathways, each for an electron spin
eigenstate, up or down. When the sum of the distinguishability between the conjugate states of all pair-flips
is large enough, the ``device'' states are ``macroscopically'' distinguishable. The
coherence between the basis states is destroyed by the measurement.\cite{Griffiths_consistent}
The pseudo-spin description will serve as the physical guide of the decoherence control by pulse sequences.

\section{Free-induction decay}
\label{S_FID}

The separation of the single-system dynamics from the ensemble dynamics
offers an opportunity to study the FID of electron spin coherence due to
dynamical entanglement, which would otherwise be concealed by
the much stronger effect of inhomogeneous broadening.

The pseudo-spin evolution is directly calculated from Eq.~(\ref{kthState}) to be
\begin{eqnarray}
\left|\psi_{k}^{\pm}(t)\right\rangle =
  \left(\cos \frac{{\chi}_{k}^{\pm} t}{2} -i \hat{\boldsymbol \sigma}_k\cdot \sin\frac{{\boldsymbol \chi}_{k}^{\pm} t}{2}\right)|\downarrow\rangle,
\end{eqnarray}
where the sine function of a vector ${\mathbf v}$ is defined as $\sin({\mathbf v})\equiv ({\mathbf v}/v)\sin(v)$.
The distinguishability between the conjugate pseudo-spin states $|\psi_k^{\pm}\rangle$ is
\begin{widetext}
\begin{eqnarray}
{\delta_k^2} = t^4\left(B_kE_k-A_kD_k\right)^2 {\rm sinc}^2\frac{{\chi}_{k}^{+} t}{2}{\rm sinc}^2\frac{{\chi}_{k}^{-} t}{2}
       +t^2\left[(B_k+A_k)\cos\frac{{\chi}_{k}^{-} t}{2}{\rm sinc}\frac{{\chi}_{k}^{+} t}{2}
               - (B_k-A_k)\cos\frac{{\chi}_{k}^{+} t}{2}{\rm sinc}\frac{{\chi}_{k}^{-} t}{2} \right]^2. \ \
\label{Eq_FID_dk}
\end{eqnarray}
\end{widetext}
This expression is used in the numerical evaluation of the electron spin coherence.

\begin{figure}[t]
\begin{center}
\includegraphics[width=8cm]{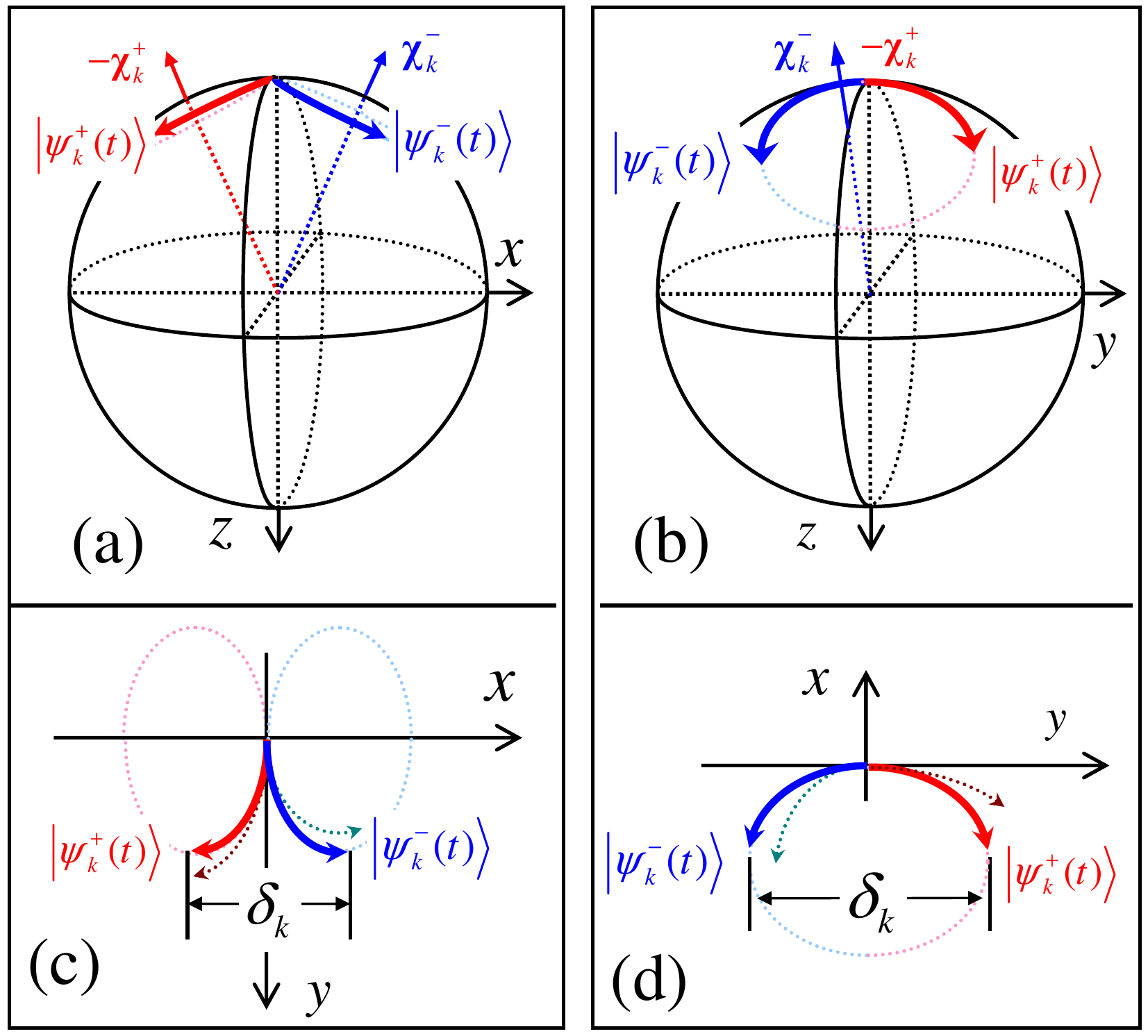}
\end{center}
\caption{(a) The conjugate Bloch vectors for a pair-flip by the intrinsic nuclear
interaction with the diagonal part neglected [${\boldsymbol \chi}_{k}^{\pm}=(B_k,0,\pm E_k)$].
(b) The same as (a) but for a non-local pair-flip [${\boldsymbol \chi}_{k}^{\pm}=\pm(A_k,0,E_k)$].
(c) and (d) are the projections of (a) and (b) to the $x$-$y$ plane, respectively,
in which the dotted trajectories show the effect of the diagonal nuclear interaction ($D_k$)
(with the deviation from the solid curves exaggerated for visibility).} \label{Fig_geometryAB}
\end{figure}

To gain some insight into the FID features, we make two approximations which are well justified.
First, the pseudo-spins are separated into two groups,
${\mathbb{G}}_A$ and ${\mathbb{G}}_B$, corresponding to non-local and local flip-flops, respectively.
For the local pair-flips ($k\in {\mathbb{G}}_B$),
the transition matrix element is dominated by the intrinsic nuclear spin interaction as
$B_k\gg A_k$ under a moderate magnetic field, while for the non-local pair-flips ($k\in {\mathbb{G}}_A$),
the extrinsic hyperfine-mediated interaction dominates.\cite{Note_AB}
Thus the coherence is factorized as
\begin{subequations}
\begin{eqnarray}
{\mathcal L}^{\rm s}_{+,-} & = & {\mathcal L}^{A}_{+,-} \times {\mathcal L}^B_{+,-}, \\
{\mathcal L}^{A/B}_{+,-} & = & \prod_{k\in\mathbb{G}_{A/B}}\left|\langle\psi^-_k(t)|\psi^+_k\rangle\right|
\nonumber \\ & \cong & \exp\left(-\frac{1}{2}\sum_{k\in\mathbb{G}_{A/B}}\delta_k^2\right).
\end{eqnarray}
\end{subequations}
As shown in Fig.~\ref{Fig_geometryAB}, the pseudo-spins in ${\mathbb{G}}_A$
and those in ${\mathbb{G}}_B$ have dramatically different precession
behaviors. The conjugate pseudo-spins for a non-local pair-flip
precess into opposite directions, since both the transition
amplitude ($A_k$) and the dominating part of energy cost ($E_k$)
are associated with opposite signs for opposite electron spin
states because of their hyperfine interaction origins.  On the
other hand, the intrinsic nuclear spin interaction is
independent of the electron spin state, and the corresponding
pseudo-spin bifurcates tangentially. In the second approximation,
we neglect the diagonal nuclear spin interaction $D_k$, which is
justified from Eq.~(\ref{Eq_FID_dk}) with the condition $D_k\ll
E_k$. This simplification can also be understood with the
geometrical picture shown in Fig.~\ref{Fig_geometryAB}: A small
change of the energy cost by the diagonal nuclear interaction
induces only a slight modification of the pseudo-spin precession
pathways, and hence a negligible change of the distance
$\delta_k$. We will see, however, the diagonal nuclear spin
interactions are important for the decoherence dynamics under
control by pulse sequences.

With the two simplifications and the conditions $E_k\gg B_k, A_k$, the distinguishability
between conjugate pseudo-spin states is
\begin{eqnarray} \delta_k^2=
\left\{
\begin{array}{ll}
   4t^2A_k^2\cos^2\frac{E_kt}{2}{\rm sinc}^2\frac{E_kt}{2}\approx 4t^2A_k^2
&  \left(k\in {\mathbb{G}}_A\right)
\\
&
\\
   t^4E_k^2B_k^2{\rm sinc}^4\frac{E_kt}{2} \approx t^4E_k^2B_k^2
&  \left(k\in {\mathbb{G}}_B\right)
\end{array}
\right. ,
\ \ \ \label{FID_distance}
\end{eqnarray}
where the short-time approximation holds for $t\ll E_k^{-1}$.
Within timescales of interest ($t\ll B_k^{-1}, A_k^{-1}$), it is always satisfied that
$\delta_k^2\ll 1$, so the coherence is approximated as
\begin{subequations}
\begin{eqnarray}
{\mathcal L}^A_{+,-}
&\cong & \exp\left(- {2}t^2 \sum_{k\in {\mathbb{G}}_A} A_k^2\cos^2\frac{E_kt}{2}{\rm sinc}^2 \frac{E_kt}{2}\right), \ \ \ \ \  \ \
\\
{\mathcal L}^B_{+,-}
&\cong & \exp\left(- \frac{1}{2}t^4 \sum_{k\in {\mathbb{G}}_B} E_k^2B_k^2{\rm sinc}^4 \frac{E_kt}{2} \right).
\end{eqnarray}
\label{FID}
\end{subequations}
As the number of pseudo-spins is large, the details of individual pseudo-spins are not important to the
decoherence but what matters is the excitation spectrum, which is defined as\cite{Note_excitationspectrum}
\begin{subequations}
\begin{eqnarray}
S_A(\varepsilon) & \equiv & \sum _{k\in{\mathbb{G}}_A} \delta\left(\varepsilon-E_k\right)A_k^2, \\
S_B(\varepsilon) & \equiv & \sum _{k\in{\mathbb{G}}_B} \delta\left(\varepsilon-E_k\right)B_k^2,
\end{eqnarray}
\label{excitationspectra}
\end{subequations}
for the pair-flips in group ${\mathbb{G}_A}$ and ${\mathbb{G}_B}$, respectively.
The decoherence in terms of the excitation spectra is
\begin{subequations}
\begin{eqnarray}
{\mathcal L}^A_{+,-}
&\cong & \exp\left(- {2}t^2 \int d\varepsilon S_A(\varepsilon)\cos^2\frac{\varepsilon t}{2}{\rm sinc}^2 \frac{\varepsilon t}{2}\right), \ \ \ \ \ \ \\
{\mathcal L}^B_{+,-}
&\cong & \exp\left(- \frac{1}{2}t^4 \int d\varepsilon S_B(\varepsilon)\varepsilon^2{\rm sinc}^4 \frac{\varepsilon t}{2} \right).
\end{eqnarray}\label{FID_spectra}
\end{subequations}

We are now ready to analyze some important features of the FID, including the
short-time behavior, the dependence on the field strength and on the QD size, and the emergence of
Markovian decay.

In the timescale $t\ll B_k^{-1}$, the pair-flips cannot be described by
energy-conserving scattering events, but should be understood in terms of quantum evolution.
The decoherence is a highly non-Markovian dynamics. Particularly, in the very initial stage ($t\ll E_k^{-1}$),
according to Eq.~(\ref{FID_distance}), the electron spin coherence is well approximated as
\begin{subequations}
\begin{eqnarray}
{\mathcal L}^A_{+,-} &\approx & \exp\left(- t^2/T^2_{2,A}\right), \\
{\mathcal L}^B_{+,-}
&\approx & \exp\left(- t^4/T^4_{2,B}\right),
\end{eqnarray}\label{FID_shorttime}
\end{subequations}
which is not an exponential decay, a typical indicator of Markovian dynamics.
The decoherence times in the short-time limit are
\begin{subequations}
\begin{eqnarray}
T_{2,A}&\approx & \left(2\sum_{k\in{\mathbb{G}}_A} A_k^2\right)^{-1/2}\sim \frac{N\Omega_e}{{\mathcal A}_{\alpha}^{2}}, \\
T_{2,B}&\approx & \left(\frac{1}{2}\sum_{k\in{\mathbb{G}}_B} E_k^2B_k^2\right)^{-1/4}\sim \frac{N^{5/12}}{B_k^{1/2}{\mathcal A}_{\alpha}^{1/2}}.
\end{eqnarray}\label{FID_T2}
\end{subequations}
Here we have used the facts
that $A_k\sim N^{-2}{\mathcal A}_{\alpha}^2\Omega_e^{-1}$, the number of spin pairs
connected by the extrinsic hyperfine-mediated interaction is about $N^2$, the number of
spin pairs connected by the intrinsic interaction is in the order of $N$, and the typical
hyperfine energy cost $E_k$ for spin flip-flops between neighboring nuclei is in the order
of $N^{-1/3}a_{n,\alpha}\sim N^{-4/3}{\mathcal A}_{\alpha}$.

\begin{figure}[t]
\begin{center}
\includegraphics[width=8cm]{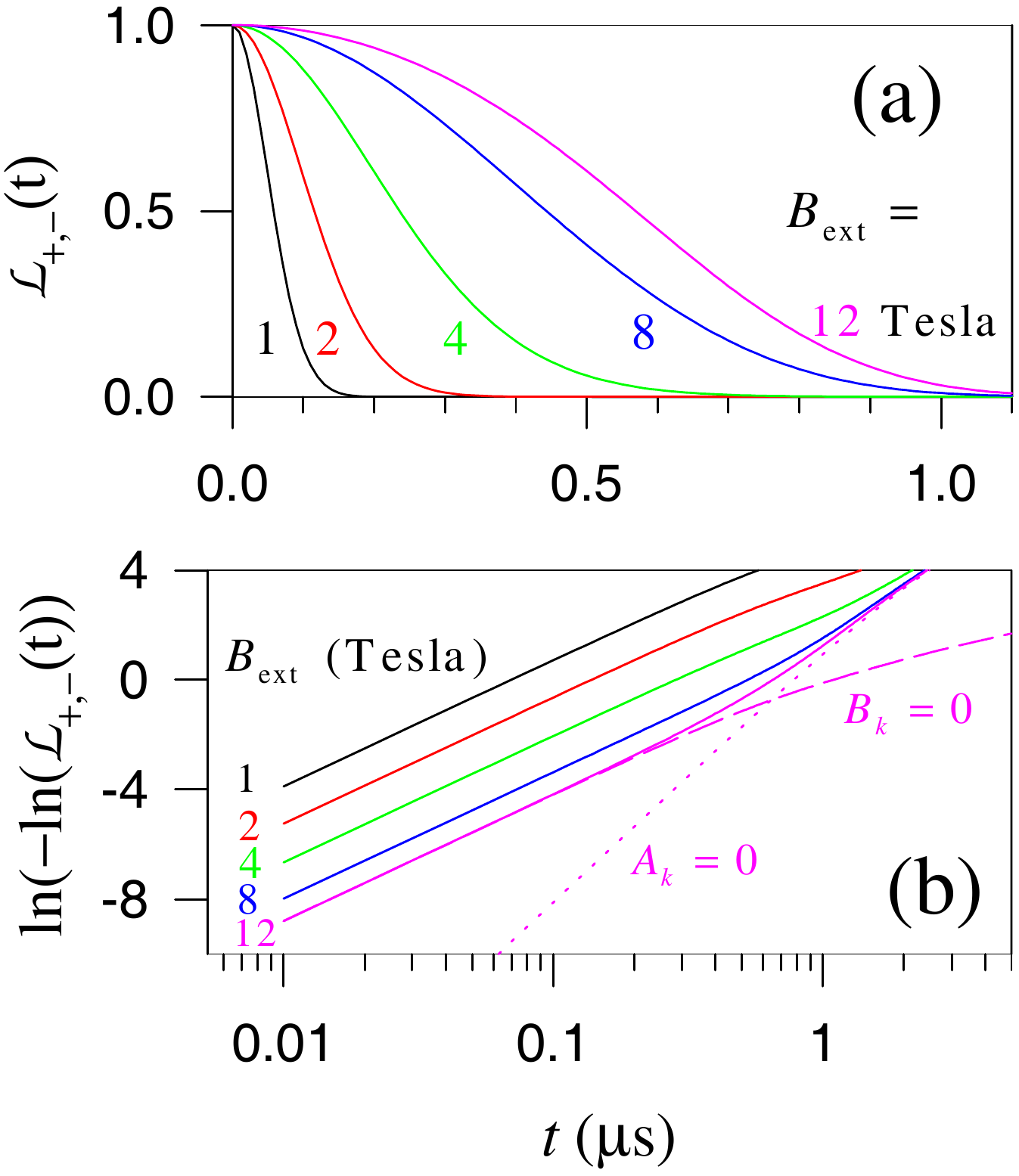}
\end{center}
\caption{(a) Electron spin coherence as functions of time for various field strengths.
(b) The logarithm plot of (a), in which the curve for $B_{\rm ext}=12$~Tesla is compared to the
contribution by the hyperfine-mediated interaction (the dashed line) and that by the intrinsic interaction
(the dotted line), respectively.
The InAs dot is of the size $33 \times 33\times 3$~nm$^{3}$ and the nuclear-spin initial state
$|{\mathcal J}\rangle$ is randomly selected from an ensemble at temperature 1~K. The field strength
is indicated by the numbers for each curve.}
\label{Fig_FIDftfield}
\end{figure}

A few features of the FID in the short-time limit ($t\ll E_k^{-1}$) are summarized as follow.
The decoherence caused by the intrinsic interaction has a quartic exponential decay profile and
that by the hyperfine-mediated interaction has a quadratic one.
So the decoherence is initially dominated by the hyperfine-mediated interaction and
then cross over to the regime dominated by the intrinsic interaction as time increases.
For a QD with $10^5$ nuclear spins and an external field of strength $10$~Tesla,
the decoherence times due to the local and non-local pair-flips, $T_B$ and $T_A$,
are estimated to be both of the order of $1$~$\mu$s. The two decoherence times have very different
dependence on the QD size (measured by $N$) and the field strength (measured by $\Omega_e$),
as can be seen from Eq.~(\ref{FID_T2}).
The hyperfine-mediate interaction could be the dominating decoherence mechanism when the QD size
is small or when the external magnetic field is weak (e.g., $B_{\rm ext}\sim 1$~Tesla).
The intrinsic interacton becomes dominating for large QD or when the field is strong
(see Fig.~\ref{Fig_FIDftfield}). In a QD of proper size and under an external field of proper strength,
the crossover from the quadratic to the quartic exponential decay presents in a visible regime, i.e.,
in a time range where the electron spin decoherence has decayed by a finite amount but not vanished yet [see
the curves for $B_{\rm ext}=12$~Tesla in Fig.~\ref{Fig_FIDftfield}~(b)].
The crossover can be tuned to occur after the coherence has vanished (as for $B_{\rm ext}<10$~Tesla in
Fig.~\ref{Fig_FIDftfield}) or before the decoherence is significant (as for $B_{\rm ext}>14$~Tesla for the
QD in Fig.~\ref{Fig_FIDftfield}, in which the data are not shown), then the decoherence in the visible regime
will be dominated by the hyperfine-mediated interaction or the intrinsic interaction, respectively.

\begin{figure}[t]
\begin{center}
\includegraphics[width=7cm]{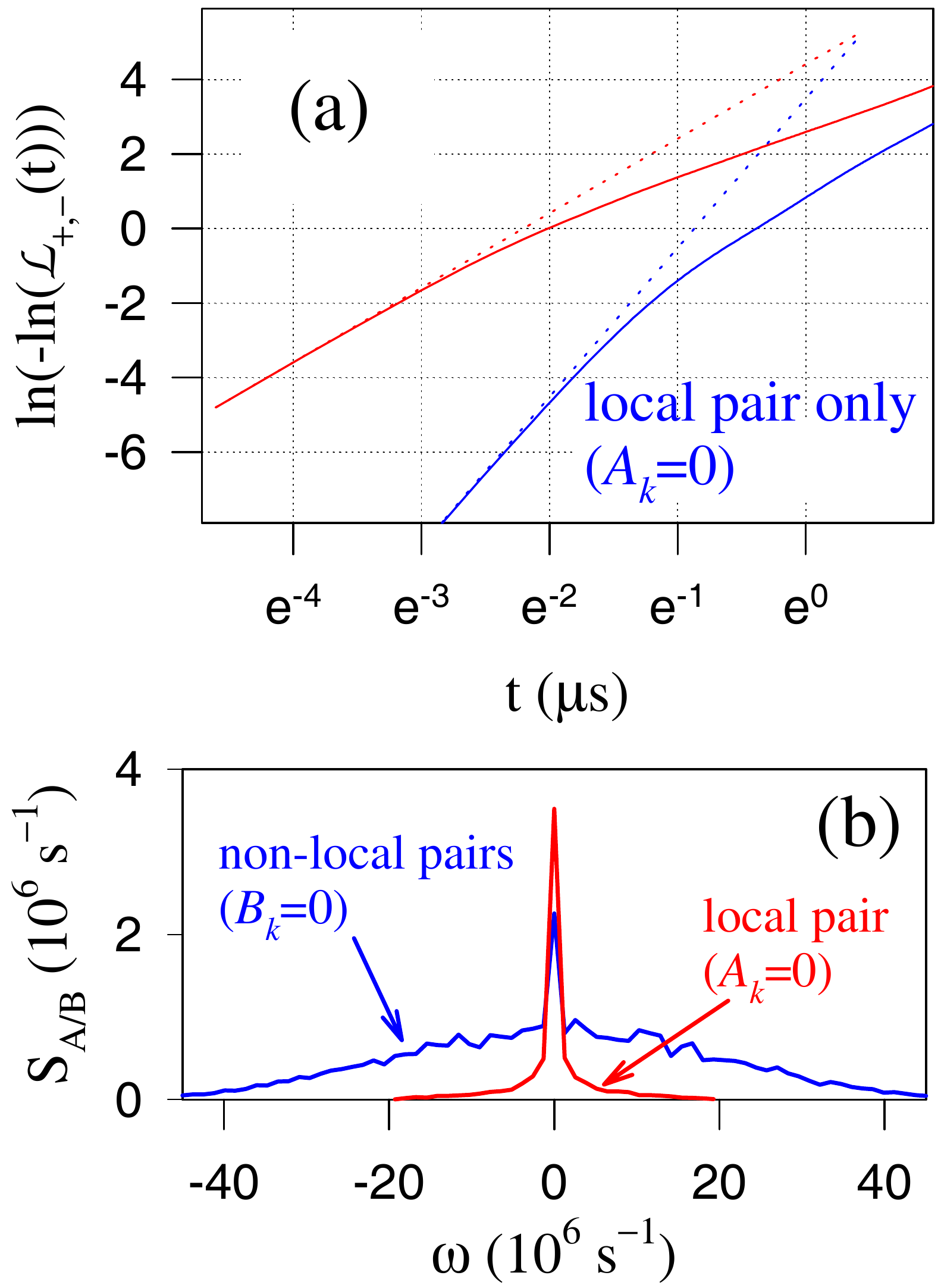}
\end{center}
\caption{(a) The electron spin coherence in a QD
of size $15\times 15\times 2.4$~nm$^{3}$, under a field of 10~Tesla and temperature of 1~K.
The lower curves are calculated with the hyperfine-mediated pair-flips neglected, and the dotted lines
are the short-time profile. The non-Markovian-to-Markovian crossover is observed.
(b) The excitation spectra for the non-local and local pair-flips. }
\label{Fig_markovian}
\end{figure}

In general, the Markovian decay will emerge with time increasing, particularly for $t> E_k^{-1}$,
when the Fermi-Golden rule starts to come into effect [as indicated by the sinc function in
Eq.~(\ref{FID}) and (\ref{FID_spectra})]. Actually, the spin coherence at time $t$
is mostly determined by the pseudo-spins with energy cost $E_k<t^{-1}$, while the effects
of pseudo-spins with higher precession frequency are cancelled out by destructive interference
(between the fast-oscillating sinc functions). In the long-time limit (while
$t\ll B_k^{-1}$ is still satisfied), the excitation spectra $S_{A/B}(\varepsilon)$ in
Eq.~(\ref{FID_spectra}) can be taken as ``flat''
in the frequency range $[-1/t,1/t]$, which covers those pseudo-spins that
contribute significantly to the decoherence. The assumption of ``flat'' spectra is just
the Markovian approximation. So for time much greater than the inverse widths of the excitation
spectra (which are roughly the inverse typical energy cost of the pair-flips, $E_k^{-1}$), the
electron spin coherence is approximated as
\begin{subequations}
\begin{eqnarray}
{\mathcal L}^{A}_{+,-} & \approx &
e^{-t S_{A}\left(t^{-1}\right)  \int 4 \cos^2 x {\rm sinc}^2 x dx } \equiv e^{-\frac{t}{T_{\infty,A}}},
\ \ \ \  \\
{\mathcal L}^{B}_{+,-} & \approx &
e^{-t S_{B}\left(t^{-1}\right)  \int  4 \sin^2 x {\rm sinc}^2 x dx } \equiv e^{-\frac{t}{T_{\infty,B}}},
\end{eqnarray}
\label{Markovian}
\end{subequations}
presenting the exponential decay, a signature of the Markovian dynamics.
The decoherence rate in the Markovian regime is determined by the weighted density of states
of the elementary excitations, i.e., the pair-flips, which, by the
definition of the excitation spectra in Eq.~(\ref{excitationspectra}), is
related to the Fermi-Golden rule by
\begin{subequations}
\begin{eqnarray}
T_{\infty,A}^{-1}=\xi_{A} \sum_k A_k^2\delta\left(t^{-1}-E_k\right), \\
T_{\infty,B}^{-1}=\xi_{B} \sum_k B_k^2\delta\left(t^{-1}-E_k\right), \label{T2B}
\end{eqnarray}
\end{subequations}
where $\xi_{A/B}$ denotes the integral in Eq.~(\ref{Markovian}) and is of the order of unity.

Under the conditions considered in this paper, the Markovian behavior will not be fully
developed before the coherence has totally vanished, but instead a crossover behavior, namely,
a reduction of the exponential index could be observed in the visible regime for certain
QD size and field strength. An example of such non-Markovian-to-Markovian crossover is shown
in Fig.~\ref{Fig_markovian} (a). The typical energy cost of local pair-flips and that of non-local
ones is in the order of ${\mathcal A}_{\alpha}N^{-4/3}$ and ${\mathcal A}_{\alpha} N^{-1}$, respectively, so the
excitation spectrum of local pair-flips is much narrower than that of non-local ones [as shown in
Fig.~\ref{Fig_markovian} (b)]. Thus we expect that the Markovian dynamics will emerge earlier
for the quadratic exponential decay (caused by non-local pair-flips) than for the quartic one
(caused by local pair-flips). To observe the non-Markovian-to-Markovian crossover before the decoherence
is complete, one should have $E_k^{-1}<T_{2,A/B}$ which is satisfied in relatively small QDs (for decoherence
contributed by local pair-lips) or under relatively strong external field (for decoherence contributed
by non-local pair-flips).

\begin{figure}[t]
\begin{center}
\includegraphics[width=8cm]{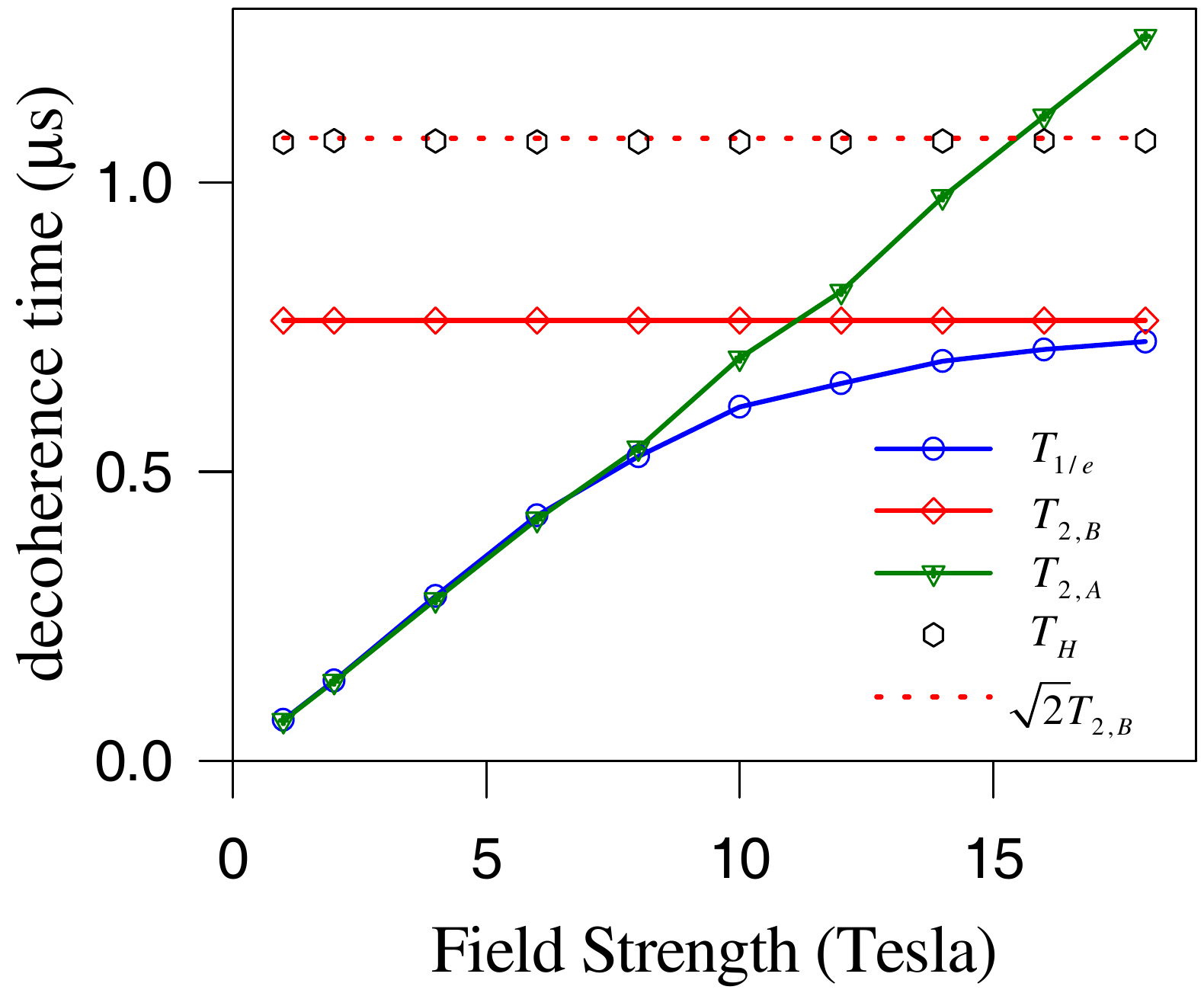}
\end{center}
\caption{Field dependence of decoherence times ($T_{1/e}$ - time for FID coherence being
$1/e$ of its initial value, $T_{2,A}$ - FID decoherence time resulting solely from hyperfine-mediated pair-flips,
$T_{2,B}$ - FID decoherence time resulting solely from the
intrinsic nuclear spin interaction,  $T_H$ - decay time of the Hahn echo signal).
The $\sqrt{2}T_{2,B}$ is plotted to compare with the Hahn echo decay time.
The QD is as in Fig.~\ref{Fig_FIDftfield}.}
\label{Fig_T2TM_field}
\end{figure}

\begin{figure}[t]
\begin{center}
\includegraphics[width=8cm]{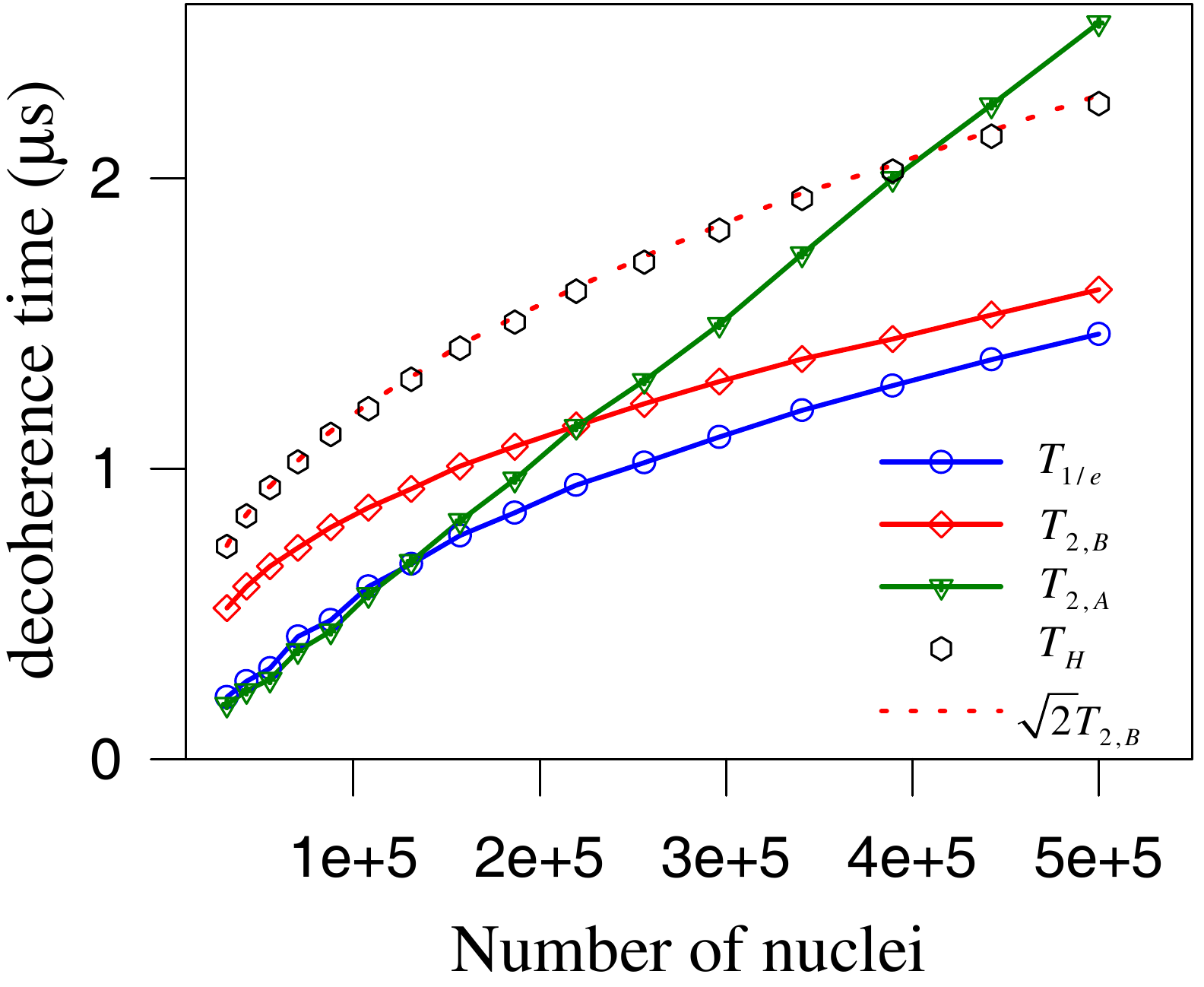}
\end{center}
\caption{QD-size dependence of decoherence times (see Fig.~\ref{Fig_T2TM_field} and the text
for definition). The $\sqrt{2}T_{2,B}$ is plotted to compare with the Hahn echo decay time.
The QD size is varied with fixed width:depth:height ratio 33:33:6, the field strength is 10 Telsa,
and the temperature is 1~K.}
\label{Fig_T2TM_size}
\end{figure}

Due to the rich crossover behaviors discussed above, the decoherence in general cannot
be characterized by a single time parameter $T_2$. We introduce a decoherence time
$T_{1/e}$ to quantify the time when the electron spin coherence has decayed to $1/e$
of its initial value.
The decoherence time $T_{1/e}$ can approach $T_{2,A}$, $T_{2,B}$, $T_{\infty,A}$, or
$T_{\infty,B}$, depending on the QD size and the field strength which set the conditions
for the crossovers. The field- and the size-dependence of different decoherence times shown
in Fig.~\ref{Fig_T2TM_field} and Fig.~\ref{Fig_T2TM_size}, respectively,
are consistent with the analysis above. For comparison, we also plot the decoherence time
measured by the Hahn echo $T_H$, which will be studied in the next Section.

\section{Hahn echo signal}
\label{S_echo}

The FID studied in the previous section is not visible in ensemble experiments in which
dephasing due to inhomogeneous broadening is faster by orders of magnitude than
decoherence by dynamical entanglement. It might be difficult in the near future to
directly observe the single-system dynamics by filtering out the inhomogeneous broadening with
projective measurement of the local Overhauser
field.\cite{Espin_HF_1_Loss,Giedke_spinmeasure,Klauser_spinmeasure,Imamoglu_WeakMeasurement} Thus the spin
echo may be used to study the electron spin decoherence problem in QDs. In Hahn echo experiments,
the static Overhauser field experienced by the electron spin effectively changes its
sign each time the electron spin is flipped by a short $\pi$-pulse. Considering the simplest
configuration in which only one pulse is applied at $t=\tau$, the precession phase accumulated
from the random local fields is eliminated at $t=2\tau$. Thus the decay of Hahn echo signals
at $t=2\tau$ is solely due to the dynamical entanglement. The decay time of Hahn echo
signals is usually used to quantify the spin decoherence time. We shall show, however, the Hahn
echo decay time cannot be equated with the decoherence time, as the decoherence by hyperfine-mediated
interaction will be virtually eliminated from the echo signal and that by intrinsic interaction
is also suppressed to some extent.

The $\pi$-rotation or flip of the electron spin for Hahn echo can be operated by a GHz microwave
pulse,\cite{ESR_silicon_T2,Abe_Si,Abe_Si29} which is only marginally fast enough for eliminating
the rapid dephasing by inhomogeneous broadening in III-V
compound QDs. In optical control of an electron spin,\cite{Chen_Raman} an arbitrary rotation of the electron spin
can be completed in the timescale of $10$~ps via exciton-mediated Raman processes. With respect to
the timescale of the electron spin decoherence, the optical pulses can be considered instantaneous.
The recent experiment on double GaAs QDs also employs a rather long DC voltage pulse to control the
singlet-triplet transition to realize the spin echo,\cite{Marcus_T2} which does not satisfy the
instantaneous pulse condition. It is certainly interesting to study the electron spin decoherence under
the control of finite-duration pulses,\cite{Lidar_DD_longpaper} but in this paper we would rather focus
on the case of instantaneous pulses.

\begin{figure}[b]
\begin{center}
\includegraphics[width=8cm]{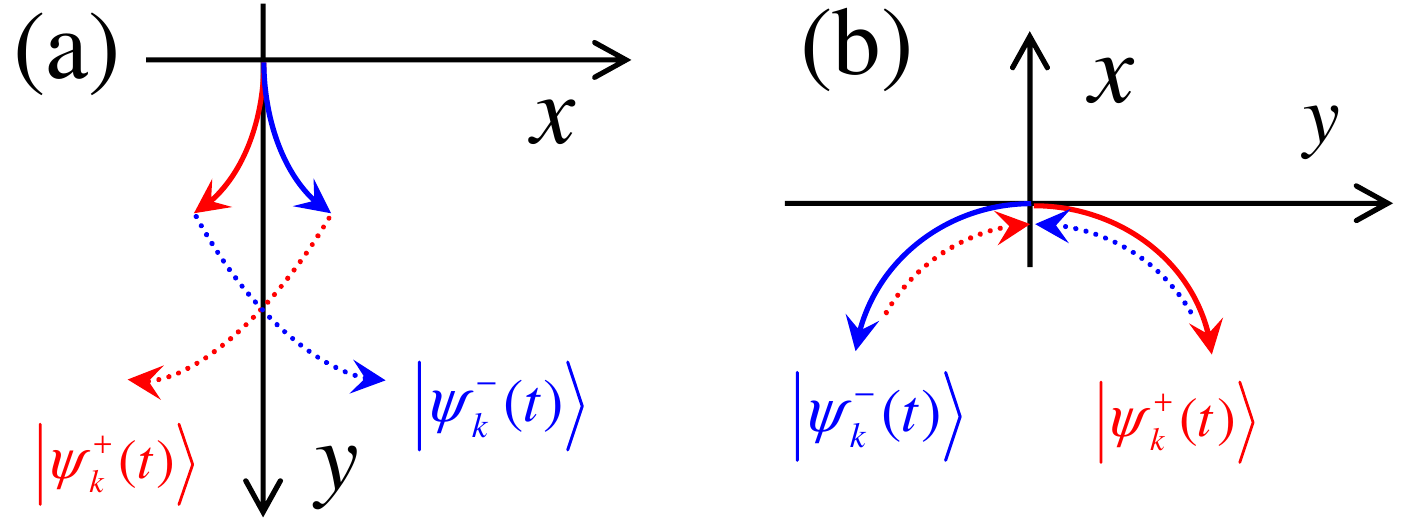}
\end{center}
\caption{(a) The trajectories (projected to the $x$-$y$ plane) of the conjugate pseudo-spins
by the intrinsic interaction. The conjugate pseudo-spins $|\psi_{k}^{\pm}\rangle$ exchange their
pseudo-fields ${\boldsymbol \chi}_{k}^{\pm}$ at $t=\tau$ when the electron spin is flipped by a short pulse.
The solid and dotted curves denote the trajectories before and after the flipping pulse, respectively.
(b) The same as (a) but for a non-local pair-flip by the hyperfine-mediated interaction.
}
\label{Fig_geometryecho}
\end{figure}

As our general rule, the Hahn echo signal is determined by the distinguishability
$\delta_k^2$ between the conjugate pseudo-spins.
In the rotating reference frame rested on the electron spin,
the pseudo-spin states after the $\pi$-pulse applied at $\tau$ are
\begin{eqnarray}
|\psi_{k}^{\pm}(t>\tau)\rangle = \hat{U}_{k}^{\mp}(t-\tau)\hat{U}_{k}^{\pm}(\tau)|\downarrow\rangle,
\end{eqnarray}
which is equivalent to the conjugate pseudo-spin states $|\psi_{k}^{\pm}(t)\rangle$ exchanging their
pseudo-fields ${\boldsymbol \chi}_{k}^{\pm}$ when the electron spin is flipped.
The trajectories of the Bloch vectors, projected to the $x$-$y$ plane,
are shown in Fig.~\ref{Fig_geometryecho}, both for local and for non-local pair-flips.
The pseudo-spin states after the flip pulse are
\begin{widetext}
\begin{eqnarray}
|\psi_{k}^{\pm}(\tau+t')\rangle &=& \Bigg[ \cos\frac{\chi_{k}^{\mp}t'}{2}\cos\frac{\chi_{k}^{\pm}\tau}{2}
-\sin\frac{{\boldsymbol \chi}_{k}^{\mp}t'}{2}\cdot\sin\frac{{\boldsymbol \chi}_{k}^{\pm}\tau}{2}
\nonumber \\
& & -i\hat{\boldsymbol \sigma}_k\cdot \left(\cos\frac{\chi_{k}^{\mp} t'}{2}\sin\frac{{\boldsymbol\chi}_{k}^{\pm} \tau}{2}+
\cos\frac{\chi_{k}^{\pm} \tau}{2}\sin\frac{{\boldsymbol\chi}_{k}^{\mp} t'}{2}
+\sin\frac{{\boldsymbol\chi}_{k}^{\mp} t'}{2} \times
\sin\frac{{\boldsymbol\chi}_{k}^{\pm} \tau}{2}\right)\Bigg]|\downarrow\rangle. \ \
\end{eqnarray}
\end{widetext}
The Hahn echo signal is
\begin{eqnarray}
{\mathcal L}^{\rm s}_{+,-}(2\tau)= \prod_{k}\left|\langle\psi_{k}^{-}(2\tau)|\psi_{k}^{+}(2\tau)\rangle\right|.
\end{eqnarray}

With the same arguments as in Sec.~\ref{S_FID}, the electron spin decoherence can be factorized into
two contributions, namely ${\mathcal L}^{\rm s}_{+,-}={\mathcal L}^{A}_{+,-}\times {\mathcal L}^{B}_{+,-}$ with
\begin{eqnarray}
{\mathcal L}^{A/B}_{+,-} \approx \prod_{k\in{\mathbb{G}}_{A/B}}\exp\left(-\delta_k^2/2\right).
\end{eqnarray}
For the local pair-flips driven by the intrinsic nuclear spin interaction, the distance $\delta_k$ at $t=2\tau$
is calculated by setting $D_k=A_k=0$:
\begin{eqnarray}
{\delta_k^2}&\cong& \frac{1}{4}(2\tau)^4E_k^2B_k^2{\rm sinc}^4\frac{E_k\tau}{2}\nonumber \\
&\approx & \frac{1}{4}(2\tau)^4E_k^2B_k^2,
\end{eqnarray}
where the approximation in the second line holds for $\tau\ll E_k^{-1}$.
The short-time behavior is
\begin{eqnarray}
{\mathcal L}^B_{+,-}(2\tau)\approx e^{-(2\tau)^4/T_{H,B}^4},
\end{eqnarray}
the same as the FID signal except that the decay time
\begin{eqnarray}
T_{H,B}=\sqrt{2}T_{2,B}.
\label{THB}
\end{eqnarray}
For non-local pair-flips driven by the hyperfine-mediated interaction, the pseudo-fields
${\boldsymbol\chi}_{k}^{\pm}$ are opposite to each other if the energy cost due to the diagonal
nuclear spin interaction is neglected ($D_k=0$). In this case, the pseudo-spins simply reverse their
precession directions when the electron spin is flipped, returning to its original states
at the echo time, as shown in Fig.~\ref{Fig_geometryecho} (b), so the decoherence by the
hyperfine-mediated pair-flips is fully eliminated if $D_k$ is negligible. As the
leading order of the distance $\delta_k$ vanishes at the echo time, the second-order effect of the
diagonal nuclear spin interaction becomes important. The distance $\delta_k$ for non-local pair-flips,
including the effect of $D_k$, is given by
\begin{eqnarray}
{\delta_k^2} &\cong& \frac{1}{4}(2\tau)^4D_k^2A_k^2 {\rm sinc}^4\frac{E_k\tau}{2}\nonumber \\
&\approx & \frac{1}{4}(2\tau)^4D_k^2A_k^2,
\end{eqnarray}
where again, the approximation in the second line holds for $\tau\ll E_k^{-1}$.
The residual decoherence due to the interplay of the hyperfine-mediated interaction and
the diagonal nuclear spin interaction has the quartic exponential form for $\tau\ll E_k^{-1}$ as
\begin{eqnarray}
{\mathcal L}^{A}_{+,-}(2\tau)\cong e^{-(2\tau)^4/T^4_{H,A}}.
\end{eqnarray}
The corresponding decoherence time is
\begin{eqnarray}
T_{H,A}=\left(\sum_{k\in{\mathbb G}_A}{ D_k^2A_k^2/8} \right)^{-1/4}\sim \frac{N^{1/2}\Omega_e^{1/2}}{{\mathcal A}_{\alpha}D_k^{1/2}},
\end{eqnarray}
proportional to the square root of the field strength.

\begin{figure}[b]
\begin{center}
\includegraphics[width=8cm]{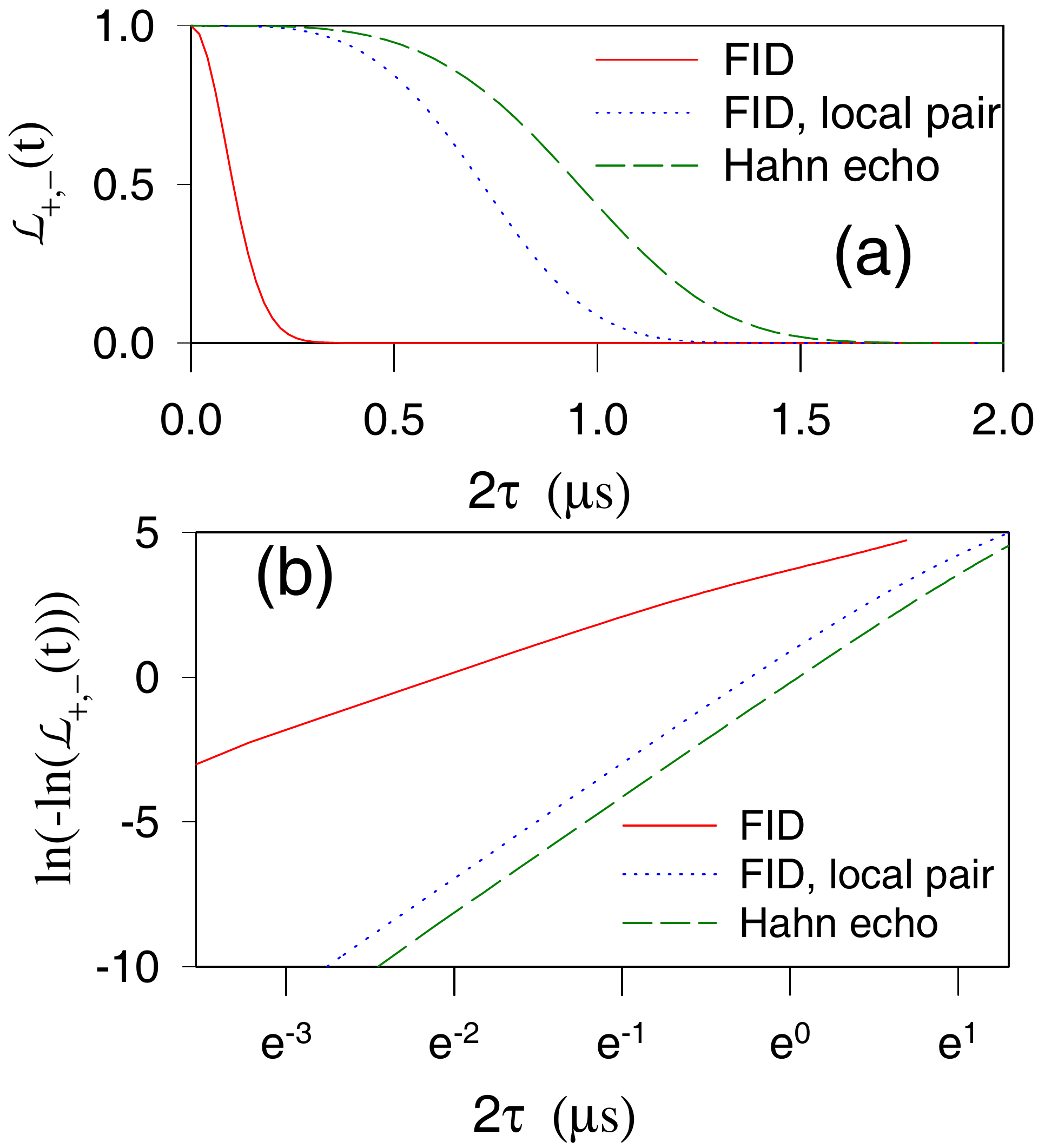}
\end{center}
\caption{(a) Comparison of the Hahn echo (dashed green line) and the FID (solid red line) signals.
The FID signal is also shown with the hyperfine-mediated pair-flips neglected (dotted  blue line).
(b) The logarithm plot of (a). The QD is as in Fig.~\ref{Fig_FIDftfield}, with $B_{\rm ext}=2$~Tesla.
}
\label{Fig_Hahnecho}
\end{figure}

Now we compare the echo signals to the FID signal in single-system dynamics (see Figs.~\ref{Fig_T2TM_field}
and \ref{Fig_Hahnecho}).
The observation is that the decoherence due to the dynamical quantum entanglement
in general is not measured by the Hahn echo signals. First, the FID and the echo signals could have qualitatively
different short-time behaviours, as the former could have quadratic exponential decay due to the hyperfine-mediated
interaction while the latter always has the quartic exponential decay. Second, the decay timescale could be dramatically
different from the FID to the Hahn echo. The difference in the decay time can be discussed in three regimes
of field strength and QD size, roughly divided by
\begin{subequations}
\begin{eqnarray}
& {\rm I}: &  \Omega_e \lesssim {\mathcal A}_{\alpha}N^{-1/6}, \label{field1} \\
& {\rm II}: & {\mathcal A}_{\alpha}N^{-1/6}\ll \Omega_e \lesssim {\mathcal A}_{\alpha}^{3/2}N^{-7/12}B_k^{-1/2}, \label{field2} \\
& {\rm III}: & {\mathcal A}_{\alpha}^{3/2}N^{-7/12}B_k^{-1/2} \ll \Omega_e, \label{field3}
\end{eqnarray}
\label{fieldscondition}
\end{subequations}
corresponding to
\begin{subequations}
\begin{eqnarray}
& {\rm I}: & T_{H,B}\gtrsim T_{H,A}, \label{time1} \\
& {\rm II}: & T_{H,A} \gg  T_{H,B} \sim T_{2,B} \gtrsim T_{2,A}, \label{time2} \\
& {\rm III}: & T_{2,A} \gg T_{2,B}. \label{time3}
\end{eqnarray}
\label{timescondition}
\end{subequations}
In regime I, the hyperfine-mediated
interaction is important both in the FID and in the Hahn echo decay.
As the excitation of non-local pair-flips is eliminated in the leading
order at the echo time [See Fig.~\ref{Fig_geometryecho} (b)], the echo decay time is
much longer than the FID decoherence time, namely,
\begin{eqnarray}
T_{2,A}/T_{H,A}\sim \sqrt{N\Omega_eD_k/{\mathcal A}_{\alpha}^2} \ll 1,
\end{eqnarray}
for QDs of reasonable sizes. The pair-correlation approximation, however, is not valid in regime I,
since the condition that the number of pair-flips in the timescale of $T_{H,A}$ be much fewer than
the number of nuclei requires $N\gg A_{\alpha}^{4/3}\Omega_e^{-2/3}D_k^{-2/3}$ for the hyperfine mediated
non-local pair-flips and $N\ll A_{\alpha}^{4/3}\Omega_e^{-2/3}D_k^{-2/3}$ for the local pair-flips driven
by the intrinsic interaction (see Appendix~\ref{Append_pairflip} for estimation of the pair-flip numbers),
which cannot be simultaneously satisfied. Theories beyond the pair-correlation approximation need to
be developed to explore this regime. In regime II defined by Eq.~(\ref{field2}), the hyperfine-mediated
interaction contributes significantly to the FID but virtually nothing to the Hahn echo decay.
In this regime, the decay time measured in echo signals is much longer than the FID decoherence time as
\begin{eqnarray}
T_{H,B}=\sqrt{2}T_{2,B}\gg T_{2,A}.
\end{eqnarray}
In regime III, the hyperfine-mediated interaction is strongly suppressed, so the FID and the
Hahn echo decay, both determined by the intrinsic nuclear spin interaction, have essentially
the same decay profile but different decay times (differing by a factor of $\sqrt{2}$).
Thus, in all the three regimes, there is substantial difference between the FID and the Hahn echo decay.
Such difference is indeed a consequence of the modified dynamics of a mesoscopic bath (the nuclear spins)
under the manipulation of the quantum object (the electron spin). In semiclassical spectral
diffusion theories, the electron spin experiences passively the ``background'' of fluctuating  local fields.
In the full quantum theory, the electron spin dynamics actively alters the mesoscopic bath dynamics
(as the nuclear spin pair-flips depends dramatically on the electron spin state). Such a view paves the path
toward manipulation of the nuclear spin dynamics in mesoscopic QDs and control of the electron spin decoherence.

The active modification of the bath dynamics by the manipulation of the quantum object also manifests itself
in the effect of the diagonal nuclear spin interaction on the Hahn echo signal.
The diagonal nuclear spin interaction contributes much less energy cost of a pair-flip than the hyperfine
interaction ($D_k\ll E_k$) and has no direct effect on the effective Overhauser field felt by the electron spin.
Thus it is expected that the diagonal interaction terms should have little effect on the the electron spin decoherence.
The calculation of the FID signal in Sec.~\ref{S_FID} indeed confirms such a semiclassical argument.
The difference between the quantum theory and its semiclassical counterparts becomes significant when the electron
spin is under control. In semiclassical theories, the electron spin only passively detects the fluctuating local field
which is essentially independent of the electron dynamics, so the diagonal nuclear spin interaction should contribute
no more than it does in the FID configuration. In the quantum theory, the nuclear spin dynamics conditioned on the
electron spin state is held responsible for the electron spin decoherence. The diagonal nuclear spin interaction
affects the nuclear spin dynamics significantly and thus contributes to the decoherence when the
nuclear spin dynamics is under active control by the manipulation of the electron spin state. In next sections, we
will see the importance of the diagonal nuclear spin interaction in determining some qualitative decoherence features.

\section{Disentanglement and recoherence}
\label{S_disentanglement}

In the previous section we have seen that the decoherence caused by the hyperfine-mediated interaction
can be nearly eliminated in Hahn echo signals.
Such elimination of decoherence is realized because
the hyperfine-mediated pair-flips are reversed when the
electron spin is flipped at $t=\tau$ [see Fig.~\ref{Fig_geometryecho} (b)] and hence
the pseudo-spin states $|{\psi}_{k}^{\pm}(t)\rangle$ return to their original positions at $t=2\tau$.
As the decoherence is caused by the entanglement between the electron spin
and the nuclear spins, the elimination of decoherence, also called ``recoherence'', can be understood
in terms of disentanglement. For non-local pair-flips under Hahn echo control, the disentanglement is realized
at $t=2\tau$ when the electron spin and the pseudo-spin are brought back to a factorized state
$C_+|+\rangle|\psi_{k}^{+}(2\tau)\rangle+C_-|-\rangle|\psi_{k}^{-}(2\tau)\rangle$
with $|\psi_{k}^{\pm}(2\tau)\rangle=|\psi_{k}^{\pm}(0)\rangle=|\downarrow\rangle$.

In general, a pseudo-spin (or the environment) is disentangled from the electron spin
(or the quantum system in contact with the environment) whenever its states
$|\psi_{k}^{\pm}(t)\rangle$ for different electron spin states $|\pm\rangle$ become identical, but need
not return to the original states $|\psi_{k}^{\pm}(0)\rangle$. When the electron spin is flipped,
the conjugate pseudo-spins exchange their pseudo-fields, so their trajectories on the
Bloch sphere will inevitably intersect, leading to the disentanglement.\cite{Note_disentanglement}
The intersection of conjugate pseudo-spins driven by the intrinsic nuclear spin interaction is
actually seen in Fig.~\ref{Fig_geometryecho} (a), which, unlike that for non-local pair-flips,
occurs at a time different from the echo time. The intersection time, in general, is different for
different pair-flips, making it impossible to fully recover the lost electron spin coherence.
Nonetheless, in the short-time limit ($\tau\ll E_k^{-1}$), the distance between conjugate
pseudo-spins can be eliminated in the leading order of time, simultaneously for the same group of
pair-flips, as have been seen for non-local pair-flips. It is straightforward to verify that
for the local pair-flips, such coincident disentanglement
does take place at $t=\sqrt{2}\tau$ in the single-pulse Hahn echo configuration [the magic number
$\sqrt{2}$ can be understood by noticing that $\delta_k\propto \tau^2$ at short time limit and
$\left(\sqrt{2}\tau\right)^2-\tau^2=\tau^2$].

With the leading order contribution vanishing, the distinguishability between conjugate pseudo-spins
at the disentanglement time ($\sqrt{2}{\tau}$), including the effect of the diagonal nuclear spin
interaction, in the next leading order of $\tau$ for local pair-flips is
\begin{eqnarray}
{\delta_k}\left(\sqrt{2}\tau\right)\sim E_kB_kD_k\tau^3.
\end{eqnarray}
Note that the residual decoherence would otherwise be of a higher order in $\tau$ if $D_k$ were set
to zero:
\begin{eqnarray}
{\delta_k}\left(\sqrt{2}\tau\right)\sim E_kB_k^3\tau^4.
\end{eqnarray}
Here we see again that the small effect of the diagonal
nuclear spin interaction (the $D_k$ term) emerges when the nuclear spin
dynamics is under control.

A semiclassical spectral diffusion theory also predicts a coherence recovery at a time earlier than
the echo time ($2\tau$) when the random field memory time ($\tau_c$) is comparable to
the pulse delay time.\cite{Shaowen} But the semiclassical theory differs qualitatively from
the quantum counterpart in at least two aspects:
(1) The recovery time depends on $\tau_c$, being about $\tau+\tau_c$ for $\tau_c<\tau$ and
    approaching $2\tau$ when $\tau_c\gg\tau$; and
(2) The exponential index of the decoherence profile in the spectral diffusion theory
    is essentially unchanged by the control.
The latter difference points directly to a fundamental shortness of the semiclassical picture: It does
not take into account the active control of the local field fluctuation by the flip of the quantum object
but considers the quantum evolution of the bath as a fluctuating field experienced passively by the quantum object.

\begin{figure}[t]
\begin{center}
\includegraphics[width=8cm]{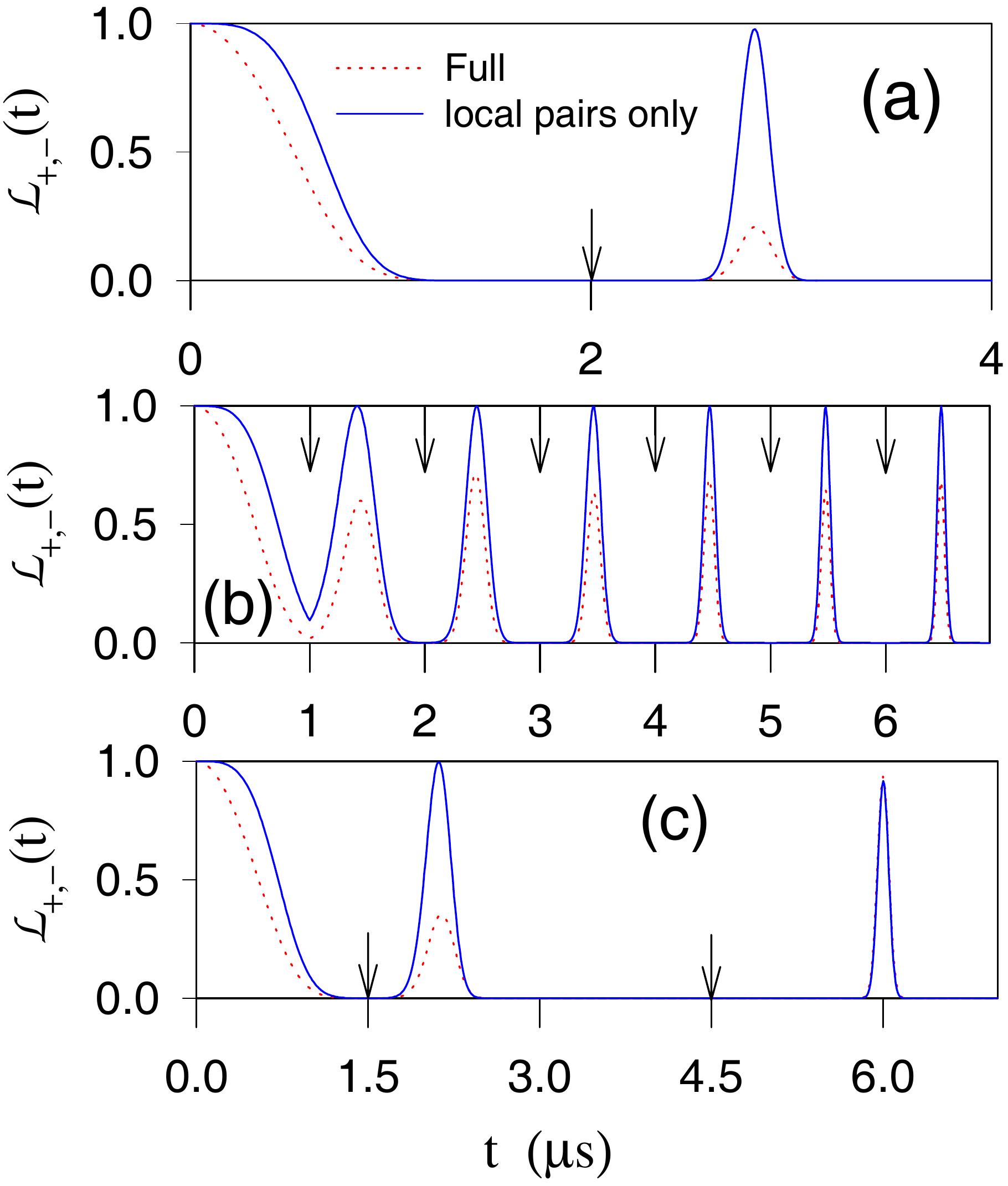}
\end{center}
\caption{(a) The electron spin coherence under the control of a short $\pi$-pulse applied at
$\tau=2$~$\mu$s (when the FID signal has vanished), the recoherence at $\sqrt{2}\tau$ is pronounced
while no signal survives at the echo-time $2\tau$.
(b) The electron spin coherence under the control of a sequence of evenly spaced pulses.
(c) The electron spin coherence under the control of a Carr-Purcell pulse sequence.
The arrows indicate positions of the short pulses. The solid blue (dotted red) lines are calculated with(out)
including the hyperfine-mediated pair-flips. The QD is as in Fig.~\ref{Fig_FIDftfield}
with $B_{\rm ext}=10$~Tesla. No ensemble average is done.}
\label{Fig_CP}
\end{figure}

Inclusion of contributions from both local and non-local pair-flips fields the pulse-controlled decoherence
in the short-time limit,
\begin{eqnarray}
{\mathcal L}^{\rm s}_{+,-}(\sqrt{2}\tau)\cong e^{-\left(\sqrt{2}\tau\right)^6/T_{{\rm eff},B}^6}\times
e^{-\left(\sqrt{2}-2\right)^2\tau^2/T_{2,A}^2},
\end{eqnarray}
where the effective decoherence time due to the residual entanglement from local pair-flips is defined by
\begin{eqnarray}
T_{{\rm eff}, B}^{-6}\sim \sum_{k\in{\mathbb G}_B} E_k^2B_k^2D_k^2.
\end{eqnarray}
The electron coherence is recovered at $\sqrt{2}\tau$, even when the pulse is
applied after the coherence has been totally lost in FID (for $\tau>T_{2,B}$).
Such recoherence in single-system dynamics and the spin echo in ensemble dynamics have
different physical bases: the former is due to quantum disentanglement, but the latter
is due to classical refocusing of random phases. The difference between the two is
already evidenced by their different occurrence time. Fig.~\ref{Fig_CP} (a) plots an example of
the real-time dependence of the electron spin coherence under a single-pulse control,
which demonstrates the recovery of the coherence even when the FID signal
has vanished for the chosen delay time. The contribution from the hyperfine-mediated pair-flips,
which are not reversed at $\sqrt{2}\tau$, makes the coherence to be only partially recovered.
Under a stronger magnetic field, a full recovery at $\sqrt{2}\tau$ can be realized as the
hyperfine-mediated interaction is fully suppressed (not shown).

\begin{figure}[t]
\begin{center}
\includegraphics[width=8cm]{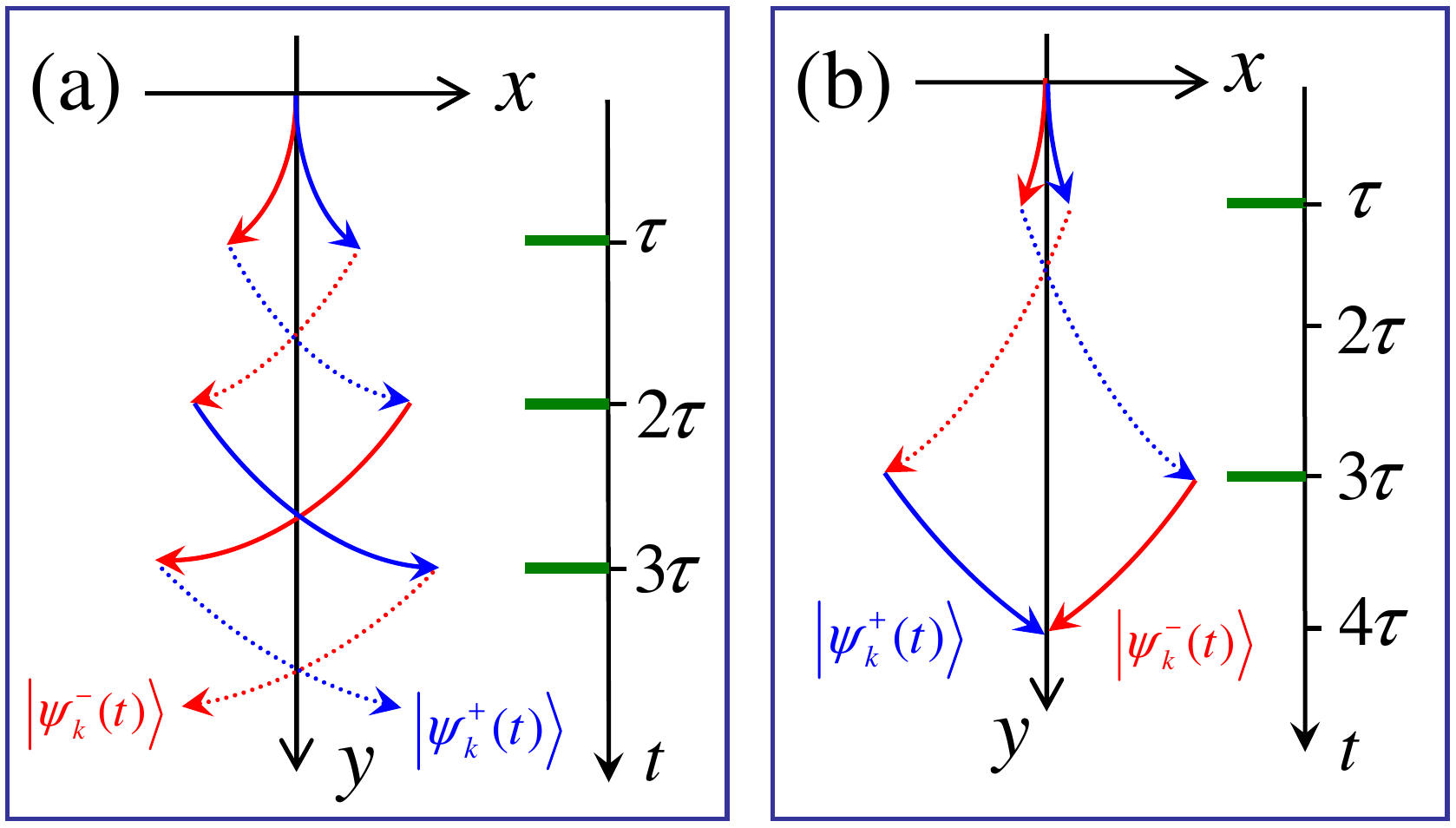}
\end{center}
\caption{Trajectories of conjugate pseudo-spins from the intrinsic nuclear spin interaction (a) under
the control of a sequence of pulses equally spaced and (b) under the control of a Carr-Purcell two-pulse sequence.
The green bars on the time axis represent the short pulses flipping the electron spin.}
\label{Fig_geometryCP}
\end{figure}

As illustrated in Fig.~\ref{Fig_geometryCP} (a), a sequence of pulses can force the trajectories of
conjugate pseudo-spins to cross into each other again and again.
In the short-time limit (the delay time $\tau\ll E_k^{-1}$), an
equally spaced pulse sequence eliminates the entanglement due to the local pair-flips
up to the $\tau^2$ terms, leading to a sequence of recoherence at
\begin{eqnarray}
t_n=\sqrt{n(n+1)}\tau, \ \ n=1, 2, \ldots.
\end{eqnarray}
One can verify these magic numbers by using the quadratic dependence of $\delta_k$ on time and checking that
$\left(\sqrt{n(n+1)}\tau\right)^2-(n\tau)^2=(n\tau)^2-\left(\sqrt{(n-1)n}\tau\right)^2$.
The decoherence caused by the hyperfine-mediated pair-flips under the control of evenly spaced pulses is bounded,
which can be understood from the reversed precession of the corresponding pseudo-spins.
Fig.~\ref{Fig_CP} (b) confirms our above analysis. By using a long sequence of
pulses at short intervals, the electron spin coherence can be preserved for an
arbitrarily long time until other
mechanisms of decoherence (such as phonon scattering) come into effect.

The recoherence in the control configurations discussed above, however, is not observable in ensemble experiments,
since the rapid dephasing due to inhomogeneous broadening will prevent any coherence being observed except
around the spin-echo time. It is desirable to design a pulse sequence to force the recoherence from
disentanglement to coincide with a spin-echo from phase-refocusing so that the disentanglement effect can be
studied in ensemble experiments. This possibility can be seen from Fig.~\ref{Fig_geometryCP} (b).
When the electron spin is flipped by two short pulses, for example, the disentanglement time after the second pulse
can be adjusted by changing the delay time between the two pulses. It is shown straightforwardly that
when the pulse sequence is designed such that the first pulse is applied at $t=\tau$ and the second at $t=3\tau$,
the disentanglement coincides with the echo time at $t=4\tau$ (by virtue of the change in $\delta_k$,
$(4\tau)^2-(3\tau)^2-[(3\tau)^2-\tau^2]+\tau^2=0$). Obviously, the hyperfine-mediated non-local pair-flips
are disentangled from the electron spin at the echo time. A two-pulse sequence so designed happens to be
the famous Carr-Purcell sequence, widely used in NMR experiments to dynamically decouple the nuclear spins.
The disentanglement studied here, however, is fundamentally different from the dynamical decoupling,
as will be discussed later in the next section.

In the rotating frame of the electron spin, the pseudo-spin states at the echo time
after a Carr-Purcell sequence is
\begin{eqnarray}
\left|{\psi}_{k}^{\pm}(4\tau)\right\rangle &=& e^{-i{\boldsymbol \theta}_{0}^{\pm}\cdot\hat{\boldsymbol \sigma}/2}
e^{-i{\boldsymbol \theta}_{0}^{\mp}\cdot\hat{\boldsymbol \sigma}}
e^{-i{\boldsymbol \theta}_{0}^{\pm}\cdot\hat{\boldsymbol \sigma}/2}|\downarrow\rangle \nonumber \\
&=& e^{-i{\boldsymbol \theta}_{1}^{\mp}\cdot\hat{\boldsymbol \sigma}/2}
e^{-i{\boldsymbol \theta}_{1}^{\pm}\cdot\hat{\boldsymbol \sigma}/2}|\downarrow\rangle \nonumber \\
&= & e^{-i{\boldsymbol \theta}_{2}^{\pm}\cdot\hat{\boldsymbol \sigma}/2}|\downarrow\rangle ,
\label{psiCP}
\end{eqnarray}
with the series of angles of rotation defined by
\begin{subequations}
\begin{eqnarray}
{\boldsymbol \theta}_{0}^{\pm} & \equiv & {\boldsymbol \chi}_{k}^{\pm}\tau,\label{theta0}\\
e^{-i{\boldsymbol \theta}_{l+1}^{\pm}\cdot\hat{\boldsymbol \sigma}/2}
&\equiv &
e^{-i{\boldsymbol \theta}_{l}^{\mp}\cdot\hat{\boldsymbol \sigma}/2}
e^{-i{\boldsymbol \theta}_{l}^{\pm}\cdot\hat{\boldsymbol \sigma}/2},
\label{angle_iteration}
\end{eqnarray}
\label{iteration}
\end{subequations}
The vector ${\boldsymbol\theta}_{l}^{\pm}$ has a geometrical interpretation as the pseudo-spin effective precession
angle ${\theta}_{l}^{\pm}$ about the axis along ${\boldsymbol\theta}_{l}^{\pm}$ at $\tau$, $2\tau$, and $4\tau$ for
$l=0$, $1$ and $2$, respectively (the pseudo-spin index $k$ is understood where no confusion is caused).
Eq.~(\ref{angle_iteration}) yields a geometrical recursion relation:
\begin{subequations}
\begin{eqnarray}
 \sin\frac{{\boldsymbol \theta}_{l+1}^{\pm}}{2} & = &
  \cos\frac{{\theta}_{l}^{\mp}}{2}\sin\frac{{\boldsymbol \theta}_{l}^{\pm}}{2}
 +\cos\frac{{\theta}_{l}^{\pm}}{2}\sin\frac{{\boldsymbol \theta}_{l}^{\mp}}{2}
 \nonumber \\ &&
+ \sin\frac{{\boldsymbol \theta}_{l}^{\mp}}{2}\times \sin\frac{{\boldsymbol \theta}_{l}^{\pm}}{2},\\
\cos\frac{{\theta}_{l+1}^{\pm}}{2} & = &
  \cos\frac{{\theta}_{l}^{\mp}}{2}\cos\frac{{\theta}_{l}^{\pm}}{2}
 - \sin\frac{{\boldsymbol \theta}_{l}^{\mp}}{2}\cdot \sin\frac{{\boldsymbol \theta}_{l}^{\pm}}{2}. \ \ \ \ \ \ \ \
\end{eqnarray}\label{angleCP}
\end{subequations}
The distinguishability between conjugate pseudo-spins at the echo time $4\tau$ in the leading order of delay time
is
\begin{eqnarray}
{\delta^2_k} \cong 2^4 \tau^6  D_k^2 \left(B_k E_k-A_k D_k\right)^2,
\end{eqnarray}
for $\tau\ll E_k^{-1}$. Note the decoherence would otherwise be of a higher order
in $\tau$ if $D_k$ is set to zero,
\begin{eqnarray}
{\delta^2_k} \cong  2^{10} \tau^8  B_k^6E_k^2.
\end{eqnarray}
This demonstrates again the role of the diagonal nuclear spin interaction. We note that the
decoherence profile in the two-pulse control configuration, in particular the exponential
index, is qualitatively different from the predictions of semiclassical theories\cite{desousa_control}
in which the bath dynamics is accounted by a randomly fluctuating force (without being
actively changed by the manipulation of the quantum object) and the resultant decoherence
exponential index is basically unchanged by the pulse control. The recoherence at the echo time
$4\tau$ in the Carr-Purcell control configuration is clearly demonstrated in Fig.~\ref{Fig_CP}~(c),
even though the signal at the Hahn echo time $2\tau$ after the first pulse is absent.

\section{Concatenated disentanglement}
\label{S_concatenation}

\subsection{Formalism and geometrical picture}

We have used the iteration of effective precession angles
to derive the pseudo-spin states under the Carr-Purcell control in Eqs.~(\ref{psiCP}-\ref{angleCP}).
Such an iteration formalism inspires us to borrow the idea of concatenation control
in dynamical decoupling recently studied for preserving coherence in quantum
computation.\cite{Viola_Random,Lidar_CDD,Lidar_DD_longpaper,Kern_DD}
To illustrate the basic idea, we can imagine that if ${\boldsymbol \chi}_{k}^{\pm}$ in Eq.~(\ref{theta0})
is replaced by the effective field corresponding to the precession angles at the Carr-Purcell echo time,
i.e., ${\boldsymbol\theta}_{2}^{\pm}/\tau$, the distance between the conjugate pseudo-spins
would be eliminated to an even higher order of the delay time. Such a replacement can be done
recursively, and a concatenated pulse sequence can be so designed to control the decoherence.
The unit cell of pulses to be concatenated does not have to be the Carr-Purcell sequence, but it could
be either simpler or more complicated, or even various types of unit sequences can be interwoven to
construct a sophisticated concatenation.

\begin{figure}[t]
\begin{center}
\includegraphics[width=8cm]{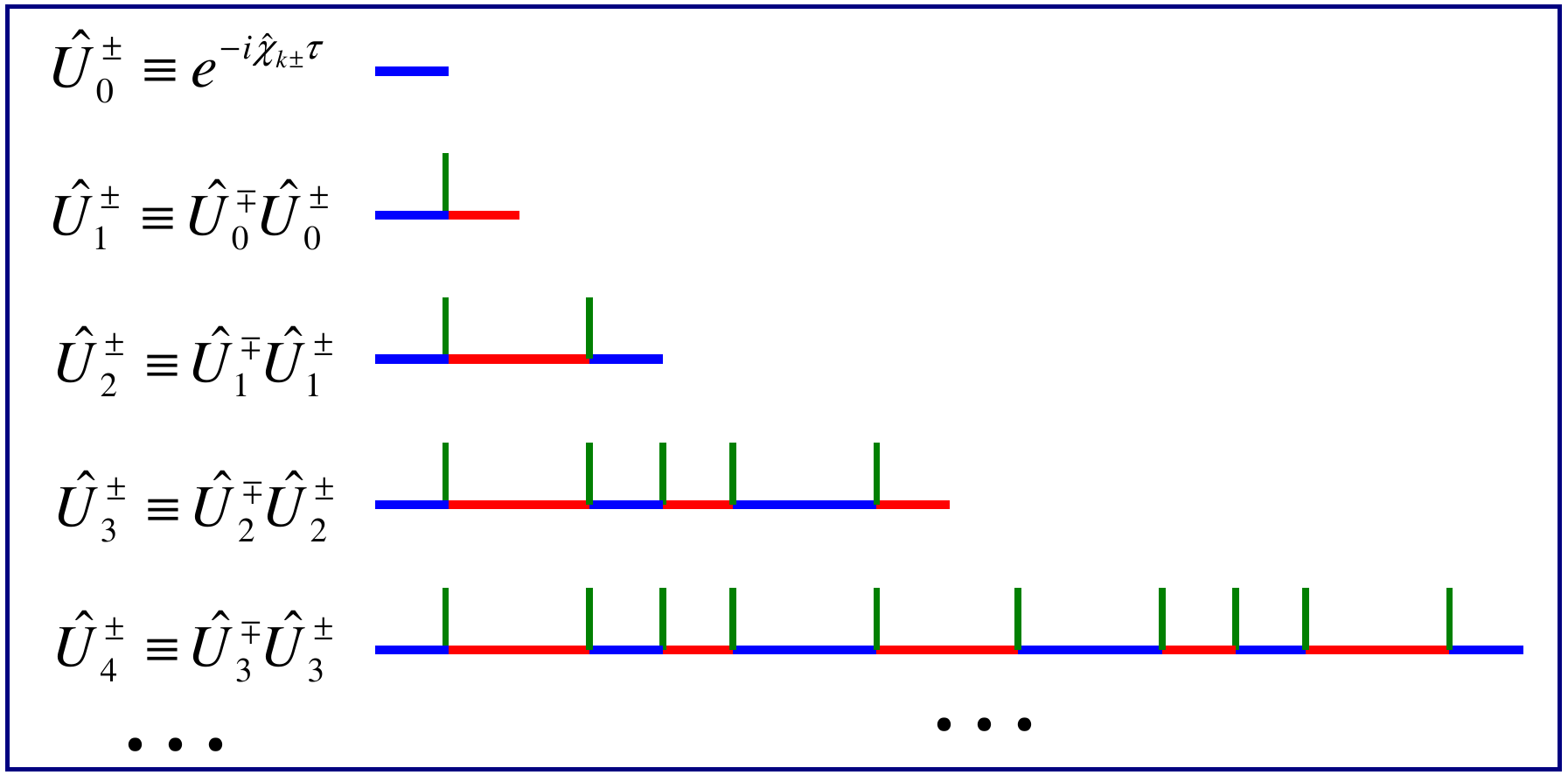}
\end{center}
\caption{Concatenated sequences of short-pulses flipping the electron spin, represented
by vertical bars.}
\label{Fig_concatenatedsequence}
\end{figure}

To exemplify the concatenated disentanglement for decoherence control, we choose the
simplest concatenation sequence which is recursively defined by Eq.~(\ref{iteration}).
Fig.~\ref{Fig_concatenatedsequence} illustrates how such a sequence can be concatenated to any
order. The 0th order is a FID evolution with no pulse control and the $(l+1)$th order sequence is constructed by two
subsequent $l$th sequences with or without one extra flipping pulse inserted in between depending on whether $l$ is
even or odd. Accordingly, the evolution propagator is iterated as
\begin{subequations}
\begin{eqnarray}
\hat{U}_{0}^{\pm} &=& e^{-i{\boldsymbol \chi}^{\pm}_k\cdot\hat{\boldsymbol \sigma}_k\tau/2}
\equiv e^{-i{\boldsymbol \theta}_{0}^{\pm}\cdot \hat {\boldsymbol \sigma}/2}, \\
\hat{U}_{l+1}^{\pm} &=& \hat{U}_{l}^{\mp} \hat{U}_{l}^{\pm} \equiv
e^{-i{\boldsymbol \theta}_{l+1}^{\pm}\cdot \hat {\boldsymbol \sigma}/2}.
\end{eqnarray}
\end{subequations}

\begin{figure}[b]
\begin{center}
\includegraphics[width=7cm]{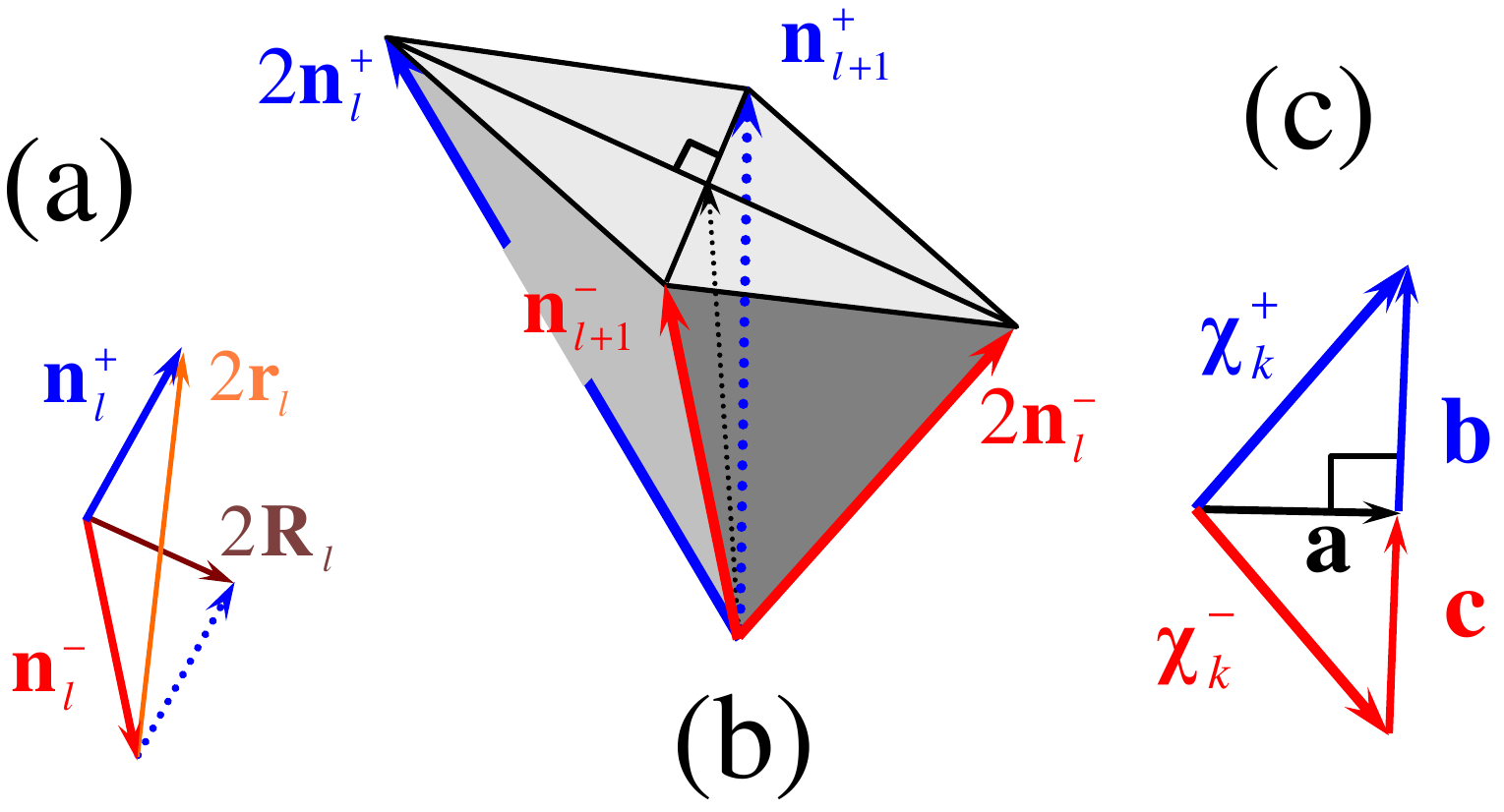}
\end{center}
\caption{(a) The effective precession vectors for conjugate pseudo-spins are
decomposed into the common component and the difference part.
(b) The iteration of the effective precession vectors from the $l$th order to the $(l+1)$th order concatenation.
(c) The pseudo-fields are decomposed into the perpendicular and the parallel components.}
\label{Fig_geometryconc}
\end{figure}

For a pictorial understanding of the concatenated control of decoherence, we rewrite the effective
precession angles as
\begin{eqnarray}
\sin\frac{{\boldsymbol \theta}_{l}^{\pm}}{2}\equiv {\mathbf n}_{l}^{\pm}\equiv
{\mathbf R}_{l}\pm  {\mathbf r}_{l}.
\label{decompose}
\end{eqnarray}
Without confusion, the vectors ${\mathbf n}_{l}^{\pm}$ will also be referred to as effective precession angles
since ${\mathbf n}_{l}^{\pm}\approx {{\boldsymbol \theta}_{l}^{\pm}}/2$ for small precession angles.
As depicted in Fig.~\ref{Fig_geometryconc} (a), the conjugate precession angles are decomposed into
the common part ${\mathbf R}_{l}$ and the difference part ${\mathbf r}_{l}$.
Then, the recursion in Eq.~(\ref{angleCP}) is rewritten as
\begin{subequations}
\begin{eqnarray}
{\mathbf r}_{0} &=&\frac{1}{2}\sin\frac{{\boldsymbol\chi}_{k}^{+}\tau}{2}-\frac{1}{2}\sin\frac{{\boldsymbol\chi}_{k}^{-}\tau}{2}, \\
{\mathbf R}_{0} &=&\frac{1}{2}\sin\frac{{\boldsymbol\chi}_{k}^{+}\tau}{2}+\frac{1}{2}\sin\frac{{\boldsymbol\chi}_{k}^{-}\tau}{2}, \\
{\mathbf R}_{1} &=&\cos\frac{{\chi}_{k}^{-}\tau}{2}\sin\frac{{\boldsymbol\chi}_{k}^{+}\tau}{2}
                      +\cos\frac{{\chi}_{k}^{+}\tau}{2}\sin\frac{{\boldsymbol\chi}_{k}^{-}\tau}{2}, \ \ \ \ \ \ \ \ \ \\
{\mathbf r}_{l} &=& 2{\mathbf R}_{l-1}\times{\mathbf r}_{l-1}, \ \ \ \ \ (l\ge 1),\\
{\mathbf R}_{l} & =& 2{\mathbf R}_{l-1}\sqrt{1-R_{l-1}^2-r_{l-1}^2}, \ \ \ \ \ (l\ge 2).
\end{eqnarray}
\label{rR}
\end{subequations}
The distance $\delta_k$ is in the same order of the difference part. Particularly, for the $l$th order concatenation,
the distance at $\tau_l\equiv 2^l\tau$ is
\begin{widetext}
\begin{eqnarray}
{\delta_l^2} &=& 1-\left|\langle\psi_k^-|\psi_k^+\rangle\right|^2=
 1-\left|\left\langle \downarrow\left|\left(\sqrt{1-n_l^2}+i{\mathbf n}_l^-\cdot\hat{\boldsymbol\sigma}\right)\left(\sqrt{1-n_l^2}-i{\mathbf n}_l^+\cdot\hat{\boldsymbol\sigma}\right)\right|\downarrow\right\rangle\right|^2
 \nonumber \\
 &=& 1-\left|\left\langle \downarrow\left| 1-n_l^2+{\mathbf n}_l^-\cdot{\mathbf n}_l^+
 -i2{\mathbf r}_l\cdot\hat{\boldsymbol\sigma}\sqrt{1-n_l^2} -i\left({\mathbf n}_l^+\times{\mathbf n}_l^-\right)\cdot\hat{\boldsymbol\sigma}\right|\downarrow\right\rangle\right|^2 \nonumber \\
 & =& 4 r_l^2\left[1-r_l^2-\left(z_{l+1}R_l-z_l\sqrt{1-R_l^2-r_l^2}\right)^2\right],
 \label{concatenateddistance}
\end{eqnarray}
\end{widetext}
for $l\ge 1$, where $n_l\equiv {n}_l^+=n_l^-=\sqrt{R_l^2+r_l^2}$ and $z_l$ is the $z$-component of ${\mathbf r}_l/r_l$.

\subsection{Short-time behaviors}

To gain some insight into the concatenated control of decoherence, first consider the
short-time limit $\tau\ll E_k^{-1}$. There, the recursion of the rotation angles has an
intuitive form as
\begin{subequations}
\begin{eqnarray}
{\mathbf R}_l &\approx& 2^l{\mathbf R}_0\approx 2^l\tau\frac{{\boldsymbol\chi}_k^++{\boldsymbol\chi}_k^-}{4}=
(2B_k,0,D_k)\frac{\tau_l}{2}, \ \ \ \ \ \ \\
{\mathbf r}_0 &\approx& \left({\boldsymbol\chi}_k^+-{\boldsymbol\chi}_k^-\right)\frac{\tau}{4}=(2A_k,0,E_k)\frac{\tau}{2}, \\
{\mathbf r}_{l} &\approx & 2^l {\mathbf R}_0\times {\mathbf r}_{l-1}, \ \ \ \ (l\ge 1),
\end{eqnarray}
\label{shorttimerecrusion}
\end{subequations}
until the precession angle $R_l$ approaches the order of unity at a threshold order of concatenation
$l_0$ given by
\begin{eqnarray}
l_0\approx -\log_2(B_k\tau).
\end{eqnarray}
The distinguishability between conjugate pseudo-spins in the leading order of delay time is
\begin{eqnarray}
{\delta_l^2}\cong 4r_l^2\left(1-z_l^2\right),
\end{eqnarray}
determined by the difference between the conjugate precession vectors.
The concatenated control can be understood in the geometrical picture shown
in Fig.~\ref{Fig_geometryconc}~(b): In the FID (the 0th order concatenation), the common
part of the precession angles $R_0$ is roughly determined by the nuclear spin interaction strength
times the time, i.e., $B_k\tau$, which is much less than 1 in the timescale of interest, and
the difference part $r_0$ is roughly the precession angle given by the hyperfine energy cost $E_k$.
In the 1st order concatenation, the common part ${\mathbf R}_1$ is
approximately in the same direction of ${\mathbf R}_0$ with the angle roughly doubled if $E_k\tau\ll 1$,
and the difference part ${\mathbf r}_1$ is perpendicular to both ${\mathbf R}_0$ and ${\mathbf r}_0$ with the
amplitude reduced by a factor of $2R_0$ from $r_0$. Staring from the 2nd order concatenation,
the common part ${\mathbf R}_l$ will be along the the same direction as the 1st order one ${\mathbf R}_1$ and the
difference part ${\mathbf r}_l$ will be alternatively in the two orthogonal directions ${\mathbf r}_1$
and ${\mathbf r}_2$, both perpendicular to the common part direction ${\mathbf R}_l$. By each further level of concatenation,
the common part is increased by a factor of two and the difference is reduced by a factor
of $2R_l$ for $l<l_0$. Thus each level of concatenation will reduce the difference between effective
conjugate precession angles by an order of delay time times the nuclear spin interaction strength
and remove the decoherence accordingly, until the controlling effect is saturated at the threshold level $l_0$.

\begin{figure}[b]
\begin{center}
\includegraphics[width=8cm]{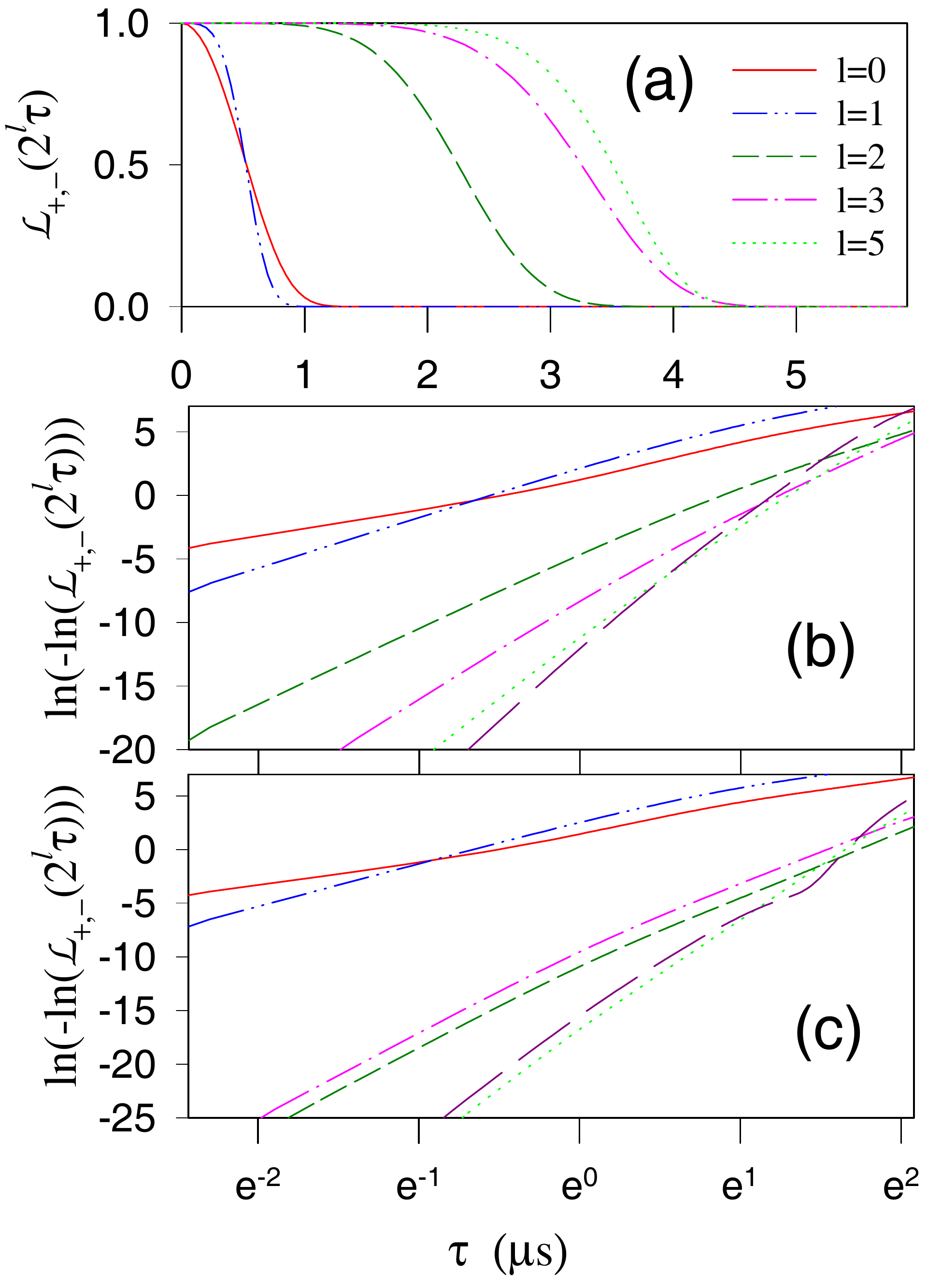}
\end{center}
\caption{(a) The electron spin coherence under the $l$th order concatenation control,
$l=0,\ldots,5$,
as functions of the pulse delay time $\tau$. (b) The logarithm plot of (a). (c) The same as (b),
but with the diagonal nuclear spin interaction put to zero. The QD is as in Fig.~\ref{Fig_CP}.}
\label{Fig_concacoh}
\end{figure}

With the short-time approximation of the precession angles given in Eq.~(\ref{shorttimerecrusion}),
the distinguishability between conjugate pseudo-spins in the leading order of $\tau$ is
\begin{eqnarray}
{\delta_l^2} & \cong & 4\tau^{2l+2}2^{l(l-1)}\left(1-z_l^2\right) \nonumber \\
& & \times \left(4B_k^2+D_k^2\right)^{l-1}\left(B_kE_k-A_kD_k\right)^2,
\end{eqnarray}
with
\begin{eqnarray}
z_l^2 =\left\{\begin{array}{ll}
0 & (l=1,3,\ldots) \\
4B_k^2\left(4B_k^2+D_k^2\right)^{-1} & (l=2,4,\ldots)
\end{array}\right.. \label{zcomponent}
\end{eqnarray}
Alternatively, for a fixed time $t$, a certain order of
concatenation control with $\tau=t/2^l$ gives
\begin{eqnarray}
{\delta_l^2} & \cong& {4t^{2l+2}}{2^{-l(l+3)}} \left(1-z_l^2\right) \nonumber \\ &&
\times \left(4B_k^2+D_k^2\right)^{l-1}\left(B_kE_k-A_kD_k\right)^2,
\end{eqnarray}
which can be suppressed to an arbitrary power of the echo time $t$ times the nuclear spin interaction strength
under the condition
\begin{eqnarray}
R_l \sim B_k t \ll 1.
\end{eqnarray}
The electron spin coherence under control is
\begin{eqnarray}
{\mathcal L}^{\rm s}_{+,-}(\tau_l)\cong\prod_ke^{-\delta_l^2/2}\cong e^{-\tau_l^{2l+2}/T_{(l)}^{2l+2}},
\end{eqnarray}
decaying exponentially with powers of $\tau$ at 2, 4, 6, 8, $\ldots$, for the concatenation order
0, 1, 2, 3, $\ldots$, respectively.

To show the effect of the diagonal nuclear spin interaction in the
decoherence control, we note that in Eq.~(\ref{zcomponent}) $z_{2m}=1$ if $D_k=0$ and thus
$\delta_k^2$ vanishes in the given order.
In this case, the distinguishability between the conjugate pseudo-spins in the next leading order for $l=2m$ is
\begin{eqnarray}
{\delta_l^2}  \cong  4R_l^2r_l^2 \cong
\tau^{2l+4} 2^{l(l+3)}E_k^2B_k^{2l+2}.
\end{eqnarray}
So when $D_k$ is put to zero, the coherence decays exponentially with powers of $\tau$ at 2, 4, 8, 8, 12, 12, $\dots$
for concatenation order 0, 1, 2, 3, 4, 5, $\ldots$, respectively.

The effective preservation of the coherence by concatenated control is demonstrated by the numerical
simulations presented in Fig.~\ref{Fig_concacoh}. Note that the coherence shown in the figure is at
the time $2^l\tau$ which doubles at every order of concatenation. The role of the diagonal part of the
nuclear spin interaction is demonstrated [compare Fig.~\ref{Fig_concacoh}~(b) and (c)].

\subsection{Controlling small parameters}

The suppression of the decoherence by concatenated control, however, does not depend on the short-time
condition $\tau\ll E_k^{-1}$. This point can be seen from Eq.~(\ref{rR}): An iteration of concatenation
suppresses the decoherence further as long as the common component of the conjugate precession angles $R_{l}\ll 1$,
without requiring $E_k\tau\ll 1$. Since $R_{0}$ is in the order of nuclear spin interaction
strength times the delay time ($B_k\tau$), the condition for a concatenated sequence to be efficient in
eliminating the decoherence is that the delay time be much shorter than the inverse strength
of the nuclear spin interaction, i.e.,
\begin{eqnarray}
\tau\ll B_k^{-1}, D_k^{-1}, A_k^{-1},
\label{notshorttime}
\end{eqnarray}
which is a much less stringent condition than $t\ll E_k^{-1}$. To calculate the electron spin coherence under
this relaxed ``short-time'' condition, we decompose the conjugate pseudo-fields in such a way that
\begin{subequations}
\begin{eqnarray}
{\boldsymbol\chi}_{k}^{+}={\mathbf a}+{\mathbf b}, \\
{\boldsymbol\chi}_{k}^{-}={\mathbf a}-{\mathbf c},
\end{eqnarray}
\end{subequations}
with ${\mathbf a} \perp{\mathbf b} \parallel{\mathbf c}$, as depicted in Fig.~\ref{Fig_geometryconc} (c).
Then, we are able to separate different timescales according to the orders of magnitude
of different components. We notice that $a$ and $|b-c|$ are of the order of nuclear spin interaction strength
and $b$ and $c$ are of the order of hyperfine energy cost $E_k$.
Within timescales given by Eq.~(\ref{notshorttime}), the precession angles in
the leading order of the nuclear spin interaction strength are
\begin{widetext}
\begin{subequations}
\begin{eqnarray}
{\mathbf r}_{0} & \cong & \sin \frac{\left({\mathbf b}+{\mathbf c}\right)\tau}{4}, \\
{\mathbf R}_{0} & \cong & \frac{\left({\mathbf b}-{\mathbf c}\right)\tau}{4}\cos \frac{\left({b}+{c}\right)\tau}{4}
                             +\frac{{\mathbf a}\tau}{2}{\rm sinc}\frac{\left({b}+{c}\right)\tau}{4},   \\
{\mathbf R}_{l} & \cong & 2^{l-2}{\left({\mathbf b}-{\mathbf c}\right)\tau}
                              +2^{l-1}{\mathbf a}\tau {\rm sinc}\frac{\left({b}+{c}\right)\tau}{2},  \ \ \ \ (l>0), \\
{r}_{l} & \cong &  2^{l(l-1)/{2}}\tau^{l+1} \frac{{ a}({b}+{c})}{4} {\rm sinc}^2\frac{(b+c)\tau}{4}
\left[\frac{\left({b}-{c}\right)^2}{4}+a^2{\rm sinc}^2\frac{(b+c)\tau}{2}\right]^{\frac{l-1}{2}} \\
&\le& 2^{l(l-1)/{2}}\tau^{l+1} \frac{{ a}}{4} \left|{\boldsymbol \chi}_k^+ -{\boldsymbol \chi}_k^-\right|
{\left|\frac{{\boldsymbol \chi}_k^+ +{\boldsymbol \chi}_k^-}{2}\right|}^{l-1} \nonumber \\
& \cong & 2^{\frac{l(l-1)}{2}} \tau^{l+1}\left|E_kB_k-A_kD_k\right| \left(4B_k^2+D_k^2\right)^{\frac{l-1}{2}}, \  \ (l>0),
\end{eqnarray}
\end{subequations}
\end{widetext}
under the condition $R_l\ll 1$. The $z$-component of the direction of ${\mathbf r}_l$ is
\begin{eqnarray}
z_l^2 \cong \left\{\begin{array}{ll}
0 & (l=2m+1) \\
\frac{4B_k^2{\rm sinc}^2\left(E_k\tau\right)}
{4B_k^2{\rm sinc}^2\left(E_k\tau\right)+D_k^2} & (l=2m+2)
\end{array}\right. .
\end{eqnarray}
The distinguishability between conjugate pseudo-spins $\delta_k^2$ is determined by Eq.~(\ref{concatenateddistance}).
The decoherence is estimated accordingly.

It is obvious that the small parameter controlling the maximum
delay time for a concatenation control to be efficient is the
nuclear spin interaction strength $\sqrt{4B_k^2+D_k^2}$ instead of
the object-bath interaction strength (i.e., the hyperfine energy
cost $E_k$). The threshold concatenation level $l_0$, beyond which
an additional level of concatenation cannot suppress the
decoherence further, is determined by the condition that the
precession angle approaches 1, i.e., $R_{l_0}\lesssim 1$. With the
precession angle given by
\begin{eqnarray}
R_l & \cong & 2^{l-1}\tau \sqrt{4B_k^2{\rm sinc}^2\left(E_k\tau\right)+D_k^2} \nonumber \\
 & \le & 2^{l-1}\tau \sqrt{4B_k^2+D_k^2},
\end{eqnarray}
the threshold concatenation level is estimated as
\begin{eqnarray}
l_0\sim -\log_2\left(\tau \sqrt{4B_k^2+D_k^2}\right),
\end{eqnarray}
also determined by the nuclear spin interaction strength.
A close examination of Fig.~\ref{Fig_concacoh} (a) reveals that the
decoherence suppression is still effective even when the pulse delay time $\tau$
is substantially longer than the typical value of the inverse hyperfine energy cost $E_k^{-1}$,
which is about 0.5~$\mu$s for the QD considered.

\subsection{Coherence stabilization}

\begin{figure}[b]
\begin{center}
\includegraphics[width=8cm]{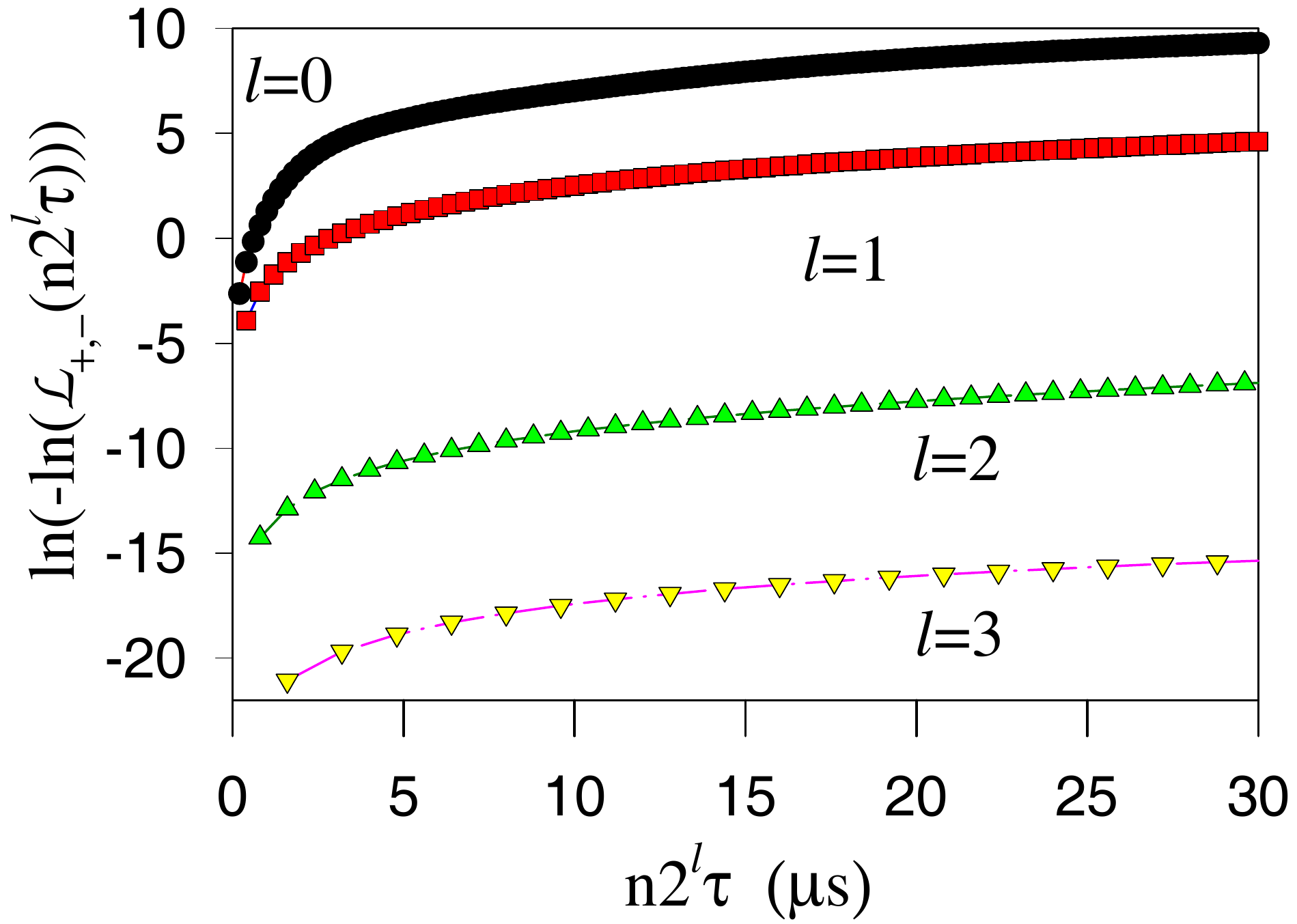}
\end{center}
\caption{The electron spin coherence under the control of
repeated units of concatenated sequences, as functions of the number of
units for a fixed pulse delay time $\tau=0.2$~$\mu$s. The QD is
under the same condition as in Fig.~\ref{Fig_concacoh}.}
\label{Fig_stable}
\end{figure}

The concatenated pulse sequence can be repeated.
The propagator of a pseudo-spin for $n$ repeated $l$th-order concatenation sequences is
\begin{eqnarray}
\hat{U}^{\pm}_{l,n}=\left(\hat{U}^{\pm}_{l}\right)^n
=\exp\left(-{i} n {\boldsymbol\theta}^{\pm}_l\cdot\hat{\boldsymbol\sigma}/2\right).
\end{eqnarray}
Note that in implementing the pulse sequence for repeated concatenation, a $\pi$-pulse should be
inserted between two unit sequences if $l$ is an odd number.
The common and difference parts of the conjugate precession angles ${\mathbf R}_{l,n}$ and ${\mathbf r}_{l,n}$
can be defined similarly to those for unit concatenation sequences. The precession vectors
${\boldsymbol\theta}^{\pm}_{l,n}\equiv n{\boldsymbol\theta}^{\pm}_l$
do not changes their directions with increasing the
number of repeated units but only the angle of rotations grows linearly with $n$. Therefore,
\begin{eqnarray}
r_{l,n}/R_{l,n}=r_l/R_l\cong r_{l-1}.
\end{eqnarray}
Since $R_{l,n}=\left|\sin \frac{{\boldsymbol\theta}^{+}_{l,n}}{2}+\sin\frac{{\boldsymbol\theta}^{-}_{l,n}}{2}\right|/2\le 1$,
the distinguishability between conjugate pseudo-spins given by Eq.~(\ref{concatenateddistance}) is bounded from above by
\begin{eqnarray}
{\delta_{l,n}^2}\le 4r_{l,n}^2\le 4\left(r_l/R_l\right)^2\cong 4r_{l-1}^2.
\end{eqnarray}
Since by each order of concatenation until the threshold level is reached,
the common precession angle $R_l$ is increased by a factor two and the difference part $r_l$ is
suppressed by a factor of $R_l\ll 1$, the ratio $r_l/R_l$ can be made arbitrarily small. In the mesoscopic system,
the number of excitation modes (pseudo-spins) is finite, so the coherence under repeated concatenation
control is bounded from below by
\begin{eqnarray}
{\mathcal L}_{+,-}\left(n\tau_l\right)\ge \exp\left(-2\sum_k r_l^2/R^2_l\right)
\cong {\mathcal L}_{+,-}\left(\tau_{l-1}\right).
\end{eqnarray}
The summation $\sum_k 2r_l^2/R_l^2$ can be made arbitrarily
small by choosing a small delay time and a sufficiently deep concatenation level,
which means the electron spin coherence is stabilized virtually without decoherence.
The coherence after $n$ units of $l$th order concatenation sequences is no less
that that after a single unit of $(l-1)$th order concatenation sequence.
The analysis above is verified by numerical simulations presented in Fig.~\ref{Fig_stable}.

\section{Summary and discussions}

We have studied the electron spin decoherence in a mesoscopic bath
of nuclear spins in semiconductor QDs. For QDs
of interest in experiments, the electron-nuclear hyperfine interaction
is much stronger than the nuclear spin interactions, so the
electron spin and those nuclear spins in direct contact with the
electron can be well isolated from the environment as a
closed quantum system in the timescale of interest, giving the
definition of a mesoscopic system.

When the longitudinal spin relaxation is absent (suppressed here
by a moderate to strong magnetic field $\gtrsim 0.1$~T and a low temperature
$\lesssim 1$~K), pure dephasing (or transverse relaxation) induced by
the nuclear spin bath is the dominant cause for the lost of
electron quantum coherence. In the quantum theory,
decoherence is a consequence of the entanglement
between the quantum object and the bath established by the evolution of bath along
bifurcated pathways in the Hilbert space for different basis states of the quantum object.
By contrast, semiclassical spectral diffusion theories ascribe the decoherence to accumulation of a
random phase from a fluctuating local field.
The nuclear spin dynamics, which is conditioned on the electron spin basis states,
can be shepherded by manipulation of the electron spin. In particular,
when the electron spin is flipped, the bifurcating nuclear spin
pathways will exchange their evolution directions in the Hilbert space
and could intersect at a later time, leading to the disentanglement of
the electron from the nuclei and hence the restoration of the coherence
even after it has been totally lost to the nuclear bath.
Such recoherence by disentanglement is fundamentally different from spin echoes which
result from re-focusing of random phases.

The solution to the many-body nuclear bath dynamics relies on
four essential conditions: (1) The QD size is in the mesoscopic
regime. (2) The external field is much stronger than the hyperfine
interaction. (3) The temperature is low for the electron spin but
high for the nuclear spins. (4) The timescale considered is
much shorter than the inverse nuclear interaction strength. Such
conditions are realistic. They make possible two important simplifications in theory.
First, the electron spin dephasing in ensemble dynamics can be factorized into two factors,
namely, one due to the inhomogeneous broadening, and the other due to
the entanglement between the electron spin and the nuclear spins.
Second, the nuclear spin dynamics in the timescale of interest can be
approximated as independent excitations of pair-wise flip-flops.
The solution of the bath quantum dynamics provides a simple geometrical
basis for the design of pulse sequences for recoherence.
The controllability of the mesoscopic bath dynamics, however, is a rather
general concept independent of the simplifications made in the present
paper.

A close examination of the single-system dynamics free of the
inhomogeneous broadening reveals a wealth of phenomena in the
electron spin decoherence. For example, a coherence recovery at a
time different from the spin echo time would have been totally
eclipsed in ensemble dynamics, since the dephasing due to
inhomogeneous broadening is usually faster by orders of magnitude
than the decoherence due to the entanglement. It is also shown that the single-pulse
Hahn echo signals do not measure the electron spin decoherence in
FID configuration: First, the decoherence caused by
hyperfine-mediated nuclear pair-flips has a strong effect on the
FID when the external field is not too large, but it is
removed from the Hahn echo signal due to disentanglement;
Second, the decoherence by the intrinsic nuclear spin interaction
is also partially suppressed, which results in a Hahn-echo decay
time approximately $\sqrt{2}$ times the FID time due to the
intrinsic interaction. The decoherence behavior in single-system
dynamics (such as the recoherence at $\sqrt{2}\tau$) could eventually be directly
observed in experiments if the inhomogeneous broadening can be filtered out, e.g.,
by a projective measurement of the local Overhauser
field.\cite{Espin_HF_1_Loss,Giedke_spinmeasure,Klauser_spinmeasure,Imamoglu_WeakMeasurement}

To observe the recoherence by disentanglement in ensemble experiments,\cite{ESR_silicon_T2,Marcus_T2}
pulse sequences can be designed to force the disentanglement to coincide with a spin echo.
The simplest solution is the Carr-Purcell sequence, which has one pulse at $\tau$ and
another at $3\tau$ and forces the disentanglement to take place at the echo time $4\tau$.
More sophisticated designs such as concatenated sequences can be employed to suppress the
decoherence to an arbitrary order of the pulse delay time in units of the inverse nuclear
interaction strength, and a repetition of the concatenated sequences can preserve the
electron spin coherence until other decoherence mechanisms
(such as phonon scattering) become significant.

The design of concatenated pulse sequences for disentanglement resembles
the dynamical decoupling schemes which have been developed in decades for
NMR spectroscopies\cite{Haeberle,Pines_DD,Slichter} and recently studied for
quantum computation.\cite{Viola_Random,Lidar_CDD,Kern_DD}
The disentanglement schemes exploit the controlling power of pulse
sequences in a subtly different way from the dynamical decoupling schemes:
The latter uses short pulses to flip the quantum object frequently
so that the object-bath interaction is dynamically averaged to zero,
while the former does not seek to eliminate the effective coupling
but rather focuses on controlling the evolution of the bath wavefunction.
For example, for local nuclear spin pair-flips in the single-pulse control configuration,
the effective interaction $\hat{S}^z_e\otimes\hat{H}_{\rm eff}$ at the disentanglement time
$\sqrt{2}\tau$, defined by $\hat{H}_{\rm eff}\equiv \hat{H}^+_{\rm eff}-\hat{H}^{-}_{\rm eff}$
and $e^{-i\hat{H}^{\pm}_{\rm eff}\sqrt{2}\tau}\equiv e^{-i\hat{H}^{\mp}(\sqrt{2}-1)\tau}e^{-i\hat{H}^{\pm}\tau}$, does
not vanish even in the leading order of the hyperfine coupling.
This is also the case for the disentanglement under the control of
equally spaced pulses. One may question why a non-zero effective
object-bath interaction could lead to vanishing decoherence.
The reason is that the control steers the wavefunction
evolution and the initial state of the bath is of some special form (namely,
the nuclear spin states at the temperature under consideration possesses no off-diagonal
coherence). The disentanglement time is in general dependent on the initial state
(a universal disentanglement scheme independent of the initial state is only possible when
the effective system-bath coupling vanishes). The dependence of the disentanglement on
the initial state is implied in the control by evenly separated pulses shown in Fig.~\ref{Fig_geometryCP}~(a):
For instance, if the nuclear spins there at $\sqrt{2}\tau$ is set as the initial state
(which possesses off-diagonal two-spin coherence), the disentanglement after the electron spin flip at $\tau_0$
would occur at $(\sqrt{3}-1)(\sqrt{2}+1)\tau_0$ (on the contrary, for initial states ${\mathcal J}\rangle$
which has no off-diagonal coherence, the disentanglement occurs at $\sqrt{2}\tau_0$).
It would be very interesting to study the decoherence and the disentanglement for spin baths with
certain initial off-diagonal coherence such as in quantum memory applications.\cite{Lukin_NucleiMemory}

Even though the concatenated disentanglement does cause the
effective electron-nuclear coupling vanish, the difference from
the dynamical decoupling is still unambiguously evidenced by their
different controlling small parameters. The controlling small
parameter in dynamical decoupling is the object-bath interaction
strength, and the pulse delay time should be much shorter than the
inverse of that strength. In the dynamical disentanglement, the
object-bath interaction strength is a quite irrelevant parameter,
but the short-time condition is given by the bath interaction.
Furthermore, in concatenated dynamical decoupling, the decoherence
is eliminated in orders of the system-interaction strength times
the pulse delay time, but in the disentanglement, the decoherence
is suppressed in orders of the bath interaction strength times the delay time.

\acknowledgments
This Work was supported by  NSF DMR-0403406, and ARO/NSA-LPS.

\begin{appendix}

\section{Factorization of single-system dynamics and inhomogeneous broadening}
\label{Append_factorization}

\begin{figure}[t]
\begin{center}
\includegraphics[width=8cm]{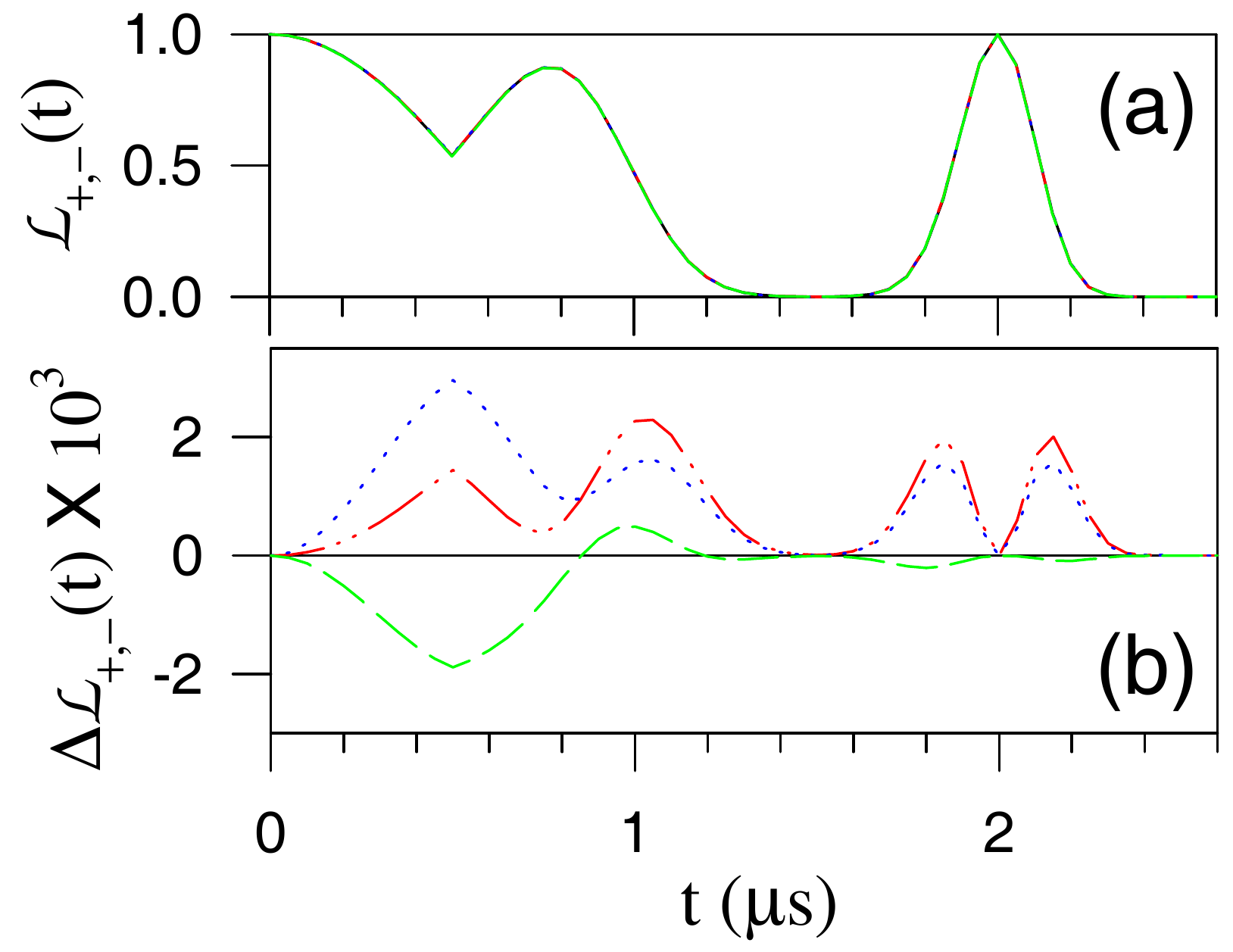}
\end{center}
\caption{(a) The electron spin coherence
under the control of a Carr-Purcell pulse sequence,
for various initial nuclear spin states (indicated by arrows in Fig.~\ref{Fig_inhom}):
$|\mathcal{J}_1 \rangle$ with local Overhauser field
$E_{\mathcal{J}_1}=-0.5\Gamma_2^*$ (dot-dashed red line),
$|\mathcal{J}_2 \rangle$ with $E_{\mathcal{J}_2}=0$ (solid black line),
$|\mathcal{J}_3 \rangle$ with $E_{\mathcal{J}_1}=0.5\Gamma_2^*$
(dotted blue line), and $|\mathcal{J}_4 \rangle$ with
$E_{\mathcal{J}_4}=\Gamma_2^*$ (dashed green line).
The four lines are indeed indistinguishable in the figure.
(b) The deviation of the coherence for various initial states
from that for the state $|{\mathcal J}_1\rangle$,
$\Delta \mathcal{L}_{+,-}(t) \equiv |\langle \mathcal{J}_n| e^{i\hat{H}_- t} e^{-i\hat{H}_+ t}
|\mathcal{J}_n\rangle| - |\langle \mathcal{J}_1| e^{i\hat{H}_- t}
e^{-i\hat{H}_+ t} |\mathcal{J}_1\rangle|$, amplified by 1000 times. The InAs
QD is of size $34\times 34\times 3$~nm$^{3}$, at a temperature of
1~K, and under an external field of 10~Tesla.}
\label{Fig_factorization}
\end{figure}

The nuclear spin dynamics is basically determined by the excitation spectra of the pseudo-spins
when the bath size is large enough. The excitation spectra for
different nuclear spin configurations $|{\mathcal J}\rangle$
should be the same except for a relative variance in the order of
$1/{\sqrt{N}}$  according to the central limit theorem in
statistics. Here we give a numerical test of the single-system
dynamics for a few initial nuclear spin configurations randomly selected
from the thermal ensemble. The results are shown in
Fig.~\ref{Fig_factorization}. Remarkably, the coherence for four
initial states randomly selected from the thermal ensemble is
visually indistinguishable, both in the FID regime
($t<0.5$~$\mu$s) and after the controlling pulses are applied
($t>0.5$~$\mu$s). The relative deviation between different curves
is consistent with the estimate by the central limit theorem.

\section{Boundary of mesoscopic nuclear spin bath}
\label{Append_boudary}

\begin{figure}[b]
\begin{center}
\includegraphics[width=8cm]{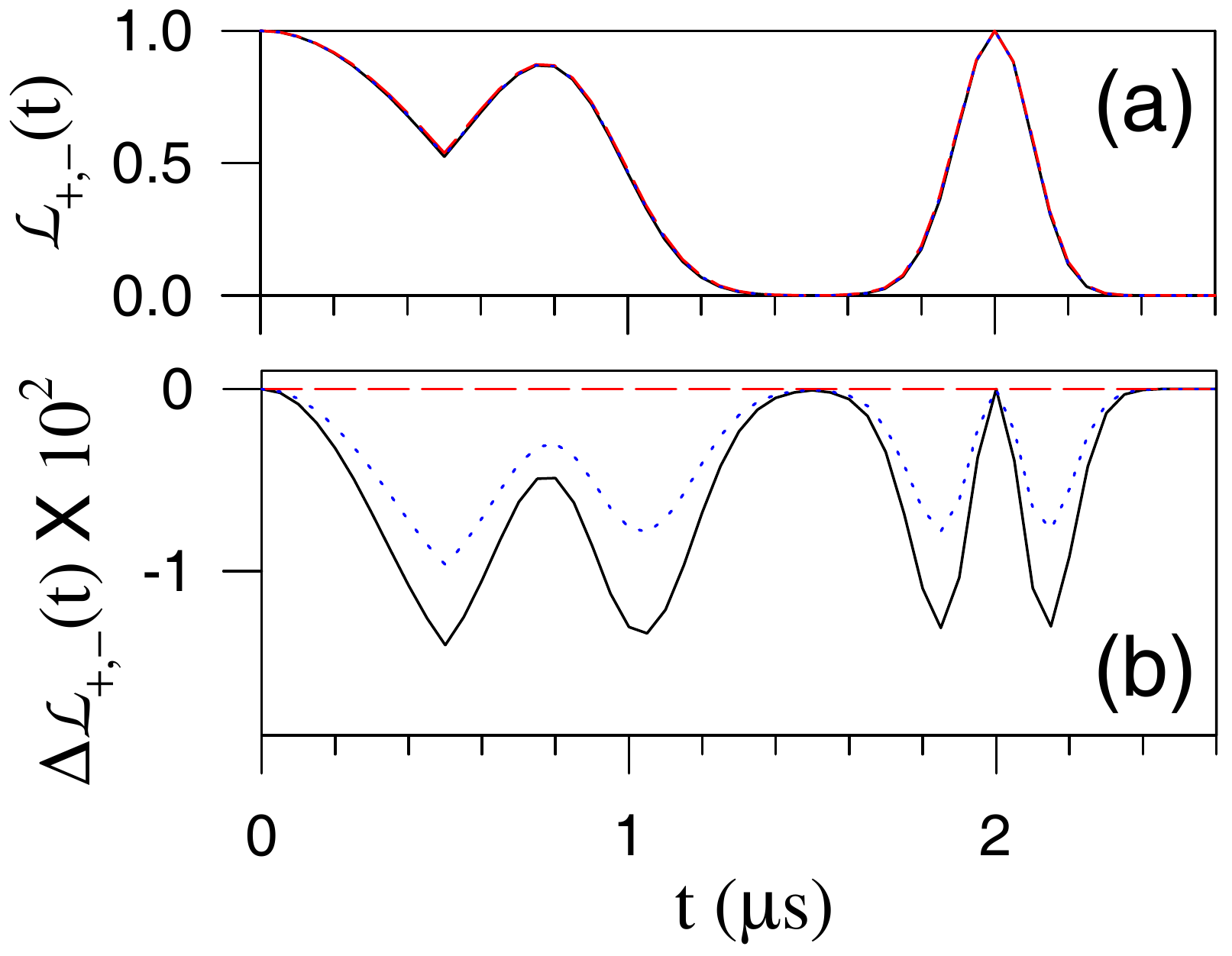}
\end{center}
\caption{(a) The electron spin coherence under
the control of a Carr-Purcell pulse sequence, for various choices of boundary of the
nuclear spin bath: For the dashed red, solid black, and dotted blue lines, the boundary is set 0, 1, and 3
layers of unit cells outside the boundary defined by the QD potential walls.
The three lines are indistinguishable in the figure.
(b) The deviation (amplified by 100 times) of the coherence for various choices of the bath boundary,
referenced to the first choice of boundary.
The QD is as in Fig.~\ref{Fig_factorization}.}
\label{Fig_boundary}
\end{figure}

Intuitively, the nuclear spins in direct contact with the electron spin can be taken as the
mesoscopic bath. The boundary of the bath is roughly defined by the condition that the hyperfine
interaction is stronger than the nuclear spin interaction for nuclei within the bath, and
otherwise for those without the bath. The interaction between the nuclear spins within and without
the bath across the boundary, however, could make this definition somehow arbitrary. One can define a larger
spin bath by including one or more layers of nuclei outside the boundary.
Since the nuclear spin interaction is much weaker than the hyperfine interaction, the cross-boundary
interaction should be unimportant in timescales of interest. Here we present the numerical test of
the dependence of the electron spin decoherence on the choice of the boundary of the mesoscopic
nuclear spin bath, particularly by examining the real-time behavior under the Carr-Purcell control.
The result in Fig.~\ref{Fig_boundary} confirms the assumption that the slight ambiguity in defining the
mesoscopic bath as a closed system is unimportant.

\section{Pair-flip numbers}
\label{Append_pairflip}

\begin{figure}[t]
\begin{center}
\includegraphics[width=8cm]{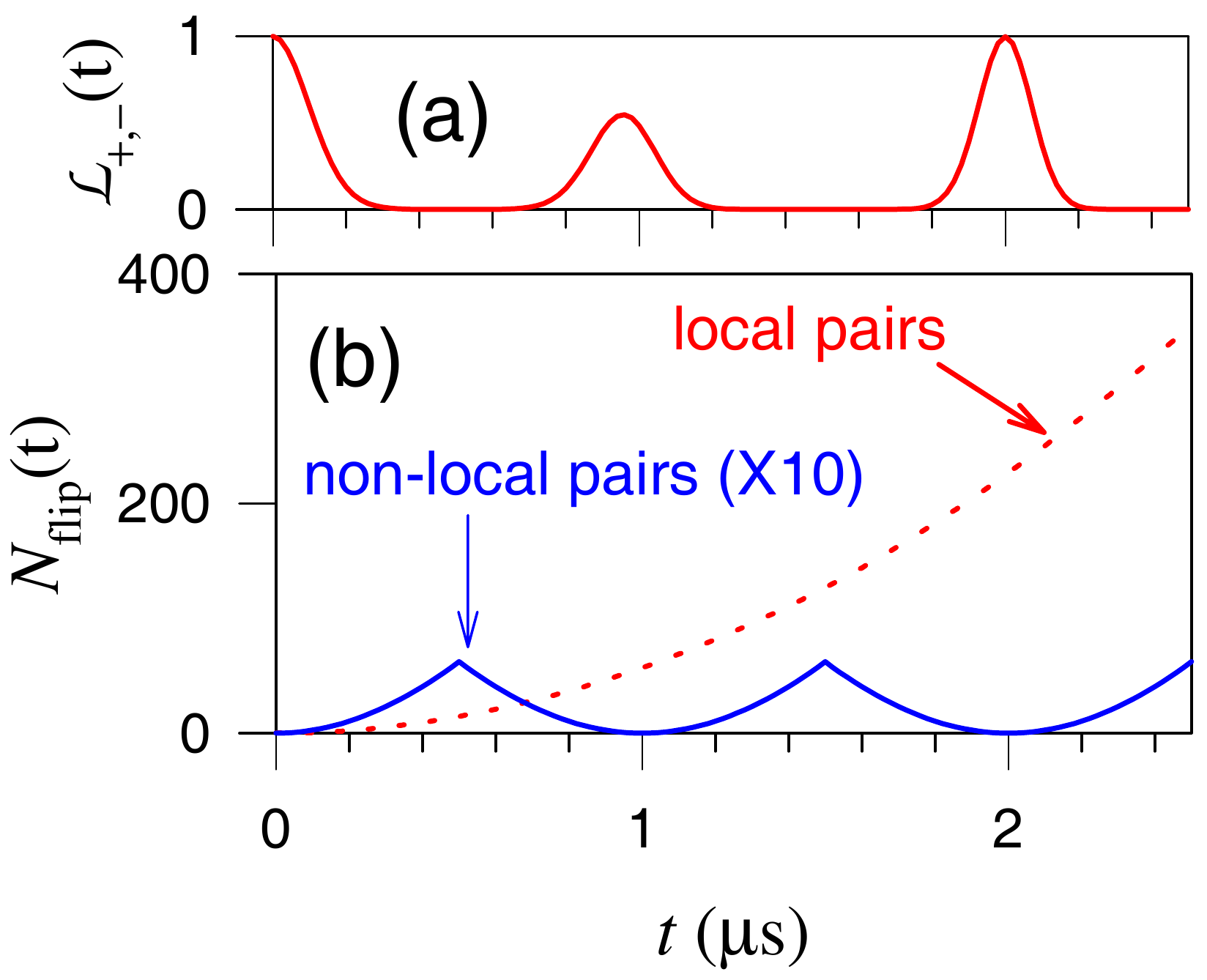}
\end{center}
\caption{(a) The electron spin coherence in the Carr-Purcell control
configuration. (b) The number of pair-flips due to the
intrinsic nuclear spin interaction (dotted red line) and that due to the hyperfine-mediated interaction
(solid blue line, amplified by 10 times). The InAs QD is as in Fig.~\ref{Fig_factorization}
with $B=2$~Tesla.}
\label{Fig_nflip}
\end{figure}

The number of pair-flips can be calculated with the pseudo-spin
model, as an {\it a posterior} check of the pair-correlation
approximation. The general formalism is\cite{Yao_Decoherence}
\begin{eqnarray}
N_{\rm flip}(t)\equiv \max_{\pm} \sum_k \left|\langle\uparrow|\psi_k^{\pm}(t)\rangle\right|^2\approx
\sum_k \left|\langle\uparrow|\psi_k^{+}(t)\rangle\right|^2. \ \ \
\end{eqnarray}
To be specific, let us consider the pair-flip number under concatenated control.
With the notations defined in Sec.~\ref{S_concatenation}, the pair-flip number under the $l$th
order concatenation control is
\begin{eqnarray}
N_{\rm flip}^{(l)}\left(\tau_l\right)=\sum _k \left[\left(n_l^+\right)_x^2+\left(n_l^+\right)_y^2\right],
\end{eqnarray}
where $\left(n_l^+\right)_x$ and $\left(n_l^+\right)_y$ are the $x$- and $y$-component of the
precession vector ${\mathbf n}_l^+$ defined in Eq.~(\ref{decompose}), respectively, and $\tau_l\equiv 2^l\tau$.
With the condition $\tau_l\ll B_k^{-1}$, the results in the leading order
of the nuclear spin interaction strength are
\begin{subequations}
\begin{eqnarray}
N_{\rm flip}^{(0)}\left(\tau\right)&\cong& N_{{\rm flip},A}^{(0)}\left(\tau\right)+N_{{\rm flip},B}^{(0)}\left(\tau\right), \\
N_{{\rm flip},A}^{(0)}\left(\tau\right) & \cong& \sum_{k\in{\mathbb G}_A} A_k^2\tau^2{\rm sinc}^2\frac{E_k \tau}{2}\lesssim \frac{{\mathcal A}_{\alpha}^4\tau^2}{N^2\Omega_e^2}, \\
N_{{\rm flip},B}^{(0)}\left(\tau\right) & \cong& \sum_{k\in{\mathbb G}_B} B_k^2\tau^2{\rm sinc}^2\frac{E_k \tau}{2}\lesssim {NB_k^2\tau^2}, \ \ \ \ \ \\
N_{\rm flip}^{(l>0)}\left(\tau_l\right)&\cong& \sum_{k\in{\mathbb G}_B} B_k^2\tau_l^2 {\rm sinc}^2\left({E_k \tau}\right)\lesssim N B_k^2\tau_l^2. \ \ \ \ \ \
\end{eqnarray}
\end{subequations}
Here the number of local pair-flips and that for hyperfine-mediated non-local pair-flips in FID configuration
are identified with the subscripts $B$ and $A$, respectively.
As illustrated in Fig.~\ref{Fig_geometryecho} (b), the pseudo-spin precession for non-local pair-flips is
reversed when the electron spin is flipped, so under the concatenated control the number of hyperfine-mediated
pair-flips will oscillate periodically, with the maximum value $N_{{\rm flip},A}^{(0)}\left(\tau\right)$
at $(2n+1)\tau$ and minimum value 0 at $2n\tau$. At the spin echo time $\tau_l$ for $l>0$, the number of pair-flips
is contributed by the intrinsic nuclear spin interaction.
The number of pair-flips is plotted in Fig.~\ref{Fig_nflip} for a typical QD in the Carr-Purcell control configuration,
showing the features discussed above.

\begin{figure}[t]
\begin{center}
\includegraphics[width=5cm]{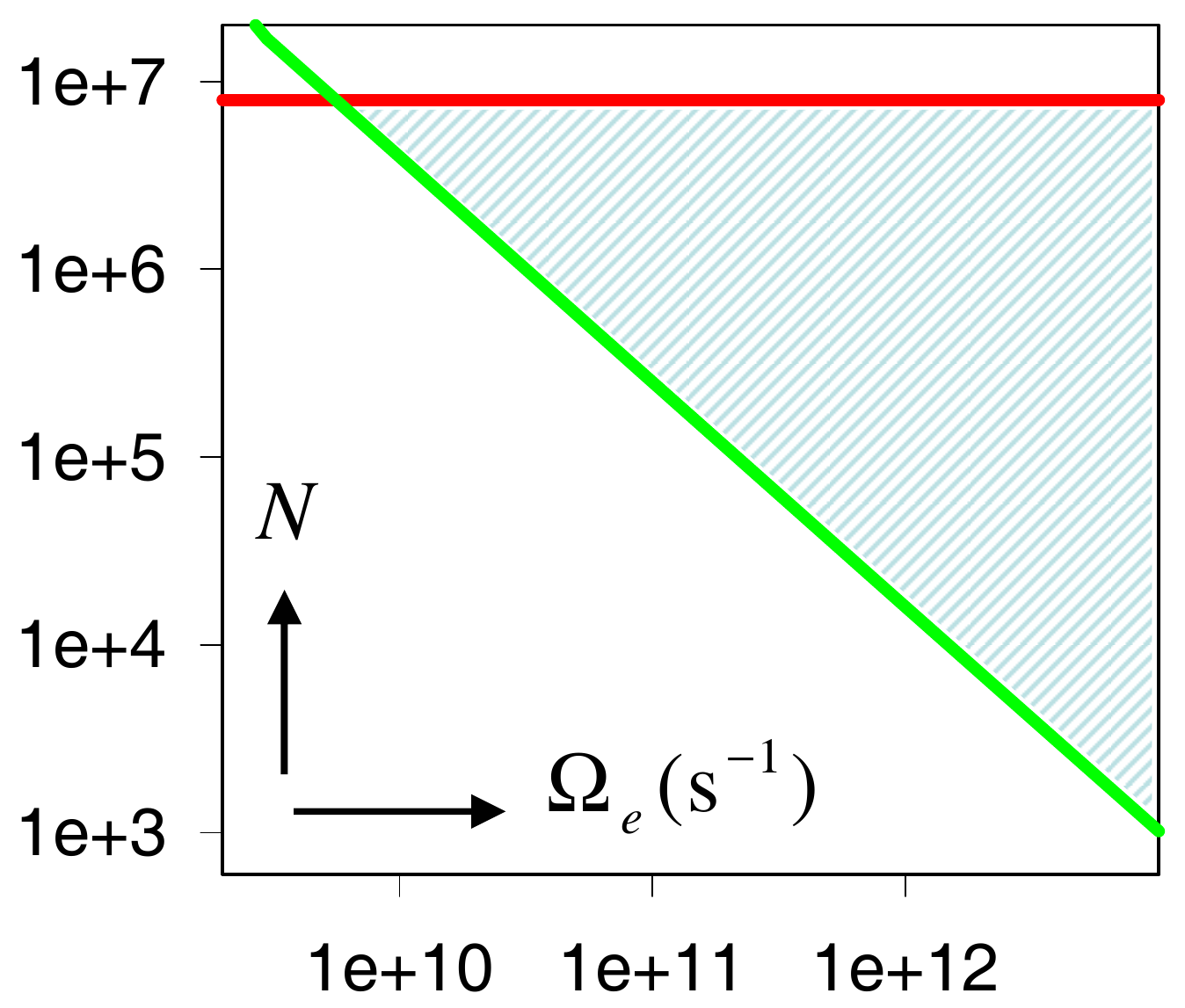}
\end{center}
\caption{Schematics for the range of the QD size and the field strength
(shadowed area) where the pair-correlation approximation is justified in the timescales
of interest in the single-pulse Hahn echo configuration. The upper and lower bounds
set limits on the number of the intrinsic local pair-flips and on that of the hyperfine-mediated
non-local pair-flips, respectively.}
\label{Fig_sizefield}
\end{figure}

To justify the pair-correlation approximation, the small pair-flip number condition
\begin{eqnarray}
N_{\rm flip}^2\ll N,
\end{eqnarray}
is to be fulfilled (see Appendix~\ref{Append_error}).
Thus, the criteria for the timescales $\tau$ and $\tau_l$ being short are given by
\begin{subequations}
\begin{eqnarray}
NB^4_k\tau_l^4\ll 1, \\
N^{-5}\Omega_e^{-4}{\mathcal A}_{\alpha}^8\tau^4\ll 1,
\end{eqnarray}
\label{validtimescales}
\end{subequations}
for consideration of the intrinsic interaction and the hyperfine-mediated interaction, respectively.
For an InAs QD containing about $10^5$ nuclei under an external field of $B_{\rm ext}\sim 1$~Tesla,
the valid timescales is estimated to be
\begin{subequations}
\begin{eqnarray}
\tau_l\lesssim 100\ \mu{\rm s}, \\
\tau\lesssim 10\ \mu{\rm s}.
\end{eqnarray}
\label{validtimescalesInAs}
\end{subequations}

A natural timescale of interest in experiments is the decay time of the Hahn echo signal,
which can be estimated from Eq.~(\ref{T2B}) and (\ref{THB}) to be
\begin{eqnarray}
T_{H,B}\sim N^{5/12}B_k^{-1/2}{\mathcal A}_{\alpha}^{-1/2}.
\end{eqnarray}
This timescale together with Eq.~(\ref{validtimescales}) sets ranges for the field
strength and the QD size where the pair-correlation approximation is valid:
\begin{subequations}
\begin{eqnarray}
N^{8/3}B_k^2{\mathcal A}_{\alpha}^{-2}\ll 1, \\
N^{-10/3}\Omega_e^{-4}B_k^{-2}{\mathcal A}_{\alpha}^6\ll 1,
\end{eqnarray}
\end{subequations}
as schematically shown in Fig.~\ref{Fig_sizefield}.

\section{Error estimation for pair-correlation approximation}
\label{Append_error}

To estimate the error of the pair-correlation
approximation, we express the exact and the approximate nuclear spin wavefunctions as
\begin{subequations}
\begin{eqnarray}
 |{\mathcal J}^{\pm}_{\rm exact}(t)\rangle &= &|{\mathcal J}^{\pm}_{\rm exact, uncorr}\rangle
 +|{\mathcal J}^{\pm}_{\rm exact,corr}\rangle, \\
|{\mathcal J}^{\pm}_{\rm PCA}(t)\rangle & = & |{\mathcal J}^{\pm}_{\rm PCA, uncorr}\rangle
 +|{\mathcal J}^{\pm}_{\rm PCA,corr}\rangle,
\end{eqnarray}
\end{subequations}
respectively, where $|{\mathcal J}^{\pm}_{\rm exact,corr}\rangle$ and $|{\mathcal J}^{\pm}_{\rm PCA,corr}\rangle$
denote the wavefunctions containing correlated pair-flips, and $|{\mathcal J}^{\pm}_{\rm exact, uncorr}\rangle$ and
$|{\mathcal J}^{\pm}_{\rm PCA, uncorr}\rangle$ are the parts containing uncorrelated pair-flips.
The evolution starts with a randomly chosen nuclear spin state $|{\mathcal J}\rangle$.
The uncorrelated part of the wavefunction in the exact solution and that in the approximation are determined by the same
Green's function $\hat{G}_0(t,t')$ and by the probability amplitudes at the initial state
$C_{{\rm PCA},{\mathcal J}}^{\pm}(t')$ and  $C_{{\rm exact},{\mathcal J}}^{\pm}(t')$ as
\begin{subequations}
\begin{eqnarray}
 |{\mathcal J}^{\pm}_{\rm exact, uncorr}\rangle &= & \int \hat{G}_0(t,t')C_{{\rm exact},{\mathcal J}}^{\pm}(t') |{\mathcal J}\rangle dt', \\
|{\mathcal J}^{\pm}_{\rm PCA, uncorr}\rangle &= & \int \hat{G}_0(t,t')C_{{\rm PCA},{\mathcal J}}^{\pm}(t') |{\mathcal J}\rangle dt'. \ \ \ \
\end{eqnarray}
\end{subequations}
After a few pair-flips have occurred while still well before the correlated part of the wavefunction
becomes significant, the probability amplitude of the initial state $|{\mathcal J}\rangle$ decays already to zero.
So the amplitudes at the initial state are essentially the same in the exact and the approximate solutions.
Consequently, the wavefunctions containing only uncorrelated pair-flips are equal in the two solutions
\begin{subequations}
\begin{eqnarray}
 |{\mathcal J}^{\pm}_{\rm exact, uncorr}\rangle = |{\mathcal J}^{\pm}_{\rm PCA, uncorr}\rangle.
\end{eqnarray}
\end{subequations}
Thus the error of the pair-correlation approximation in calculating the electron spin
coherence correlation is estimated to be
\begin{eqnarray}
\delta{\mathcal L}_{+,-}^{\rm s} &\sim &  \left\langle {\mathcal J}^{-}_{\rm exact,corr}
\left| {\mathcal J}^{+}_{\rm exact,corr}\right\rangle\right. \nonumber \\
&& - \left\langle {\mathcal J}^{-}_{\rm PCA,corr} \left| {\mathcal J}^{+}_{\rm PCA,corr}\right\rangle\right.
\le  P_{\rm corr},
\end{eqnarray}
where $P_{\rm corr}$ is the probability of having correlated pair-flips. When the number of
pair-flips $N_{\rm flip}\ll N$, $P_{\rm corr}\sim 1-\exp\left(-qN^2_{\rm flip}/N\right)$
($q$ is a factor close to the number of nearest neighbors of a nuclear spin).

\end{appendix}


\begin{thebibliography}{81}
\bibitem[{Not({\natexlab{a}})}]{Note_on_product}
\bibinfo{note}{It is also interesting to consider the electorn spin decoherence
  and the nuclear spin dynamics for a system with certain initial
  electron-nuclear entanglement, which will be covered in our future
  publications.}
\bibitem[{Not({\natexlab{b}})}]{Note_on_InitialState}
\bibinfo{note}{The eletron spin being in a mixed state means that it is
  entangled with its ``environment''. Sicnce we have assumed the electron and
  the nuclei are initially in a produc state, the mixedness must come from
  other sources, such as the photon noise in optical pump and control of the
  electron spin.}
\bibitem[{Not({\natexlab{c}})}]{Note_on_StrongFieldAssumption}
\bibinfo{note}{In low fields, it may be possible to treat the off-diagonal
  hyperfine interaction as elementary excitations within the pair-correlation
  approximation as well as the nuclear spin pair-flips are. This
  generalization, however, is beyond the scope of this paper.}
\bibitem[{Not({\natexlab{d}})}]{Note_AB}
\bibinfo{note}{The boundary between the two groups is not exact but this is not
  important since the pair-flips around the boundary are much weaker in the
  strength of the transition matrix elements than those in ${\mathbb{G}}_B$ and
  are much fewer in number than in ${\mathbb{G}}_A$.}
\bibitem[{Not({\natexlab{e}})}]{Note_excitationspectrum}
\bibinfo{note}{The definition of the excitation spectrum at vanishing
  frequency, however, is tricky as the density of states at $\epsilon=0$ can be
  made arbitrarily large by enclosing nuclear spins out of the range of the
  electron wavefunction, which contribute to local pair-flips costing zero
  hyperfine energy ($E_k=0$) and inducing no decoherence. To remove such
  indefiniteness, we indeed define the regularized excitation spectra to be $$
  {\tilde S}_{A/B}(\epsilon)\equiv \frac{
  \int^{\epsilon+\epsilon_0}_{\epsilon-\epsilon_0} \omega^2 {S}_{A/B}(\omega)
  d\omega} {\int^{\epsilon+\epsilon_0}_{\epsilon-\epsilon_0}\omega^2 d\omega} ,
  $$ where $\epsilon_0$ is chosen such that $t^{-1}\gg \epsilon_0 > B_k$.}
\bibitem[{Not({\natexlab{f}})}]{Note_disentanglement}
\bibinfo{note}{A pseudo-spin is disentangled from the electron spin after a
  full precession cycle in free evolution. But the pseudo-field strength
  $\chi^{\pm}_k$ is different for different pseudo-spins, so is the
  disentanglement time, making it impossible for the recoherence to occur in
  QDs of many nuclear spins without applying a controlling pulse. For small
  QDs, however, spontaneous recovery of coherence in FID may be observed.}
\end{thebibliography}

\end{document}